%% file: paper.tex
\documentclass[acmsmall,screen,nonacm]{acmart}

\input{preamble}
\languagecoloringfalse

\begin{document}

\title
{\texorpdfstring{\Ziv*}{Ziv}: A Language for Correctly Rounded Math Libraries}

\author{Lucas Yunkyu Lee}
\email{lucaslee@postech.ac.kr}
\affiliation{%
  \institution{POSTECH}
  \city{Pohang}
  \country{South Korea}
}

\author{Sungjin Hwang}
\email{sungjin.hwang@postech.ac.kr}
\affiliation{%
  \institution{POSTECH}
  \city{Pohang}
  \country{South Korea}
}

\author{Wonyeol Lee}
\email{wonyeol.lee@postech.ac.kr}
\affiliation{%
  \institution{POSTECH}
  \city{Pohang}
  \country{South Korea}
}

\authorsaddresses{Authors' Contact Information:
  Lucas Yunkyu Lee, \href{mailto:lucaslee@postech.ac.kr}{lucaslee@postech.ac.kr};
  Sungjin Hwang, \href{mailto:sungjin.hwang@postech.ac.kr}{sungjin.hwang@postech.ac.kr};
  Wonyeol Lee, \href{mailto:wonyeol.lee@postech.ac.kr}{wonyeol.lee@postech.ac.kr}.
  All authors are affiliated with POSTECH in Pohang, South Korea.
}

\input{sections/sec00-abstract}

\makeatletter
\patchcmd{\@mkauthors@i}
{\medskip}{\vspace{2pt}}{}{} %
\patchcmd{\@mkauthors@i}
{\medskip}{\vspace{3pt}}{}{} %
\makeatother

\setcitestyle{citesep={,\kern-0.1em}}

\vspace*{-24pt} %
\maketitle
\enlargethispage*{1.0\baselineskip} %

\input{sections/sec01-intro}            %
\clearpage
\include{sections/sec02-overview}         %
\include{sections/sec03-language}         %
\include{sections/sec04-stdlib}           %
\languagecoloringtrue
\include{sections/sec05-compiler}         %
\languagecoloringfalse
\input{sections/sec06-implementation}   %
\input{sections/sec07-evaluation}       %
\input{sections/sec08-related}          %
\input{sections/sec09-ai-data}
\bibliographystyle{ACM-Reference-Format}
\bibliography{paper}
\clearpage
\appendix
\makeatletter
\patchcmd{\paragraph}{\@parfont}{\bfseries\itshape}{}{}
\let\ACM@origparagraph\paragraph
\makeatother
\input{appendices/app-lang-def-ziv}
\clearpage
\input{appendices/app-lang-def-irc0}
\clearpage \languagecoloringtrue
\input{appendices/app-comp-def}

\clearpage
\input{appendices/app-lang-res}
\clearpage
\input{appendices/app-comp-res}
\clearpage \languagecoloringfalse

\end{document}

%% file: preamble.tex
\usepackage{mathpartir}
\usepackage{empheq}
\usepackage[inline]{enumitem}
\usepackage{adjustbox}
\usepackage{stmaryrd}
\usepackage{afterpage}
\usepackage{lipsum}
\usepackage{array}
\usepackage{booktabs}
\usepackage{tabularray}
\UseTblrLibrary{booktabs}
\usepackage{comment}
\usepackage{graphicx}
\usepackage{multicol}
\usepackage{multirow}
\usepackage{siunitx}
\usepackage[textsize=tiny]{todonotes}
\usepackage{mathtools}
\usepackage{doi}
\usepackage{url}
\usepackage[subrefformat=parens]{subcaption}
\usepackage{accents} %

\allowdisplaybreaks
\hypersetup{bookmarksdepth=3}
\newcommand{\codepathcapital}[1]{%
  \appto{\UrlSpecials}{\do#1{\penalty0\hbox{#1}}}%
}
\DeclareUrlCommand\codepath{%
  \urlstyle{tt}%
  \forcsvlist{\codepathcapital}{%
    A,B,C,D,E,F,G,H,I,J,K,L,M,
    N,O,P,Q,R,S,T,U,V,W,X,Y,Z}%
}

\numberwithin{equation}{section}

\makeatletter
\let\ACM@origparagraph\paragraph
\makeatother

\usepackage{listings}
\definecolor{codegray}{rgb}{0,0.6,0}
\definecolor{codegreen}{rgb}{0,0.6,0}
\lstdefinestyle{CC}{
  language=C,
  basicstyle=\ttfamily\scriptsize,
  numbers=left,
  numbersep=2.25pt,
  xleftmargin=8pt,
  framesep=1pt,
  columns=fixed,
  mathescape=true,
  escapechar={@},
  commentstyle=\color{codegray},
}

\usepackage{tikz}
\usepackage{tikz-qtree}
\usetikzlibrary{arrows}
\usetikzlibrary{arrows.meta}
\usetikzlibrary{backgrounds}
\usetikzlibrary{calc}
\usetikzlibrary{decorations.pathreplacing}
\usetikzlibrary{intersections}
\usetikzlibrary{fit}
\usetikzlibrary{patterns}
\usetikzlibrary{positioning}
\usetikzlibrary{shapes}
\usetikzlibrary{shadows}
\usetikzlibrary{shapes.geometric}
\usetikzlibrary{tikzmark}

\usepackage{zref-clever}
\zcsetup{nameinlink=false,abbrev=true}
\zcLanguageSetup{english}{cap,rangesep=\textendash}
\zcRefTypeSetup{equation}{
  nocap,
  Name-sg=Eq.,
  Name-pl=Eqs.,
  name-sg=,
  name-pl=,
}

\zcRefTypeSetup{algorithm}{Name-sg-ab=Alg.,name-sg-ab=Alg.}
\usepackage{zref-sec}

\zcRefTypeSetup{appendix}{
  nocap,
  nocap,
  Name-sg = {\zcsectionmarkcapTrue Appendix~},
  Name-pl = {\zcsectionmarkcapTrue Appendices~},
  name-sg = {\zcsectionmarkcapTrue Appendix~},
  name-pl = {\zcsectionmarkcapTrue Appendices~},
  namesep = {},
  refbounds = {,\zcsectionprefix,,},
  +refbounds-rb = {,\zcsectionrangeprefix,,},
  +refbounds-re = {,,,},
  pairsep = { and~},
  listsep = {, },
  lastsep = {, and~},
  rangesep = {\textendash},
}

\zcRefTypeSetup{limitation}{
  name-sg={}, name-pl={},
  Name-sg={}, Name-pl={},
  refbounds={}{}{}{}
}
\zcRefTypeSetup{drawback}{
  name-sg={}, name-pl={},
  Name-sg={}, Name-pl={},
  refbounds={}{}{}{}
}

\zcRefTypeSetup{table}{
  name-sg-ab = tab.,
  name-pl-ab = tabs.,
  Name-sg-ab = Tab.,
  Name-pl-ab = Tabs.
}

\NewDocumentCommand{\cref}{m}{\zcref{#1}}

\NewDocumentCommand{\Cref}{m}{\zcref[S]{#1}}

\DeclareSIUnit{\ulp}{ulp}

\usepackage{newfloat}
\usepackage[newfloat]{minted}

\makeatletter
\ExplSyntaxOn
\regex_const:Nn \c__ziv_split_hex_float_regex
{
  (?:
  \c{PYG}\{l\+m\+mh\}\{((?:\c{PYGZhy}\{\})?0[xX][0-9a-fA-F]+)\}
  |
  \c{PYG}\{l\+m\+mi\}\{(\c{PYGZhy}\{\}0)\}\c{PYG}\{n\}\{([xX])\}
  )
  (?:\c{PYG}\{(?:l\+m\+mf|p)\}\{(\.[0-9a-fA-F]*)\})?
  \c{PYG}\{n\}\{([0-9a-fA-F]*[pP])
  (?:
  ([0-9]+[fFlL]?)\}
  |
  \}\c{PYG}\{o\}\{(\+)\}\c{PYG}\{l\+m\+m[if]\}\{([0-9]+[fFlL]?)\}
  |
  \}\c{PYG}\{l\+m\+m[if]\}\{(\c{PYGZhy}\{\}[0-9]+[fFlL]?)\}
  )
}
\cs_new_protected:Npn \ZivFormatCHexFloats
{
  \cs_set_eq:NN \__ziv_original_c_line: \FV@@PreProcessLine
  \cs_set_protected:Npn \FV@@PreProcessLine
  {
    \regex_replace_all:NnN \c__ziv_split_hex_float_regex
    {\c{PYG}\cB\{l+m+mf\cE\}\cB\{\1\2\3\4\5\6\7\8\9\cE\}}
    \FV@Line
    \__ziv_original_c_line:
  }
}
\ExplSyntaxOff
\makeatother
\setminted[c]{formatcom=\ZivFormatCHexFloats}

\zcsetup{countertype={FancyVerbLine=line}}
\NewDocumentCommand{\codelinenumber}{m}{%
  \begingroup
  \zcsetup{currentcounter=FancyVerbLine}%
  \renewcommand{\theFancyVerbLine}{\arabic{FancyVerbLine}}%
  \raisebox{\baselineskip}[0pt][0pt]{%
    \phantomsection\label{#1-\arabic{FancyVerbLine}}}%
  \arabic{FancyVerbLine}%
  \endgroup
}
\SetupFloatingEnvironment{listing}{name=Listing}

\usepackage{tcolorbox}
\tcbuselibrary{breakable,skins,minted,raster}
\NewDocumentEnvironment{codes}{ m O{} }{%
  \begin{tcbraster}[
    raster columns=#1,
    raster equal height=rows,
    raster column skip=3pt,
    raster before skip=4pt,
    raster after skip=4pt,
    before skip=0pt,after skip=0pt,
    breakable=false,
    #2
  ]%
}{\end{tcbraster}\prevdepth=.3\baselineskip}
\tcbset{codebox/.style={
      enhanced jigsaw,
      sharp corners,
      boxrule=0pt,
      leftrule=0.5mm,
      colframe=black!55,
      colback=black!0, %
      frame hidden, %
      borderline west={0.5mm}{0pt}{black!55},
      boxsep=0pt,
      left=4pt,right=4pt,top=2pt,bottom=2pt,
      before skip=4pt,after skip balanced=4pt,
    }}
\NewTCBListing{code}{ !O{text} !O{} !O{} !O{} }{%
  codebox,
  listing only,
  listing engine=minted,
  minted language={#1},
  minted options={fontsize=\scriptsize,breaklines,autogobble,#4},
  colbacktitle=black!10,
  coltitle=black!80,
  fonttitle=\scriptsize\bfseries,
  toptitle=2pt,bottomtitle=2pt,
  title={#2},
  #3
}
\zcsetup{
  countertype    = { sublisting = listing },
  counterresetby = { sublisting = listing },
}

\usepackage{algorithm}
\usepackage[indLines=true,noEnd=true,commentColor=darkgray,endLComment={}]{algpseudocodex}
\makeatletter\AddToHook{env/algorithmic/begin}{\zcsetup{currentcounter=ALG@line}}\makeatother
\zcsetup{countertype={ALG@line=line}}
\zcRefTypeSetup{line}{nocap}

\algrenewcommand\algorithmicrequire{\textbf{Input:}}
\algrenewcommand\algorithmicensure{\textbf{Output:}}
\algrenewcommand\algorithmicthen{}

\usepackage{keytheorems}
\newkeytheorem{theorem}[style=acmplain,numberwithin=section] %
\newkeytheorem{lemma}[style=acmplain,sibling=theorem] %
\newkeytheorem{proposition}[style=acmplain,sibling=theorem] %
\newkeytheorem{corollary}[style=acmplain,sibling=theorem] %
\newkeytheorem{definition}[style=acmdefinition,sibling=theorem] %
\newkeytheorem{remark}[style=acmdefinition,sibling=theorem] %
\newkeytheorem{example}[style=acmdefinition,sibling=theorem] %
\newkeytheorem{assumption}[style=acmdefinition,numberwithin=section] %

\DeclareCaptionLabelFormat{continued}{Fig.~#2~(continued)}

\zcRefTypeSetup{continuedfigure}{
  Name-sg = Figure,
  Name-pl = Figures,
  name-sg = Fig.,
  name-pl = Figs.,
  refbounds = {,,~(cont.),},
}

\makeatletter
\define@key{mpr}{right}
{\setbox0\hbox{$\,\TirName{#1}$}\relax
  \advance\hsize by -\wd0
  \setbox\mpr@right\hbox{\unhbox\mpr@right\unhbox0}}
\define@key{mpr}{Right}
{\setbox\mpr@right\hbox{
    \unhbox\mpr@right\rlap{$\,\TirName{#1}$}}}
\makeatother

\newcommand{\tcr}[1]{{#1}}

\newcommand{\msmall}{\fontsize{9.5pt}{11.4pt}\selectfont}

\DeclarePairedDelimiter{\abs}{\lvert}{\rvert}

\NewDocumentCommand{\e}{}{e}

\newlength{\equationcodebaseline}
\newcommand{\equationcodestrut}{%
  \rule[-.3\equationcodebaseline]{0pt}{\equationcodebaseline}}

\makeatletter
\NewDocumentEnvironment{equationcode}{O{} +b}{%
  \begin{tcolorbox}[codebox,colback=black!4,#1]
    \setlength{\equationcodebaseline}{.8\baselineskip}%
    \setlength{\jot}{0pt}%
    \setlength{\lineskip}{0pt}%
    \let\strut@\equationcodestrut
    \noindent\small\(\begin{aligned}#2\end{aligned}\)\par
  \end{tcolorbox}%
}{\ignorespacesafterend}
\makeatother

\NewDocumentEnvironment{equationblock*}{+b}{%
  \begin{array}[t]{@{}>{\displaystyle\equationcodestrut}l@{}l@{}}#1\end{array}%
}{}
\NewDocumentEnvironment{equationblock}{+b}{%
  \{\quad\begin{equationblock*}#1 & \quad\}\end{equationblock*}%
}{}

\makeatletter
\newcommand{\underbracecpt}[1]{%
  \mathop{\vtop{\m@th\ialign{##\crcr
        $\hfil\displaystyle{#1}\hfil$\crcr
        \noalign{\kern1\p@\nointerlineskip}\upbracefill\crcr
        \noalign{\kern1\p@}}}}\limits}
\makeatother

\NewDocumentCommand{\shortto}{}{%
  \mathbin{\clipbox{0.3em 0pt 0pt 0pt}{$\rightarrow$}}%
}

\newsavebox{\figdefsbox}
\newenvironment{figdefs}[1][3pt]{%
  \par
  \setlength{\fboxsep}{6pt}%
  \setlength{\fboxrule}{0.4pt}%
  \begin{lrbox}{\figdefsbox}%
  \begin{minipage}{\dimexpr\linewidth-2\fboxsep-2\fboxrule\relax}%
  \raggedright
  \setlength{\abovedisplayskip}{#1}%
  \setlength{\abovedisplayshortskip}{#1}%
  \setlength{\parskip}{0.5\baselineskip}%
}{%
  \end{minipage}%
  \end{lrbox}%
  \noindent\makebox[\linewidth][c]{%
    \fbox{\usebox{\figdefsbox}}%
  }\par
}
\newlength{\MyMathparLineskip}

\newif\iflanguagecoloring
\languagecoloringtrue

\definecolor{Zivbase}{RGB}{230,159,0}
\colorlet{Zivhighlight}{Zivbase!24}
\newcommand{\Zplain}[1]{#1}
\newcommand{\Zhlpadding}{1pt}
\newcommand{\Zhlbox}[1]{%
  \begingroup
  \setlength{\fboxsep}{\Zhlpadding}%
  \let\Zhl\Zplain
  \colorbox{Zivhighlight}{\hspace{0.3pt}#1\hspace{0.3pt}}%
  \endgroup}
\newcommand{\Zhlmath}[2]{\Zhlbox{$\mathsurround=0pt #1#2$}}
\DeclareRobustCommand{\Zhl}[1]{%
  \iflanguagecoloring
    \ifmmode
      \mathpalette\Zhlmath{#1\vphantom{f_8}}%
    \else
      \Zhlbox{#1\vphantom{$f_8$}}%
    \fi
  \else
    #1%
  \fi}
\newsavebox{\Zdisplaybox}
\newcommand{\Zrenderdisplay}{%
  \begingroup
  \setlength{\fboxsep}{1pt}%
  \iflanguagecoloring
    \rlap{\smash{\makebox[\wd\Zdisplaybox][c]{%
          \colorbox{Zivhighlight}{%
            \makebox[\linewidth][c]{\phantom{\usebox{\Zdisplaybox}}}}}}}%
  \fi
  \usebox{\Zdisplaybox}%
  \endgroup}
\newcommand{\Zalignbox}[1]{%
  \begingroup
  \sbox{\Zdisplaybox}{#1}%
  \Zrenderdisplay
  \endgroup}

\newenvironment{Zalign*}{%
  \let\Zhl\Zplain
  \empheq[box=\Zalignbox]{align*}%
}{\endempheq}
\newenvironment{Zgather*}{%
  \let\Zhl\Zplain
  \empheq[box=\Zalignbox]{gather*}%
}{\endempheq}
\makeatletter
\newenvironment{Zmathpar}[1][]{%
  $$\mpr@savepar\parskip 0em\hsize\linewidth\centering
    \setbox\Zdisplaybox\vbox\bgroup
    \let\Zhl\Zplain
    \mpr@prebindings\mpr@paroptions #1%
    \ifmmode$\else\noindent$\displaystyle\fi
  \MathparBindings
}{%
  \unskip\ifmmode$\fi\egroup
  \Zrenderdisplay$$\ignorespacesafterend}
\makeatother

\definecolor{Eextbase}{RGB}{204,121,167}
\colorlet{Eexthighlight}{Eextbase!68}
\definecolor{Eextborder}{RGB}{112,54,94}
\newcommand{\Eextplain}[1]{#1}
\newcommand{\Eextinner}[1]{%
  \begingroup
  \setlength{\fboxsep}{1pt}%
  \colorbox{Eexthighlight}{\hspace{0.3pt}#1\hspace{0.3pt}}%
  \endgroup}
\newcommand{\Eextbox}[1]{%
  \begingroup
  \setlength{\fboxsep}{0.35pt}%
  \setlength{\fboxrule}{0.35pt}%
  \let\Eext\Eextplain
  \fcolorbox{Eextborder}{white}{\Eextinner{#1}}%
  \endgroup}
\newcommand{\Eextmath}[2]{\Eextbox{$\mathsurround=0pt #1#2$}}
\DeclareRobustCommand{\Eext}[1]{%
  \iflanguagecoloring
    \ifmmode
      \mathpalette\Eextmath{#1}%
    \else
      \Eextbox{#1}%
    \fi
  \else
    #1%
  \fi}

\definecolor{IRbase}{RGB}{0,114,178}
\colorlet{IRhighlight}{IRbase!22}
\newcommand{\Iplain}[1]{#1}
\newcommand{\Ihlbox}[1]{%
  \begingroup
  \setlength{\fboxsep}{1pt}%
  \let\Ihl\Iplain
  \colorbox{IRhighlight}{\hspace{0.3pt}#1\hspace{0.3pt}}%
  \endgroup}
\newcommand{\Ihlmath}[2]{\Ihlbox{$\mathsurround=0pt #1#2$}}
\DeclareRobustCommand{\Ihl}[1]{%
  \iflanguagecoloring
    \ifmmode
      \mathpalette\Ihlmath{#1\vphantom{f_8}}%
    \else
      \Ihlbox{#1\vphantom{$f_8$}}%
    \fi
  \else
    #1%
  \fi}
\newsavebox{\Idisplaybox}
\newcommand{\Irenderdisplay}{%
  \begingroup
  \setlength{\fboxsep}{1pt}%
  \iflanguagecoloring
    \rlap{\smash{\makebox[\wd\Idisplaybox][c]{%
          \colorbox{IRhighlight}{%
            \makebox[\linewidth][c]{\phantom{\usebox{\Idisplaybox}}}}}}}%
  \fi
  \usebox{\Idisplaybox}%
  \endgroup}
\newcommand{\Ialignbox}[1]{%
  \begingroup
  \sbox{\Idisplaybox}{#1}%
  \Irenderdisplay
  \endgroup}

\newenvironment{Ialign*}{%
  \let\Ihl\Iplain
  \empheq[box=\Ialignbox]{align*}%
}{\endempheq}
\newenvironment{Igather*}{%
  \let\Ihl\Iplain
  \empheq[box=\Ialignbox]{gather*}%
}{\endempheq}
\makeatletter
\newenvironment{Imathpar}[1][]{%
  $$\mpr@savepar\parskip 0em\hsize\linewidth\centering
    \setbox\Idisplaybox\vbox\bgroup
    \let\Ihl\Iplain%
    \mpr@prebindings\mpr@paroptions #1%
    \ifmmode$\else\noindent$\displaystyle\fi
  \MathparBindings
}{%
  \unskip\ifmmode$\fi\egroup
  \Irenderdisplay$$\ignorespacesafterend}
\makeatother

\definecolor{Cbase}{RGB}{230,159,0}
\colorlet{Chighlight}{white}
\newcommand{\Cplain}[1]{#1}
\newcommand{\Chlbox}[1]{%
  \begingroup
  \setlength{\fboxsep}{1pt}%
  \let\Chl\Cplain
  \colorbox{Chighlight}{\hspace{0.3pt}#1\hspace{0.3pt}}%
  \endgroup}

\DeclareRobustCommand{\Chl}[1]{%
  \iflanguagecoloring
    #1%
  \else
    #1%
  \fi}
\newsavebox{\Cdisplaybox}
\newcommand{\Crenderdisplay}{%
  \begingroup
  \setlength{\fboxsep}{1pt}%
  \iflanguagecoloring
    \rlap{\smash{\makebox[\wd\Cdisplaybox][c]{%
          \colorbox{Chighlight}{%
            \makebox[\linewidth][c]{\phantom{\usebox{\Cdisplaybox}}}}}}}%
  \fi
  \usebox{\Cdisplaybox}%
  \endgroup}
\newcommand{\Calignbox}[1]{%
  \begingroup
  \sbox{\Cdisplaybox}{#1}%
  \Crenderdisplay
  \endgroup}

\newenvironment{Calign*}{%
  \let\Chl\Cplain
  \empheq[box=\Calignbox]{align*}%
}{\endempheq}
\newenvironment{Cgather*}{%
  \let\Chl\Cplain
  \empheq[box=\Calignbox]{gather*}%
}{\endempheq}
\makeatletter
\newenvironment{Cmathpar}[1][]{%
  $$\mpr@savepar\parskip 0em\hsize\linewidth\centering
    \setbox\Cdisplaybox\vbox\bgroup
    \let\Chl\Cplain%
    \mpr@prebindings\mpr@paroptions #1%
    \ifmmode$\else\noindent$\displaystyle\fi
  \MathparBindings
}{%
  \unskip\ifmmode$\fi\egroup
  \Crenderdisplay$$\ignorespacesafterend}
\makeatother

\newcommand{\defas}{\coloneqq}
\newcommand{\asdef}{\eqqcolon}
\newcommand{\shortbar}{\,\raisebox{-0.10ex}{\rule{0.08em}{1.50ex}}\,}
\newcommand{\concat}{\mathbin{+\!\!+}}
\newcommand{\newbar}{%
  \text{\smash[b]{\scalebox{1.2}[0.85]{\trimbox{0pt .55ex}{\textup{\textminus}}}}}
}
\newcommand{\mybar}[1]{%
  \mathchoice
  {\accentset{\displaystyle\newbar}{#1}}
  {\accentset{\textstyle\newbar}{#1}}
  {\accentset{\scriptstyle\newbar}{#1}}
  {\accentset{\scriptscriptstyle\newbar}{#1}}
}
\renewcommand{\bar}[1]{\mybar{#1}}
\newcommand{\primtype}{\tau}
\newcommand{\comptype}{T}

\NewDocumentCommand{\Znorm}{s}{%
  \IfBooleanTF{#1}{%
    \textup{\textsf{\textsc{ZivNorm}}}%
  }{%
    \Zhl{\textup{\textsf{\textsc{ZivNorm}}}}%
  }%
}
\newcommand{\Zkw}[1]{\Zhl{\textup{\textsf{#1}}}}

\NewDocumentCommand{\Ziv}{s}{%
  \IfBooleanTF{#1}{%
    \textup{\textsf{\textsc{Ziv}}}%
  }{%
    \Zhl{\textup{\textsf{\textsc{Ziv}}}}%
  }%
}
\newcommand{\ZivStd}{\textsf{\textsc{\textit{ZivStd}}}}
\NewDocumentCommand{\Std}{m}{\textsf{\textit{#1}}}
\NewDocumentCommand{\StdFn}{m}{\mathop{\Std{#1}}}

\newcommand{\ZT}[1]{\Zhl{#1}}
\newcommand{\ZE}[1]{\Zhl{#1}}
\newcommand{\ZS}[1]{\Zhl{#1}}
\newcommand{\ZP}[1]{\Zhl{#1}}
\newcommand{\ZV}[1]{\Zhl{#1}}
\newcommand{\ZCtx}[1]{\Zhl{#1}}

\newcommand{\ZTreal}{\Zhl{\Zkw{Real}}}
\newcommand{\ZTfloat}[1]{\Zhl{\Zkw{Float}_{#1}}}
\newcommand{\ZTint}[1]{\Zhl{\Zkw{Int}_{#1}}}
\newcommand{\ZTuint}[1]{\Zhl{\Zkw{UInt}_{#1}}}
\newcommand{\ZTbool}{\Zhl{\Zkw{Bool}}}

\newcommand{\ZTdyadic}[2]{\Zhl{\Zkw{Dyadic}[{#1,#2}]}}
\newcommand{\ZTmulwrd}[1]{\Zhl{\Zkw{MultiWord}[{#1}]}}
\newcommand{\ZTtable}[2]{\Zhl{\Zkw{Table}[{#1,#2}]}}
\newcommand{\ZTtuple}[1]{\Zhl{\Zkw{Tuple}[{#1}]}}
\NewDocumentCommand{\ZTfun}{s m m}{
  \IfBooleanTF{#1}{
    {{#2}\,{\shortto}\,{#3}}
  }{
    \Zhl{{#2}\,{\shortto}\,{#3}}
  }
}

\NewDocumentCommand{\ZEop}{e{_}}{%
  \Zhl{\mathit{op}\IfValueT{#1}{[{#1}]}}}
\newcommand{\ZEnew}[2]{\Zhl{\Zkw{new}[{#1}](#2)}}
\newcommand{\ZEget}[1]{\Zhl{\Zkw{get}(#1)}}
\newcommand{\ZEtrue}{\Zkw{true}}
\newcommand{\ZEfalse}{\Zkw{false}}
\newcommand{\ZEval}[2]{\Zhl{\textup{\textsf{val}}_{#1}(#2)}}

\newcommand{\ZSskip}{\Zhl{\varepsilon}}
\newcommand{\ZSassn}[2]{\Zhl{#1=#2}}
\newcommand{\ZSseq}[2]{\Zhl{#1;#2}}
\newcommand{\ZSret}[1]{\Zhl{\Zkw{return}\ #1}}
\newcommand{\ZSif}[3]{\Zhl{\Zkw{if}\ #1\ \{#2\}\ \{#3\}}}
\newcommand{\ZSwhile}[2]{\Zhl{\Zkw{while}\ #1\ \{#2\}}}

\newcommand{\ZPnil}{\Zhl{\varepsilon}}
\newcommand{\ZPseq}[2]{\Zhl{#1;\ #2}}
\newcommand{\ZPdef}[4]{\Zhl{\Zkw{def}\ #1(#2):#3\ \{#4\}}}

\NewDocumentCommand{\ZJtype}{o m m m}{%
  \Zhl{#2\IfNoValueTF{#1}{\vdash}{\vdash_{#1}}#3:#4}}
\NewDocumentCommand{\ZJeval}{o m m m}{%
  \Zhl{#2\vdash #3\Downarrow\IfValueT{#1}{^{#1}}#4}}

\newcommand{\ZJok}{\Zhl{\textup{\textsf{ok}}}}
\newcommand{\ZJret}{\Zhl{\textup{\textsf{ret}}}}

\NewDocumentCommand{\ZVal}{e{_}}{%
  {\mathit{Val}\IfValueT{#1}{_{#1}}}}
\newcommand{\ZVar}{\Zhl{\mathit{Var}}}
\newcommand{\ZOp}{\Zhl{\mathit{Op}}}
\newcommand{\ZFun}{{\mathit{Fun}}}
\newcommand{\Zsem}[1]{{\llbracket #1\rrbracket}} %
\newcommand{\Zdom}{\mathrm{dom}}
\newcommand{\Zerase}{\mathit{erase}}
\newcommand{\Zmerge}{\mathit{merge}}

\newcommand{\Zhasret}{\mathit{ret}} %
\newcommand{\Ztrans}{\mathcal{T}_{\mathrm{norm}}}
\newcommand{\Zfields}{\mathit{fields}}

\newcounter{infrule}
\zcRefTypeSetup{infrule}{
  nocap,
  Name-sg  = Rule,
  Name-pl  = Rules,
  name-sg  = rule,
  name-pl  = rules,
}

\newsavebox{\infrulebox}
\NewDocumentCommand{\rulerefbox}{m m m}{%
  \sbox{\infrulebox}{$\mathsurround=0pt\displaystyle#3$}%
  \renewcommand{\theinfrule}{\protect\mbox{\protect#1{#2}}}%
  \stepcounter{infrule}%
  \AddToHookNext{makelinktarget}{%
    \setlength{\normalbaselineskip}{\dimexpr\ht\infrulebox+2pt\relax}}%
  \refstepcounter{infrule}%
  \usebox{\infrulebox}%
}

\newcommand{\Zrule}[1]{\Zhl{\textsc{#1}}}
\NewDocumentCommand{\Zinfer}{O{} m m m}{%
  \rulerefbox{\Zrule}{#2}{%
    \inferrule*[right={\Zrule{#2}},#1]{#3}{#4}}}

\NewDocumentCommand{\IR}{s}{%
  \IfBooleanTF{#1}{%
    \textup{\textsf{\textsc{ZivIR}}}%
  }{%
    \Ihl{\textup{\textsf{\textsc{ZivIR}}}}%
  }%
}
\newcommand{\Ikw}[1]{\Ihl{\textup{\textsf{#1}}}}

\newcommand{\IT}[1]{\Ihl{#1}}
\newcommand{\IE}[1]{\Ihl{#1}}
\newcommand{\IS}[1]{\Ihl{#1}}
\newcommand{\IP}[1]{\Ihl{#1}}

\newcommand{\ITfloat}[1]{\Ihl{\Ikw{Float}_{#1}}}
\newcommand{\ITint}[1]{\Ihl{\Ikw{Int}_{#1}}}
\newcommand{\ITuint}[1]{\Ihl{\Ikw{UInt}_{#1}}}
\newcommand{\ITbool}{\Ihl{\Ikw{Bool}}}
\newcommand{\ITfun}[2]{\Ihl{{#1}\mspace{1.5mu}{\shortto}\mspace{1.5mu}{#2}}}

\newcommand{\IEval}[1]{\Ihl{#1}}
\newcommand{\IEvall}[2]{\Ihl{\textup{\textsf{val}}_{#1}(#2)}}
\newcommand{\IEvar}[1]{\Ihl{#1}}
\newcommand{\IEop}{\Ihl{\mathit{op}}}
\newcommand{\IEapp}[2]{\Ihl{#1(#2)}}
\newcommand{\IEnew}[1]{\Ihl{\Ikw{new}(#1)}} %
\newcommand{\IEget}[1]{\Ihl{\Ikw{get}(#1)}} %
\newcommand{\IEtrue}{\Ihl{\Ikw{true}}}
\newcommand{\IEfalse}{\Ihl{\Ikw{false}}}

\newcommand{\ISskip}{\Ihl{\varepsilon}}
\newcommand{\ISassn}[2]{\Ihl{#1=#2}}
\newcommand{\ISseq}[2]{\Ihl{#1;#2}}
\newcommand{\ISret}[1]{\Ihl{\Ikw{return}\ #1}}
\newcommand{\ISif}[4]{\Ihl{#1=\Ikw{if}\ #2\ \{#3\}\ \{#4\}}}

\newcommand{\IPdef}[4]{\Ihl{\Ikw{def}\ #1(#2):#3\ \{#4\}}}

\newcommand{\IJctx}[1]{%
  \if\relax\detokenize{#1}\relax\else\Ihl{#1}\fi}
\newcommand{\IJtype}[3]{\Ihl{#1\vdash #2:#3}}
\newcommand{\IJeval}[3]{\Ihl{#1\vdash #2\Downarrow #3}}
\newcommand{\IJok}{\Ihl{\textup{\textsf{ok}}}}

\NewDocumentCommand{\IVal}{o}{%
  {\mathit{Val}\IfValueT{#1}{_{#1}}}}
\newcommand{\Isem}[1]{\Ihl{\llbracket #1\rrbracket}}

\newcommand{\Irule}[1]{\Ihl{\textsc{#1}}}
\NewDocumentCommand{\Iinfer}{O{} m m m}{%
  \rulerefbox{\Irule}{#2}{%
    \Ihl{\inferrule*[right={\Irule{#2}},#1]{#3}{#4}}}}

\NewDocumentCommand{\Czero}{s}{%
  \IfBooleanTF{#1}{%
    \ensuremath{\textup{\textsf{\textsc{C}}}_\textup{\textsf{0}}}%
  }{%
    \Chl{\ensuremath{\textup{\textsf{\textsc{C}}}_\textup{\textsf{0}}}}%
  }%
}
\newcommand{\Ckw}[1]{\Chl{\textup{\textsf{#1}}}}
\newcommand{\CE}[1]{\Chl{#1}}
\newcommand{\CS}[1]{\Chl{#1}}
\newcommand{\CP}[1]{\Chl{#1}}

\newcommand{\CTfloat}[1]{\Chl{\Ckw{Float}_{#1}}}
\newcommand{\CTint}[1]{\Chl{\Ckw{Int}_{#1}}}
\newcommand{\CTuint}[1]{\Chl{\Ckw{UInt}_{#1}}}
\newcommand{\CTbool}{\Chl{\Ckw{Bool}}}
\newcommand{\CTfun}[2]{\Chl{{#1}\mspace{1.5mu}{\shortto}\mspace{1.5mu}{#2}}}

\newcommand{\CEvar}[1]{\Chl{#1}}
\newcommand{\CEval}[2]{\Chl{\textup{\textsf{val}}_{#1}(#2)}}
\newcommand{\CEop}{\Chl{\mathit{op}}}
\newcommand{\CEapp}[2]{\Chl{#1(#2)}}
\newcommand{\CEget}[2]{\Chl{#1[#2]}}
\newcommand{\CEtrue}{\Chl{\Ckw{true}}}
\newcommand{\CEfalse}{\Chl{\Ckw{false}}}

\newcommand{\CSskip}{\Chl{\varepsilon}}
\newcommand{\CSdecl}[2]{\Chl{#1\ #2{;}}}

\newcommand{\CSassn}[2]{\Chl{#1=#2{;}}}
\newcommand{\CSnew}[4]{\Chl{#1\ #2[#3]=\{#4\}{;}}}
\newcommand{\CSseq}[2]{\Chl{#1\ #2}}
\newcommand{\CSret}[1]{\Chl{\Ckw{return}\ #1{;}}}
\newcommand{\CSif}[3]{\CS{\Ckw{if}\ (#1)\ \{#2\}\ \Ckw{else}\ \{#3\}}}

\newcommand{\CPdef}[4]{\Chl{#3\ #1(#2)\{#4\}}}

\newcommand{\CJctx}[1]{\Chl{#1}}
\newcommand{\CJtype}[3]{\Chl{#1\vdash #2:#3}}
\newcommand{\CJok}{\Chl{\textup{\textsf{ok}}}}
\newcommand{\CJeval}[3]{\Chl{#1\vdash #2\Downarrow #3}}

\NewDocumentCommand{\CVal}{e{_}}{%
  {\mathit{Val}\IfValueT{#1}{_{#1}}}}
\newcommand{\Csem}[1]{\Chl{\llbracket #1\rrbracket}}
\newcommand{\Ctrans}{\mathcal{T}_{\mathrm{emit}}}

\newcommand{\Cuninit}{\Chl{\textup{\textsf{undef}}}}

\newcommand{\Crule}[1]{\Chl{\textsc{#1}}}
\NewDocumentCommand{\Cinfer}{O{} m m m}{%
  \rulerefbox{\Crule}{#2}{%
    \Chl{\inferrule*[right={\Crule{#2}},#1]{#3}{#4}}}}

\newcommand{\Tkw}[1]{\textup{\textsf{#1}}}
\NewDocumentCommand{\TVal}{e{_}}{%
  {\widehat{\mathit{Val}}\IfValueT{#1}{_{#1}}}}
\newcommand{\Tbfields}{\mathit{pfields}}
\newcommand{\Tcon}[1]{\Tkw{ctval}\ifblank{#1}{}{(#1)}} %
\newcommand{\Tref}[2]{\Tkw{rtvar}(\Ihl{#2 : #1})} %
\newcommand{\Tcomp}[2]{\langle #2 \rangle_{#1}}
\newcommand{\Tmat}{\mathit{bind}}
\newcommand{\Tlookup}{\mathit{lookup}}
\newcommand{\Tjoin}{\mathit{merge}}
\newcommand{\Tenc}{\mathit{unwrap}}
\newcommand{\Tdec}{\mathit{wrap}}
\newcommand{\Trule}[1]{\textsc{#1}}
\NewDocumentCommand{\Tinfer}{O{} m m m}{%
  \rulerefbox{\Trule}{#2}{%
    \inferrule*[right={\Trule{#2}},#1]{#3}{#4}}}

\newcommand{\TTtable}[2]{{\Tkw{Table}[{#1,#2}]}}
\newcommand{\ZJcomp}[3]{#1\vdash\Zhl{#2}\leadsto #3}

\setlist[itemize,1]{wide,leftmargin=*}
\setlist[enumerate,1]{wide,leftmargin=*}

\makeatletter
\patchcmd{\NAT@test}{\else \NAT@nm }{\else \NAT@nmfmt{\NAT@nm}}{}{}
\makeatother
\setcitestyle{square,citesep={,},notesep={;\,}}

%% file: sections/sec00-abstract.tex
\begin{abstract}
  Developing performant, correctly rounded math libraries is challenging.
  Such implementations often rely on extended-precision arithmetic for correct rounding
  and precomputed constants for performance.
  Existing implementation languages, however, provide limited support for
  (i) high-level abstractions for such arithmetic and
  (ii) staged computation to combine precomputation and runtime computation in one program.
  As a result, existing implementations often miss abstraction opportunities
  and hard-code constants precomputed using separate programs.
  These practices make it difficult to write, understand, and modify these libraries.

  We present \Ziv{}, a programming system for developing correctly rounded math libraries.
  It consists of a small language, a rich standard library, and a provably sound compiler targeting C.
  \Ziv{} supports
  (i) high-level numerical abstractions %
  via types and operations for extended-precision numbers, and
  (ii) staged computation %
  via compile-time evaluation of expressions known to be input-independent.
  These features enable high-level code that unifies precomputation and runtime computation.
  We embed \Ziv{} in Julia
  and evaluate it by reimplementing 19 functions from two state-of-the-art libraries targeting correct rounding;
  we achieve competitive performance with $2.63\times$ reductions in lines of code.
  We also implement three new functions and automatically explore 824,256 variants,
  achieving $96.5\times$ average speedup over an arbitrary-precision library.
\end{abstract}

%% file: sections/sec01-intro.tex
\vspace{-1pt} %
\section{Introduction}
\label{sec:intro}

A \emph{math library} is software that implements widely used mathematical functions
such as exponential and trigonometric functions, %
and it is a core component of programming languages~\mbox{\cite{c-libm,java-libm,python-libm,julia-libm}}.
A~math library is \emph{correctly rounded}
if each function returns the exact mathematical result rounded to the target floating-point format~\cite{ieee-754-2019}:
for every library function $\mathtt{f}$ and floating-point input $\mathtt{x}$,
we have $\mathtt{f}(\mathtt{x}) = \mathit{rnd}(f(\mathtt{x}))$.
Here, $f$ is the exact mathematical function corresponding to $\mathtt{f}$,
and $\mathit{rnd}$ rounds real numbers to floating-point numbers (or \emph{floats})
under a given rounding mode.

Correct rounding is considered the gold standard for math libraries
and is becoming more important in both research~\cite{crsurvey} and practice~\cite{coremath,llvmlibc}.
By its definition, a correctly rounded library has two benefits:
it achieves the \emph{best accuracy} allowed by the target format and rounding mode;
and it is \emph{reproducible}, always returning the same output across platforms.
Reflecting these benefits,
the C standard recently reserved \texttt{cr\_*} identifiers (e.g., \texttt{cr\_exp}) %
for correctly rounded functions~\cite{c-std-2024}, %
and the next IEEE floating-point standard is expected to require correct rounding for core functions~\cite{ieee-754-2029}. %

Developing practical correctly rounded libraries, however, is generally challenging~\cite{fphandbook,elementaryfns}.
To be correctly rounded, implementations must be \emph{highly accurate};
this often requires extended-precision computations (e.g., multi-word arithmetic).
To be practical, implementations must be \emph{performant};
this often requires a range of optimizations (e.g., precomputing constants).
Each requirement complicates implementations, and meeting both makes them extremely complex.
For example, CORE-MATH~\cite{coremath} and LLVM-libc~\cite{llvmlibc}, two state-of-the-art libraries targeting correct rounding,
use {161K and 61K} lines of C and C++ code, respectively, to implement {193 and 117} functions. %

Our goal is to make the development of correctly rounded math libraries easier from a programming languages perspective.
To this end, we present the \Ziv*{} programming system.
We focus on functions over fixed-precision floating-point formats using 64 or more bits,
for which exhaustive testing~\cite{rlibm} is infeasible.
We now discuss key challenges (\cref{sec:intro:existing-approach})
and our approach (\cref{sec:intro:our-approach}).

\vspace{-1pt}
\subsection{Existing Approaches and Key Challenges}
\label{sec:intro:existing-approach}

\input{figs/fig-intro-code}

\bgroup\widowpenalty=0 %
\noindent
Existing math libraries targeting correct rounding \cite{coremath,llvmlibc,crlibm,libmcr,mathlib}
are typically written in C or C++ and share two implementation features.
First, high-level computations (e.g., fixed-point arithmetic)
are often expressed using low-level operations (e.g., integer arithmetic).
Second, numerical constants (e.g., table entries) are precomputed by separate programs
using external tools (e.g., Sage~\cite{sage}) and then hard-coded.
\cref{fig:intro-code:sota} illustrates these features using excerpts from state-of-the-art libraries.

These features reflect two main limitations of the implementation languages.
First, these languages have limited support for \emph{high-level numerical abstractions}:
types and operations that directly express extended-precision computations.
For example, \cref{fig:intro-code:sota} (top) performs basic arithmetic on 128-bit fixed-point numbers.
Due to the lack of built-in support for these numbers,
even these simple computations require a complex series of integer operations with seemingly arbitrary constants.
\egroup

Second, these languages have limited support for \emph{staged computation}:
evaluating some code at compile time and the rest at runtime.
For example, generating the constants in \cref{fig:intro-code:sota} (bottom)
requires computing real functions at high precision and using intricate polynomial approximation methods.
Since these computations are difficult to perform at compile time,
developers precompute such constants using separate programs (often not publicly available)
and hard-code the results.

These limitations lead to three difficulties, as illustrated by \cref{fig:intro-code:sota}.
First, \emph{writing} a new function implementation is difficult.
Expressing high-level computations in low-level code and precomputing and hard-coding constants
are labor-intensive and error-prone.
Second, \emph{understanding} an existing implementation is difficult.
Its high-level computations and mathematical definitions of constants
are often challenging and sometimes impossible to infer from the code alone.
Third, \emph{modifying} an existing implementation is difficult.
Even after the computations and definitions are understood,
a small high-level change can require extensive low-level revisions
and recomputation of constants.

\vspace{-1pt}
\subsection{Our Approach and Main Contributions}
\enlargethispage*{1.0\baselineskip} %
\label{sec:intro:our-approach}

We introduce \Ziv*{}, a programming system for developing correctly rounded math libraries.
\Ziv\ comprises a language, standard library, and compiler,
and is developed around two design principles:

\begin{enumerate}[label=\textbf{DP\arabic*.},ref=DP\arabic*,leftmargin=*,wide=0pt]
  \item\label{dp:abstraction} \textit{Provide high-level \textbf{abstractions}
          for techniques common in correctly rounded math libraries.}
  \item\label{dp:staging} \textit{Support \textbf{staging}
          to unify extended-precision precomputation and efficient runtime computation.}

\end{enumerate}

The \Ziv*{} language provides a deliberately small set of composable constructs,
from which we build \ZivStd{}, a rich standard library of numerical operations and algorithms.
The \Ziv*{} compiler translates programs involving these constructs and library functions into C code
via {partial evaluation and symbolic tracing}.
To establish the correctness of this translation, we formally define the language and compiler,
and prove that successful compilation preserves the semantics of \Ziv*{} programs.

\bgroup\widowpenalty=0 %
The \Ziv*{} language and \ZivStd{} library provide \emph{high-level numerical abstractions} (\ref{dp:abstraction}),
addressing the first limitation discussed in \cref{sec:intro:existing-approach}.
The \Ziv*{} language provides types for extended-precision numbers,
including exact reals and fixed-point numbers,
and the \ZivStd{} library provides operations for these types.
Together, these abstractions allow developers %
to express numerical computations directly without writing low-level code.
For example, \cref{fig:intro-code:ziv} (top) uses 128-bit fixed-point types
to represent \texttt{vj} and \texttt{Sj[k]},
and uses the fixed-point operations \texttt{+} and \texttt{*} to compute \texttt{vj}.
These operations internally handle the complicated integer arithmetic and bit shifts
shown in \cref{fig:intro-code:sota} (top).

The \Ziv*{} language and compiler support \emph{staged computation} (\ref{dp:staging}),
addressing the second limitation discussed in \cref{sec:intro:existing-approach}.
\Ziv*{}'s type system soundly tracks the dependencies of expressions on runtime inputs,
and the \Ziv*{} compiler evaluates at compile time all expressions known to be input-independent.
These features together allow developers to write both precomputation and runtime computation in a single \Ziv*{} program,
without requiring separate programs or hard-coded constants.
For example, \cref{fig:intro-code:ziv} (bottom) defines a table using exact-real and floating-point expressions
such as $\texttt{Float64}(\tan(i \pi/ 256))$.
The type system checks that this table is input-independent,
so the compiler automatically precomputes its entries and embeds them in the generated C code
(\cref{fig:intro-overview}, purple).
\egroup

To evaluate how \Ziv*{} addresses the difficulties identified in \cref{sec:intro:existing-approach}, we conduct three case studies.

\begin{enumerate}[leftmargin=*, labelindent=0pt, itemindent=0pt, label=\labelitemi,]
  \item
        We use \Ziv*{} to reimplement 19 representative functions from two state-of-the-art math libraries, CORE-MATH and LLVM-libc.
        Our implementations return the same outputs and achieve competitive performance,
        while reducing source lines of code by factors of 2.17 and 3.59, respectively.
  \item
        We use \Ziv*{} to implement three functions not supported by existing libraries targeting correct rounding
        (except arbitrary-precision libraries).
        We achieve correct rounding on $10^9$ random inputs
        with 22.4--696.8$\times$ speedups over MPFR, a state-of-the-art arbitrary-precision library.
  \item
        We use \Ziv*{} to automatically explore 824,256 variants of the three functions
        by varying working precision, table sizes, etc. %
        We find variants 1.6--2.4$\times$ faster than our hand-written implementations without sacrificing accuracy,
        as well as many less accurate but much faster variants.
\end{enumerate}

\noindent
These results show that \Ziv*{} is effective for expressing existing implementations concisely,
developing practical implementations for new functions, and exploring myriad implementation variants.

\input{figs/fig-intro-overview}

\paragraph{Contributions.}
In summary, this paper makes the following main contributions.
\begin{enumerate}[leftmargin=*, labelindent=0pt, itemindent=0pt,
    label=\labelitemi,
  ]
  \item \textbf{The \Ziv*{} Language (\cref{sec:overview,sec:lang}).}
        We present \Ziv*, a statically typed imperative language for developing correctly rounded math libraries.
        It provides high-level numerical abstractions (\ref{dp:abstraction})
        and supports type-guided staged computation (\ref{dp:staging}).
        It is deliberately small, yet highly expressive and extensible.
        We formalize its syntax, type system, and semantics, and prove type preservation.

  \item \textbf{The \ZivStd{} Library (\cref{sec:stdlib}).}
        We identify higher-level abstractions common in math libraries,
        including extended-precision arithmetic and its variants;
        range reduction and polynomial approximation/evaluation methods;
        and table and optimization functions (\ref{dp:abstraction}).
        We systematically organize and implement them in \ZivStd{},
        a rich standard library for math library development.

  \item \textbf{The \Ziv*{} Compiler (\cref{sec:compiler}).}
        We present a compiler that translates \Ziv* programs into C code
        via {partial evaluation and symbolic tracing}.
        We formalize compilation and prove that
        it preserves program semantics and precomputes all expressions known to be input-independent.
        This lets developers express both precomputation and runtime computation in a single program (\ref{dp:staging}).

  \item \textbf{Evaluation (\cref{sec:impl,sec:eval}).}
        We implement \Ziv*{} as an embedded language in Julia.
        We show that \Ziv*{} can express 19 functions from two state-of-the-art math libraries,
        while achieving 2.17--3.59$\times$ reductions in source lines of code.
        We also implement three new functions and systematically explore 824,256 variants,
        achieving 22.4–696.8$\times$ speedups over an arbitrary-precision library.
\end{enumerate}

\enlargethispage*{1.0\baselineskip} %

%% file: figs/fig-intro-code.tex
\begin{figure}[!t]
  \captionsetup[subfigure]{skip=3pt}  %
  \captionsetup{skip=1pt}             %
  \tcbset{figcode/.style={
      before skip=0pt,
      after skip=0pt,
      toptitle=1.0pt,
      bottomtitle=1.0pt,
      top=1.0pt,
      bottom=1.0pt,
  }}
  \setminted{baselinestretch=0.93}
  \begin{subfigure}[t]{0.485\textwidth}
    \vspace{0pt}
    \centering
    \begin{code}[c][LLVM-libc: \texttt{log}][figcode][]
      size_t i3 = ((int64_t)v2 + (0x284ull << 52)) >> 55;
      Int128 v3 = ((((Int128)S3[i3] << 55) + v2) << 21)
                   + (Int128)S3[i3] * v2;
      size_t i4 = ((int)(v3 >> 68) + 131) >> 1;
      Int128 v4 = ((((Int128)S4[i4] << 69) + v3) << 28)
                   + (Int128)S4[i4] * v3;
    \end{code}
    \begin{code}[c][CORE-MATH: \texttt{atan} \vphantom{\texttt{log}}][figcode][texcomments=true]
      const double TABLE[129][2] = { {0x0p+0, 0x0p+0},
        {0x1.9224e047e368ep-7, 0x1.a3ca6c727c59dp-62},
        {0x1.92346247a91f0p-6, 0x1.138b0ef96a186p-64},
        /* ... \textbf{126 more rows} ... */ };
      const double COEFFS[5] = {0x1p+0,
        -0x1.5555555555555p-2, 0x1.99999999998c1p-3,
        -0x1.249249176aec0p-3, 0x1.c711fd121ae80p-4};
    \end{code}
    \caption{Existing implementations.}
    \label{fig:intro-code:sota}
  \end{subfigure}
  \hfill
  \begin{subfigure}[t]{0.505\textwidth}
    \vspace{0pt}
    \centering
    \begin{code}[julia][\Ziv*{}: \texttt{log} from LLVM-libc][figcode][]
      i3 = round(Int, ldexp(v2, 21), RoundNearestTiesUp) + 80
      v3 = (S3[i3] + v2) + S3[i3] * v2
      i4 = round(Int, ldexp(v3, 28), RoundNearestTiesUp) + 65
      v4 = (S4[i4] + v3) + S4[i4] * v3
      # vj, Sj[k] have 128-bit fixed-point types.

    \end{code}
    \begin{code}[julia][\Ziv*{}: \texttt{atan} from CORE-MATH \vphantom{\texttt{log}}][figcode][escapeinside=||]
      TABLE = tabulate(i -> i==128 ? (0.0, 0.0) :
        let |$\theta$| = sym"$i*pi/256", t = Float64(tan(|$\theta$|));
            (t, Float64(atan(Sym(t))-|$\theta$|))
        end; shape=129)
      COEFFS = Float64.(gridremez(
        x -> x==0 ? 1 : atan(sqrt(x))/sqrt(x);
        degree=4, domain=(0, XMAX^2)))
    \end{code}
    \caption{Corresponding implementations in \Ziv*{}.}
    \label{fig:intro-code:ziv}
  \end{subfigure}
  \caption{Edited excerpts from state-of-the-art math libraries~\cite{llvmlibc,coremath} targeting correct rounding.}
  \label{fig:intro-code}
  \vspace{-8pt}
\end{figure}

%% file: figs/fig-intro-overview.tex
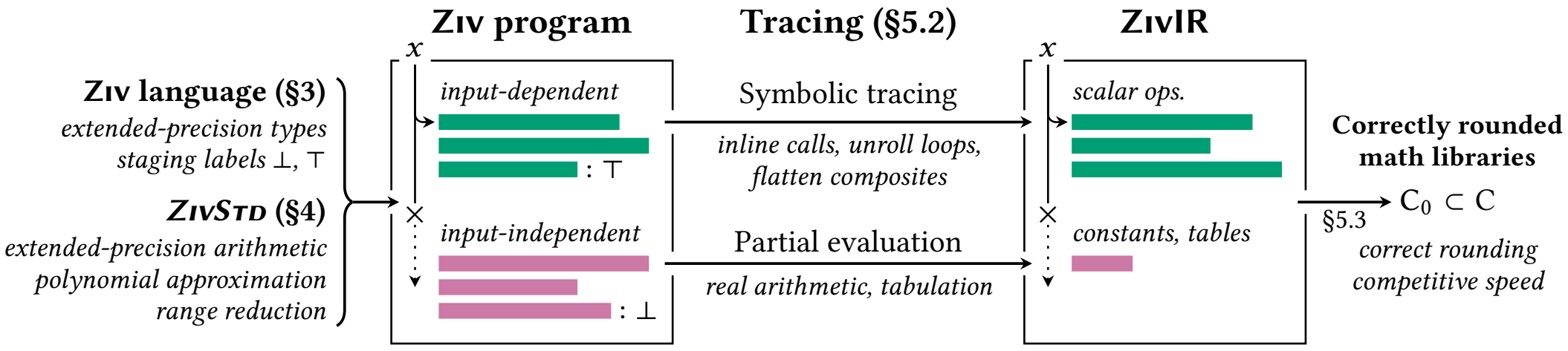
\begin{figure}[!t]
  \centering
  \hypersetup{hidelinks}
  \definecolor{overviewdependent}{RGB}{0,158,115}
  \definecolor{overviewindependent}{RGB}{204,121,167}
  \pgfmathsetlengthmacro{\overviewPanelWidth}{0.18*\linewidth}
  \pgfmathsetlengthmacro{\overviewPanelHeight}{0.125*\textheight}
  \pgfmathsetlengthmacro{\overviewTraceGap}{0.225*\linewidth}
  \pgfmathsetlengthmacro{\overviewFeaturesWidth}{0.21*\linewidth}
  \pgfmathsetlengthmacro{\overviewFeatureGap}{0.04*\linewidth}
  \pgfmathsetlengthmacro{\overviewOutputGap}{0.06*\linewidth}
  \def\overviewPadding{6pt}
  \def\overviewInputGutter{6pt}
  \def\overviewLineSkip{6pt}
  \def\overviewFlowThickness{0.7pt}
  \def\overviewFrameThickness{0.5pt}
  \pgfmathsetlengthmacro{\overviewCodeWidth}{
    \overviewPanelWidth - 2*\overviewPadding - \overviewInputGutter}

  \begin{tikzpicture}[
      font=\footnotesize\selectfont,
      every node/.style={inner sep=0pt, outer sep=0pt, align=center},
      heading/.style={font=\small\bfseries},
      detail/.style={font=\scriptsize\selectfont\itshape},
      region label/.style={detail, anchor=north west, align=left,
        text height=1.6ex, text depth=0.4ex},
      flow/.style={-{Stealth[length=3pt,width=3pt]}, line width=\overviewFlowThickness},
      connector/.style={flow, shorten <=3pt, shorten >=3pt},
      evaluation/.style={flow, preaction={draw=white, line width=6pt, -}},
      code frame/.style={line width=\overviewFrameThickness},
      code line/.style={line width=4pt, line cap=butt},
      pics/code rows/.style n args={2}{code={
        \foreach \width [count=\row from 0] in {#2} {
          \draw[code line,#1] (0,{-\row*\overviewLineSkip})
            -- ++({\width*\overviewCodeWidth},0) coordinate (-end-\row);
        }
      }},
      pics/code panel/.style n args={4}{code={
        \node[anchor=west,minimum width=\overviewPanelWidth,
          minimum height=\overviewPanelHeight] (-frame) {};
        \coordinate (-input) at ([xshift=\overviewPadding]-frame.north west);
        \coordinate (-code-left) at ([xshift=\overviewInputGutter]-input);
        \node[region label] (-dependent-label)
          at ([yshift=-3pt]-code-left) {#1};
        \node[region label] (-independent-label)
          at ([yshift=-3pt]-code-left |- -frame.center) {#2};
        \coordinate (-dependent) at
          ([yshift=-4pt]-code-left |- -dependent-label.south);
        \coordinate (-independent) at
          ([yshift=-4pt]-code-left |- -independent-label.south);
        \pic (-dependent) at (-dependent) {code rows={overviewdependent}{#3}};
        \pic (-independent) at (-independent) {code rows={overviewindependent}{#4}};
        \draw[code frame] ([xshift=-4pt]-input) -- (-frame.north west)
          -- (-frame.south west) -- (-frame.south east) -- (-frame.north east)
          -- ([xshift=4pt]-input);
        \node[anchor=south] at ([yshift=1pt]-input) {$x$};
        \draw[line width=\overviewFrameThickness] ([yshift=-1pt]-input) -- (-input |- -independent-label.north);
        \draw[flow, dotted, line width=\overviewFrameThickness] (-input |- -independent-label.north)
          -- ([yshift=-\overviewLineSkip]-input |- -independent);
        \draw[flow, line width=\overviewFrameThickness, rounded corners=2pt]
          ([yshift=-1pt]-input) |- ([xshift=-1pt]-dependent);
        \node[fill=white,inner sep=0.5pt]
          at (-input |- -independent-label.north) {$\times$};
      }},
    ]
    \node[detail,minimum width=\overviewFeaturesWidth,align=right] (features) {
      {\footnotesize\upshape\bfseries \Ziv*{} language (\cref{sec:lang})}\\[1pt]
      extended-precision types\\staging labels $\bot$, $\top$\\[5pt]
      {\footnotesize\upshape\bfseries \ZivStd{} (\cref{sec:stdlib})}\\[1pt]
      extended-precision arithmetic\\polynomial approximation\\range reduction};
    \draw[decorate,decoration={brace,amplitude=4pt},line width=\overviewFlowThickness]
      ([xshift=2pt]features.north east) -- ([xshift=2pt]features.south east);
    \coordinate (features-brace) at ([xshift=6pt]features.east);

    \pic (source) at ([xshift=\overviewFeatureGap]features.east)
      {code panel={input-dependent}{input-independent}{0.86,1,0.66}{1,0.66,0.82}};
    \node[heading,above=5pt of source-frame] (source-heading) {\Ziv*{} program};
    \node[right=2pt of source-dependent-end-2] {$:\top$};
    \node[right=2pt of source-independent-end-2] {$:\bot$};
    \draw[evaluation] (features-brace) -- ([xshift=2pt]source-frame.west);

    \pic (ir) at ([xshift=\overviewTraceGap]source-frame.east)
      {code panel={scalar ops.}{constants, tables}{0.86,0.66,1}{0.29}};
    \node[heading,anchor=base] at (ir-frame.center |- source-heading.base) {\IR*};

    \coordinate (tracing) at ($(source-frame.east)!0.5!(ir-frame.west)$);
    \node[heading,anchor=base] at (tracing |- source-heading.base)
      {Tracing (\cref{sec:compiler:trace})};
    \draw[evaluation] ([xshift=-2pt]source-frame.east |- source-dependent)
      -- ([xshift=2pt]ir-frame.west |- ir-dependent)
      node[midway,above=3pt] {Symbolic tracing}
      node[midway,below=3pt,detail] {inline calls, unroll loops,\\flatten composites};
    \draw[evaluation] ([xshift=-2pt]source-frame.east |- source-independent)
      -- ([xshift=2pt]ir-frame.west |- ir-independent)
      node[midway,above=3pt] {Partial evaluation}
      node[midway,below=3pt,detail] {real arithmetic, tabulation};

    \node[anchor=west] (c) at
      ([xshift=\overviewOutputGap]ir-frame.east)
      {$\Czero* \subset \mathrm{C}$};
    \draw[evaluation] ([xshift=-2pt]ir-frame.east) -- ([xshift=-2pt]c.west)
      node[midway,below=2pt,font=\scriptsize] {\cref{sec:compiler:emit}};
    \node[font=\scriptsize\bfseries,above=6pt of c] {Correctly rounded\\math libraries};
    \node[detail,below=6pt of c] {correct rounding\\ competitive speed};
  \end{tikzpicture}
  \caption{Overview of the \Ziv*{} programming system.}
  \label{fig:intro-overview}
  \vspace{-8pt} %
\end{figure}

%% file: sections/sec02-overview.tex
\section{Overview: Correctly Rounded Math Libraries and \texorpdfstring{\Ziv}{Ziv}}\label{sec:overview}

Implementing performant correctly rounded functions centers on
deciding \emph{how} (\zcref{sec:overview:how}), \emph{where} (\zcref{sec:overview:where}),
and \emph{when} (\zcref{sec:overview:when}) to use more precision.
We give an overview of how correctly rounded math libraries make these decisions
and how \Ziv\ simplifies their implementation,
using \Ziv's concrete syntax.

\subsection{How to Use More Precision: Extended-Precision Types}\label{sec:overview:how}

The first decision is \emph{how} to use more precision
than a machine float supports.
We focus on three representations that provide extended precision:
real, multi-word, and dyadic numbers.
\Ziv~supports these representations in the core language (\cref{sec:lang})
and operations on them in the standard library (\cref{sec:stdlib}).

\paragraph{Real Numbers}\label{sec:overview:how:real}

Math function implementations require high-precision constants,
e.g., for range reduction and polynomial evaluation (\cref{sec:overview:where}).
Computing these constants at runtime would be inefficient,
and precomputation with machine floats may not provide enough precision.
Instead, real arithmetic can be performed at compile time,
with real-valued results rounded to finite precision
and folded into the program for runtime use.
To support this,
\Ziv~provides a real type
with a restricted syntax that ensures all computations
on it are evaluated only at compile time (\cref{sec:lang:syntax}).
The language and standard library also support
real arithmetic and rounding to finite-precision types.

\paragraph{Multi-Word Numbers}\label{sec:overview:how:multiword}

\emph{Multi-word numbers}~\cite{doubleword}
comprise multiple machine floating-point words,
representing the unevaluated sum of these words.
Each successive word stores a correction too small
to be retained in the preceding words,
which would be lost if the sum was evaluated.
For example, evaluating $1 + 2^{-58}$ into a 64-bit float rounds to $1$,
but the pair $(1, 2^{-58})$ preserves all bits.

Arithmetic on multi-words uses \emph{error-free transformations}~\cite{fpsurvey},
operations that compute both a rounded result and its rounding error.
For example, given suitable float inputs $(a,b)$,
\emph{fast2sum}~\cite{doubleword,fast2sum} returns $(s,r)$ such that
$s = a \oplus b$ and $s+r=a+b$ exactly,
where $\oplus$ denotes floating-point addition.
The C code below implements double-word addition using two fast2sum operations.
\Ziv\ expresses the same intent more clearly,
using its double-word type \texttt{Float64x2}
and an infix \texttt{+} operator.

\begin{codes}{2}
  \begin{code}[c][Double-Word Addition \hfill C]
    double x_hi = 1.0,     x_lo = 0x1p-58;
    double y_hi = 0x1p-53, y_lo = 0.0;
    double s    = x_hi + y_hi, r = y_hi - (s - x_hi);
    double lo   = (x_lo + y_lo) + r;
    double z_hi = s + lo,  z_lo = lo - (z_hi - s);
  \end{code}
  \begin{code}[julia][Double-Word Addition \hfill \Ziv{}]
    x = Float64x2(1.0, 0x1p-58)
    y = Float64x2(0x1p-53, 0.0)
    z = x + y
  \end{code}
\end{codes}

\paragraph{Dyadic Numbers}\label{sec:overview:how:dyadic}

\emph{Floating-point dyadic numbers}~\cite{mpfr}
are stored as an integer triplet $(s, m, e)$,
each corresponding to the sign, significand, and exponent of the number.
The represented value is $(-1)^s \times m \times 2^e$.
For instance, using a 128-bit unsigned integer for $m$ provides 128 bits of precision.
Operations on floating-point dyadic numbers are implemented with integer arithmetic.

\emph{Fixed-point dyadic numbers} are also integer-based,
but the exponent $e$ is fixed by the type.
This makes the representation a pair $(s, m)$ of the sign and significand,
with value $(-1)^s \times m \times 2^e$.
In the fixed-point arithmetic example below,
C uses manual fixed-point alignment
and requires the programmer to track the exponent of each value.
Instead, \Ziv\ stores and tracks exponents in its \texttt{Dyadic} type,
and supports fixed-point arithmetic operators
as well as construction from floats.

\begin{codes}{2}
  \begin{code}[c][Fixed-Point Arithmetic \hfill C]
    uint128_t x = (uint128_t)3 << 15;        // * 2^-16
    uint128_t y = (uint128_t)1 << 58;        // * 2^-60
    uint128_t z = (x<<60) + (y<<16) + x * y; // * 2^-76
  \end{code}
  \begin{code}[julia][Fixed-Point Arithmetic \hfill \Ziv{}]
    x = Dyadic{128,Val{-16}}(1.5)
    y = Dyadic{128,Val{-60}}(0.25)
    z = x + y + x * y; @assert z isa Dyadic{128,Val{-76}}
  \end{code}
\end{codes}

\subsection{Where to Use More Precision: Implementation Steps}\label{sec:overview:where}

The next decision is \emph{where} to use more precision.
Implementations must use extended-precision types selectively
to balance accuracy and performance.
This decision arises throughout the three implementation steps of
correctly rounded math libraries:
\emph{range reduction}, \emph{polynomial evaluation}, and \emph{reconstruction}~\cite{crsurvey}.
We use one possible implementation of the exponential $\e^x$ as an example.

\paragraph{Range Reduction}\label{sec:overview:where:reduction}

As floats have large dynamic ranges,
inputs to functions may be extremely large,
making direct computation difficult.
\emph{Range reduction} addresses this problem by
utilizing mathematical identities
to bring the input into a small range
where better approximation is possible.

\NewDocumentCommand{\reduceConst}{}{\smash{\frac{\ln 2}{2^{12}}}}

For $e^x$,
we seek to reduce $x$ to some $r$ such that $\abs{r}$ is small, %
aiming to approximate $e^r$ instead of $e^x$ directly.
To achieve this, we choose an integer
$k$ such that $\reduceConst k$ is the nearest multiple
of $\reduceConst$ to $x$.
Then, defining $r$ as the remainder
$x - \reduceConst k$ gives the bound $\abs{r} \leq \smash{\frac{\ln 2}{2^{13}}}$.
In exact real arithmetic:
\[
  \e^x = 2^{k/2^{12}}\e^r
  \qquad\text{for }
  x = \tfrac{\ln 2}{2^{12}} k + r,
  \qquad\text{where }
  k=\mathrm{round}\big(x \big{/} \tfrac{\ln 2}{2^{12}}\big).
\]
The reduced exponential $\e^r$ can be approximated
with a polynomial on $\smash{\bigl[-\frac{\ln 2}{2^{13}}, \frac{\ln 2}{2^{13}}\bigr]}$,
and the reduced input $r$ can be accurately computed with extended-precision arithmetic,
including storing $\reduceConst$ as a multi-word.
However, directly evaluating $\smash{2^{k/2^{12}}}$ requires
computing an expensive fractional power.

Instead, \emph{table-based methods}~\cite{tablebased,tableexp} replace
runtime computations with lookups of precomputed values.
We rewrite $\smash{2^{k/2^{12}} = 2^i 2^{j/2^{12}}}$ where $k=2^{12}i+j$
for integers $i$ and $j$,
and tabulate the fractional power $T[j] \approx \smash{2^{j/2^{12}}}$.
The below example stores $T$ as double-words to ensure sufficient precision.
C requires the programmer to compute table values externally
and hard-code them as constants.
By contrast, the \Ziv\ code uses \texttt{tabulate},
which performs tabulation by applying a function to each index at compile time
(\cref{sec:stdlib:technique}).
Additionally, complete implementations of common range reduction algorithms are provided,
including the Cody-Waite~\cite{codywaite} and Payne-Hanek~\cite{paynehanek} methods (\zcref{sec:stdlib:technique}).

\begin{codes}{2}
  \begin{code}[c][Table-Based Range Reduction \hfill C][][escapeinside=||,texcomments=true]
    static const double TAB[4096][2] = {
      {0x1.0000000000000p+0, 0x0.0000000000000p+00},
      {0x1.000b175effdc7p+0, 0x1.ae8e38c59c72ap-54},
      /* ... \textbf{4094 more lines} ... */ };
    double a_hi = ldexp(TAB[j][0], i),  // reconstruction
           a_lo = ldexp(TAB[j][1], i);
  \end{code}
  \begin{code}[julia][Table-Based Range Reduction \hfill \Ziv{}][][escapeinside=||]
    TAB = tabulate(j -> Float64x2(sym"2^($j * 2^-12)"),
                   2^12)
    a = ldexp(TAB[j], i)  # reconstruction
  \end{code}
\end{codes}

\paragraph{Polynomial Evaluation}\label{sec:overview:where:poly}

After range reduction,
we approximate and evaluate the function on the reduced interval using a polynomial.
We use \emph{polynomial approximation} to refer only
to the precomputation of coefficients.
Methods of approximation include the Taylor and various minimax polynomials~\cite{sollya,fpminimax}.
Coefficients can be computed and stored with extended precision
as necessary.

The runtime computation using the precomputed coefficients
is known as \emph{polynomial evaluation}.
Below, we approximate $\e^r$ with its cubic Taylor polynomial
and evaluate it with Horner's method~\cite{taocp}, using double-word arithmetic.
In C, the polynomial is approximated with an external tool
and hard-coded into the program, and evaluation is done manually.
\Ziv\ performs all these steps in a single program.
Approximation with \texttt{taylor} and
coefficient rounding into \texttt{Float64x2} are done at compile time.
At runtime, the approximated polynomial is evaluated with \texttt{horner} (\zcref{sec:stdlib:technique}).

\begin{codes}{2}
  \begin{code}[c][Taylor Approximation and Evaluation \hfill C]
    static const double C[4][2] = {
      {1.0, 0.0}, {1.0, 0.0}, {0.5, 0.0},
      {0x1.5555555555555p-3, 0x1.5555555555555p-57} };
    double y_hi = C[3][0], y_lo = C[3][1], ry_hi, ry_lo;
    for (int k = 2; k >= 0; k--)
      dw_mul(r_hi,r_lo, y_hi,y_lo, &ry_hi,&ry_lo),
      dw_add(C[k][0],C[k][1], ry_hi,ry_lo, &y_hi,&y_lo);
  \end{code}
  \begin{code}[julia][Taylor Approximation and Evaluation \hfill \Ziv{}]
    C = Float64x2.(taylor(exp, 3, 0))
    y = horner(r, C)
  \end{code}
\end{codes}

\paragraph{Reconstruction}\label{sec:overview:where:recon}

The final \emph{reconstruction} step combines the results from
the preceding steps to produce the final output.
As a result, the necessary calculations follow naturally
from the range reduction scheme.
In the case of the exponential,
if $P(r)$ is a polynomial approximation of $\e^r$,
we can perform reconstruction as $\e^x \approx 2^i \otimes T[j] \otimes P(r)$,
using the table $T$ defined during range reduction.
This computation incurs rounding errors,
which can be reduced by using extended precision.

\subsection{When to Use More Precision: Precomputation and Ziv's Method}\label{sec:overview:when}

The last decision remaining is \emph{when} to use more precision.
By precomputation at compile time, higher precision can be used at no runtime cost.
Even at runtime, extra precision is not always required;
only a small number of cases determinable by Ziv's method necessitate
higher precision.

\paragraph{Precomputation}

\emph{Precomputation} allows input-independent values to be computed at
high precision at no runtime cost by inlining constants into programs.
For the exponential, precomputable values include
the constant $\reduceConst$,
the table entries $T[j]$, and the polynomial coefficients.

Unlike the traditional approach of hard-coded constants,
\Ziv\ integrates precomputation and
runtime code into a single program.
The \Ziv\ compiler evaluates input-independent code
and embeds its results in the compiled program,
leaving only input-dependent code for runtime (\cref{sec:compiler:trace}).

\paragraph{Ziv's Method}

While we have so far assumed that the precision for each computation is chosen in advance,
the precision required for correct rounding actually depends on the input.
\emph{Ziv's method}~\cite{zivs} uses an error bound to test whether
an approximation suffices for correct rounding on a given input.
This allows efficient implementations by using higher precision only when necessary,
dividing runtime execution into a \emph{fast path} and an \emph{accurate path}
depending on the input.
The fast path uses less precision and lower polynomial degrees,
and returns only when the rounding test passes.
Otherwise, the accurate path recomputes with more precision
and higher polynomial degrees.

\subsection{End-to-End Example: The Exponential Function}
\label{sec:overview:e2e}

We now combine the techniques of \zcref{sec:overview:how,sec:overview:where,sec:overview:when}
in an implementation of CORE-MATH's~\cite{coremath} exponential.
\Cref{fig:overview-exp} compares simplified versions of CORE-MATH's \texttt{exp},
in original C and in \Ziv.

\makeatletter
\let\cparagraph\paragraph
\patchcmd{\cparagraph}
{-.5\baselineskip \@plus -2\p@ \@minus -.2\p@}
{-1pt}
{}{\PackageError{paper}{Failed to patch cparagraph}{}}
\patchcmd{\cparagraph}
{\@parfont}
{\normalfont\itshape\bfseries}
{}{}
\makeatother

\vspace{4pt}
\cparagraph{\textbf{A}: Special Cases.}
First, special input cases are filtered to return immediately,
leaving the remaining inputs for the fast path.
C compares input bit patterns against hard-coded integers,
while \Ziv\ expresses these cases with predicates and functions.
The optimization functions \texttt{argmax} and \texttt{argmin}  (\zcref{sec:stdlib:fp}) are used
to compute underflow and overflow thresholds in \Ziv\ \zcref{exp-line-11,exp-line-12}.

\cparagraph{\textbf{B}: Range Reduction.}
The fast path reduces the input \texttt{x} to \texttt{dx}
using a table-based method similar to that of \cref{sec:overview:where}.
C again uses hard-coded constants and tables,
while \Ziv\ makes their definitions clear in code.
In \zcref{exp-line-18,exp-line-19},
the \Ziv\ code creates a real value with \texttt{sym}
and rounds it to a double-word at compile time.
Here, \Ziv's double-word types can hold non-machine types,
with the high part having the type \texttt{Dyadic\{29,Big\}},
a floating-point dyadic with 29 bits of precision.
The \texttt{tabulate} calls of \zcref[range]{exp-line-1,exp-line-5} also use
real expressions to generate lookup tables holding double-words.

\cparagraph{\textbf{C}: Polynomial Evaluation.}
The reduced argument is used to approximate $\e^{\texttt{dx}}$ as $1+\texttt{dx}\cdot\texttt{p}$,
where $\texttt{p}$ is the value of a polynomial at $\texttt{dx}$.
The C code evaluates externally approximated coefficients
with a hand-written Estrin's scheme~\cite{estrin}.
In \zcref[range]{exp-line-22,exp-line-24} of \Ziv,
both coefficient approximation with \texttt{gridremez}~\cite{remez}
and rounding to \texttt{Float64} occur at compile time,
yielding precisely the same cubic polynomial as the C version.
At runtime, \zcref{exp-line-25} evaluates the polynomial using \texttt{estrin} (\zcref{sec:stdlib:technique}).

\cparagraph{\textbf{D}: Reconstruction.}
To reconstruct the full exponential,
the evaluated polynomial is multiplied by
the tabulation factor \texttt{t} using double-word arithmetic.
Here, \texttt{t} is obtained by multiplying the
table elements \texttt{TAB1[idx1]} and \texttt{TAB2[idx2]}
with the double-word multiplication operator $\texttt{*}$ in \zcref{exp-line-27} of the \Ziv\ code.
Double-word multiplication by the evaluated polynomial in \zcref{exp-line-28} instead uses
the variant $\texttt{*}_{\texttt{h}}$,
which omits error-free transformations for higher speed at the cost of accuracy.

\cparagraph{\textbf{E}: Ziv's Method.}
At the end of the fast path,
the reconstructed result undergoes Ziv's rounding test.
For correct subnormal rounding, the test is applied separately depending on whether
the expected output is subnormal (\Ziv\ \zcref[range]{exp-line-33,exp-line-36})
or normal (\zcref[range]{exp-line-38,exp-line-40}).
If the test passes,
the result is correctly rounded and is returned.
Otherwise, \texttt{exp\_accurate} recomputes at higher precision in
\zcref{exp-line-36,exp-line-40},
restarting from range reduction.
We omit this accurate path from the figure.

\input{figs/fig-overview-exp.tex}

%% file: figs/fig-overview-exp.tex
\begin{figure}
  \definecolor{expBlue}{HTML}{0072B2}
  \definecolor{expOrange}{HTML}{E69F00}
  \definecolor{expGreen}{HTML}{009E73}
  \definecolor{expPurple}{HTML}{CC79A7}
  \definecolor{expVermilion}{HTML}{D55E00}

  \NewDocumentCommand{\ribbon}{m m m}{%
    \coordinate (ct) at ([yshift=6pt]pic cs:exp-c-#2-start);
    \coordinate (cb) at ([yshift=-2pt]pic cs:exp-c-#2-end);
    \coordinate (zt) at ([yshift=6pt]pic cs:exp-z-#2-start);
    \coordinate (zb) at ([yshift=-2pt]pic cs:exp-z-#2-end);
    \path[fill=#1!15]
    (frame.north west |- ct) -- (frame.north east |- ct)
    .. controls +(10pt,0) and +(-10pt,0) .. (zw |- zt)
    -- (ze |- zt) -- (ze |- zb) -- (zw |- zb)
    .. controls +(-10pt,0) and +(10pt,0) .. (frame.north east |- cb)
    -- (frame.north west |- cb) -- cycle;
    \draw[draw=#1,opacity=1,line width=2pt]
    (frame.north west |- ct) -- (frame.north west |- cb)
    node[midway,right,fill=none,inner sep=2pt,font=\footnotesize\bfseries,text=#1!50!black] {#3};
    \draw[draw=#1,opacity=1,line width=2pt]
    (ze |- zt) -- (ze |- zb)
    node[midway,left,fill=none,inner sep=2pt,font=\footnotesize\bfseries,text=#1!50!black] {#3};
  }

  \NewDocumentCommand{\ribbons}{}{%
    \iftikzmark{exp-z-zivs-end}{%
      \coordinate (zw) at ([xshift=20pt]frame.north east);
      \coordinate (ze) at ($(zw)+(frame.north east)-(frame.north west)$);
      \ribbon{expPurple}{special}{A}
      \ribbon{expBlue}{reduction}{B}
      \ribbon{expGreen}{poly}{C}
      \ribbon{expOrange}{recon}{D}
      \ribbon{expVermilion}{zivs}{E}
    }{}%
  }

  \tcbset{exp comparison/.style={
        opacityback=0,
        leftrule=0pt,
        tikz={remember picture},
        no borderline,
      }}

  \begin{codes}{2}[raster column skip=20pt]
    \renewcommand{\theFancyVerbLine}{\codelinenumber{exp-line}}
    \begin{code}[c][CORE-MATH (Original C, Simplified)][%
      exp comparison,underlay={\ribbons},
    ][%
      numbers=left,numbersep=8pt,escapeinside=||,texcomments=true,
    ]
      static const double TAB1[64][2] = {
        { 0x0.0000000000000p+00, 0x1.0000000000000p+0},
        {-0x1.19083535b085ep-56, 0x1.02c9a3e778061p+0},
        { 0x1.d73e2a475b466p-55, 0x1.059b0d3158574p+0},
        |\setcounter{FancyVerbLine}{65}|/* ... \textbf{61 more lines} ... */ };
      static const double TAB2[64][2] = {
        { 0x0.0000000000000p+00, 0x1.0000000000000p+0},
        { 0x1.ae8e38c59c72ap-54, 0x1.000b175effdc7p+0},
        {-0x1.7b5d0d58ea8f4p-58, 0x1.00162f3904052p+0},
        |\setcounter{FancyVerbLine}{130}|/* ... \textbf{61 more lines}... */ };

      double exp(double x) {
        |\tikzmark{exp-c-special-start}|union {double f; uint64_t u;} reinterp = {.f = x};
        uint64_t ax_u = reinterp.u & (~(uint64_t)0 >> 1);
        if (ax_u <= 0x3c90000000000000ull) return 1.0 + x;
        if (ax_u >= 0x40862e42fefa39f0ull) {
          if (ax_u > 0x7ff0000000000000ull) return x + x;
          if (ax_u == 0x7ff0000000000000ull) {
            if (reinterp.u >> 63) return 0.0;
            else return x; }
          if (!(reinterp.u >> 63)) {
            volatile double z = 0x1p1023;
            return z * z; }
          if (ax_u >= 0x40874910d52d3052ull) {
            volatile double z = 0x1p-1022;
            return z * z; }
        }|\tikzmark{exp-c-special-end}|

        |\tikzmark{exp-c-reduction-start}|double k_f = roundeven(x * 0x1.71547652b82fep+12);
        int64_t k = k_f, idxe = k >> 12;
        int64_t idx1 = (k >> 6) & 0x3f, idx2 = k & 0x3f;
        const double R_hi = 0x1.62e42ffp-13,
                     R_lo_neg = 0x1.718432a1b0e26p-47;
        double dx = (x - R_hi * k_f) + R_lo_neg * k_f;|\tikzmark{exp-c-reduction-end}|

        |\tikzmark{exp-c-poly-start}|double dx2 = dx * dx;
        static const double POLY[4] = {0x1p+0, 0x1p-1,
          0x1.55555557e54ffp-3, 0x1.55555553a12f4p-5};
        double p = (POLY[0] + dx * POLY[1]) +
          dx2 * (POLY[2] + dx * POLY[3]);|\tikzmark{exp-c-poly-end}|

        |\tikzmark{exp-c-recon-start}|double t1_hi=TAB1[idx1][1], t1_lo=TAB1[idx1][0];
        double t2_hi=TAB2[idx2][1], t2_lo=TAB2[idx2][0];
        double t_lo, t_hi = muldd(t1_hi, t1_lo,
          t2_hi, t2_lo, &t_lo);
        double f_hi = t_hi, f_lo = t_lo + t_hi * dx * p;|\tikzmark{exp-c-recon-end}|

        |\tikzmark{exp-c-zivs-start}|const double ERROR = 1.64e-19;
        if (reinterp.u > 0xc086232bdd7abcd2ull) {
          reinterp.u = (1 - idxe) << 52;
          double e;
          double g_hi = fasttwosum(reinterp.f, f_hi, &e),
                 g_lo = f_lo + e;
          double upper = g_hi + (g_lo + ERROR),
                 lower = g_hi + (g_lo - ERROR);
          if (upper != lower)
            return exp_accurate(x);
          f_hi = to_subnormal(lower);
        } else {
          double upper = f_hi + (f_lo + ERROR),
                 lower = f_hi + (f_lo - ERROR);
          if (upper != lower)
            return exp_accurate(x);
          f_hi = ldexp_fast(lower, idxe);
        }
        return f_hi;|\tikzmark{exp-c-zivs-end}|
      }
    \end{code}
    \begin{code}[julia][\Ziv*{} (Simplified)][%
      exp comparison,
    ][
      numbers=right,numbersep=8pt,escapeinside=||
    ]
      const F_53_52 = DoubleWord{Float64,Dyadic{52,Big}}
      const TAB1 = tabulate(i -> Float64x2(
        F_53_52(sym"2^($i * 2^-6)")), 64)
      const TAB2 = tabulate(i -> Float64x2(
        F_53_52(sym"2^($i * 2^-12)")), 64)

      @ziv function exp(x::Float64)
        |\tikzmark{exp-z-special-start}|isnan(x) && return x
        x == Inf && return Inf
        x == -Inf && return 0.
        x <= argmax(x->exp(x)==0, Float64) && return 0.
        x >= argmin(x->exp(x)==Inf, Float64) && return Inf
        abs(x) <= 0x1p-54 && return 1.0 + x|\tikzmark{exp-z-special-end}|

        |\tikzmark{exp-z-reduction-start}|k = round(Int64, x * Float64(sym"2^12 / log(2)"))
        k_f, idxe = Float64(k), k >> 12
        idx1, idx2 = (k>>6)&Int64(0x3f), k&Int64(0x3f)
        R = DoubleWord{Dyadic{29,Big},Float64}(
          sym"log(2) * 2^-12")
        dx = x - Float64(R.hi) * k_f - R.lo * k_f|\tikzmark{exp-z-reduction-end}|

        |\tikzmark{exp-z-poly-start}|POLY = Float64.(gridremez(exp, 4; 
          domain=log(2)/2^13 .* (-1, 1), error=:rel),
          RoundNearest)[2:end]
        p = estrin(dx, POLY)|\tikzmark{exp-z-poly-end}|

        |\tikzmark{exp-z-recon-start}|t = TAB1[idx1] * TAB2[idx2]
        f = t *|${}_\texttt{h}$| DoubleWord(1., dx * p)|\tikzmark{exp-z-recon-end}|

        |\tikzmark{exp-z-zivs-start}|ERROR = 1.64e-19
        if (x <= argmax(
          x -> exp(x) <= maxsubnormal(Float64), Float64))
          g = f + ldexp_fast(1.0, -1022 - idxe)
          lower, upper = g.hi .+ (g.lo ± ERROR)
          upper == lower && return to_subnormal(lower)
          return exp_accurate(x)
        end
        lower, upper = f.hi .+ (f.lo ± ERROR)
        upper == lower && return ldexp_fast(lower, idxe)
        return exp_accurate(x)|\tikzmark{exp-z-zivs-end}|
      end
    \end{code}
  \end{codes}%

  \caption{
    Exponential function of CORE-MATH~\cite{coremath} in original C and \Ziv, simplified.
    Includes code blocks
    \textcolor{expPurple!60!black}{\textbf{A}}:~special cases,
    \textcolor{expBlue!60!black}{\textbf{B}}: range reduction,
    \textcolor{expGreen!60!black}{\textbf{C}}: polynomial evaluation,
    \textcolor{expOrange!60!black}{\textbf{D}}: reconstruction,
    and
    \textcolor{expVermilion!60!black}{\textbf{E}}: Ziv's method.
  }\label{fig:overview-exp}
\end{figure}

%% file: sections/sec03-language.tex
\section{The \texorpdfstring{\Ziv*}{Ziv}~Language}
\label{sec:lang}

\input{figs/fig-ziv-syntax-frag}

We now formally introduce the \Ziv*{} language for developing correctly rounded math libraries.
\Ziv*{} is a statically typed imperative language
that supports extended-precision computation via high-level abstractions (\ref{dp:abstraction})
and staged computation via types (\ref{dp:staging}).
We keep \Ziv*{}'s core deliberately small,
including only the constructs we identify as necessary
for concisely implementing the techniques used
in correctly rounded math libraries (\zcref{sec:overview}).
Nevertheless, these constructs are expressive and composable enough
for us to develop a rich standard library (\cref{sec:stdlib})
and diverse math function implementations (\cref{sec:eval})
in \Ziv.
In this section, we present the syntax (\cref{sec:lang:syntax}),
type system (\cref{sec:lang:type}), and semantics (\cref{sec:lang:sem}) of \Ziv*{}.
All theorems in this paper are formalized in Lean {(see \cref{sec:lang-res,sec:comp-res}).} %

\vspace*{-3pt} %
\subsection{Syntax}
\label{sec:lang:syntax}

\enlargethispage*{1.0\baselineskip} %

\Cref{fig:ziv-syntax-frag} presents \Ziv*{}'s core syntax;
its full definition is given in \cref{sec:lang-def-ziv:syntax}.
We discuss the key constructs below.
We use $n$ to denote an integer
and $\bar{a}$ to denote a finite sequence $a_1, \ldots, a_k$ ($k \geq 0$).

\paragraph{Labels}

To support staging, \Ziv*{} distinguishes
\emph{input-independent} expressions, which do not depend on runtime inputs,
from \emph{potentially input-dependent} expressions, which may depend on runtime inputs.
A type \emph{label} $\ell \in \{\bot, \top\}$ records this distinction:
${\bot}$ indicates no input dependency, whereas ${\top}$ indicates possible input dependency.
Expressions with only $\bot$ labels in their types are always evaluated at compile time
(\cref{sec:compiler:trace}, \Cref{thm:trace-bot-static-main});
all other expressions may require runtime evaluation.
The labels are ordered by ${\bot \sqsubseteq \top}$,
and we write ${\ell \sqcup \ell'}$ for the least upper bound (or join) of
${\ell}$ and ${\ell'}$.

\paragraph{Types}

A \emph{primitive type} $\primtype^\ell$
is either a standard scalar type such as $\ZTfloat{n}^\ell$, or the type $\ZTreal^\bot$ of exact real numbers.
The subscript $n \geq 1$ denotes the bit width.
We require $\ZTreal$ to be $\bot$-labeled (i.e., $\ZTreal^\top$ is forbidden)
so that expressions of this type are evaluated at compile time,
\mbox{as discussed in \cref{sec:overview:how}.}

A \emph{type} $\ZT{\comptype}$ is either
a primitive type $\ZT{\primtype^\ell}$,
a \emph{composite type}, %
or a function type $\ZTfun{\bar\comptype}{\comptype}$.
The composite type $\smash{\ZTdyadic{\primtype_1^{\ell_1}\!, \primtype_2^{\ell_2}\!}{n}}$
denotes \emph{fixed-point dyadic} numbers (\zcref{sec:overview:how}) with
sign type $\smash{\primtype_1^{\ell_1}}$, significand type $\smash{\primtype_2^{\ell_2}}$,
and fixed exponent $n$.
The type $\smash{\ZTdyadic{\primtype_1^{\ell_1}\!, \primtype_2^{\ell_2}\!}{\primtype_3^{\ell_3}}}$
denotes \emph{floating-point dyadic} numbers
with sign type $\smash{\primtype_1^{\ell_1}}$, significand type $\smash{\primtype_2^{\ell_2}}$,
and exponent type $\smash{\primtype_3^{\ell_3}}$.
The type $\ZTmulwrd{\bar\comptype}$ denotes \emph{multi-word} numbers
represented as unevaluated sums of floating-point or dyadic components of types $\ZT{\bar\comptype}$.
The type $\ZTtable{\comptype^\bot \!}{\bar n}$ denotes \emph{lookup tables} (\zcref{sec:overview:where})
with shape $\bar{n}$ and element type $\comptype^\bot$,
whose labels are all $\bot$.
This restriction on $\comptype^\bot$ ensures that all table entries are precomputed at compile time.
In contrast, the type $\ZTtuple{\bar\comptype}$ denotes {tuples} with element types $\ZT{\bar\comptype}$
\mbox{that need not be $\bot$-labeled.}

\paragraph{Expressions}

An expression $E$ is either a value $v$, variable $x$,
operator/function application, or composite value construction/projection.
Here, $v$ is a primitive/composite value or function name~$f$.

The expression $\ZE{\ZEop^\ell[{\ZTfun{\bar\primtype}{\primtype'}}](\bar E)}$
applies the built-in \emph{primitive operator} $\ZEop$ to arguments $\bar E$,
where $\ZTfun{\bar\primtype}{\primtype'}$ specifies the unlabeled argument and result types,
and $\ell$ specifies the maximum allowed label on the argument types.
Primitive operators include standard boolean, integer, or floating-point operators
${\ZEop}_{\mathsf{Bool}}, \ldots, {\ZEop}_{\mathsf{Float}}$;
real operators ${\ZEop}_{\mathsf{Real}}$;
and conversion operators ${\ZEop}_{\mathsf{Cast}}$
(e.g., $\Zkw{cast}$ and $\Zkw{reint}$, which resemble C++'s \texttt{static\_cast} and \texttt{reinterpret\_cast}).
We require operators with $\ZTreal$ arguments (e.g., ${\ZEop}_{\mathsf{Real}}$) to be $\bot$-labeled
so that their applications are evaluated at compile~time.

The expression $\ZE{E(\bar E')}$ evaluates $E$ to a function name $f$ and calls $f$ with arguments $\bar E'$,
where $f$ names a user-defined function in the program.
The expression $\ZEnew{\comptype}{\bar{E}}$ creates a composite value of type $\comptype$ from $\bar{E}$,
and $\ZEget{E,E'}$ selects the $E'$-th component of the composite value $E$ evaluates to.

\paragraph{Statements and Programs}

A statement $S$ is either a skip, assignment, return, sequential composition, conditional, or loop.
A program $P$ is a list of definitions of \emph{user-defined functions},
each of the form $\Zkw{def}\ f(\bar x \,{:}\, \bar\comptype) \,{:}\, {\comptype'} \{S\}$.
A user-defined function can be a main function implementing a correctly rounded function (e.g., \texttt{exp}),
a standard library function (\zcref{sec:stdlib}) providing a general operation (e.g., dyadic addition),
or a helper function specific to a few implementations (e.g., \texttt{exp} table generation).

User-defined functions and function calls make \Ziv*{} highly extensible and composable:
users can define new operations using core constructs and existing functions,
then compose the resulting functions to build further abstractions.
For instance, operations on composite values beyond $\Zkw{new}$ and $\Zkw{get}$,
such as dyadic/multi-word arithmetic (\zcref{sec:overview:how}),
are implemented as user-defined functions in the standard library (\cref{sec:stdlib})
rather than provided as primitive operators in the core language.

\subsection{Type System}
\label{sec:lang:type}

\input{figs/fig-ziv-typing-frag}

\Cref{fig:ziv-typing-frag} shows an excerpt of \Ziv*{}'s type system;
its full definition is given in \cref{sec:lang-def-ziv:typing}.
The type system tracks data and control dependencies on runtime inputs to guide staging,
and uses flow-sensitive typing to allow variable types to change on reassignment.
We write $[n] \defas \{1, 2, \ldots, n\}$.

\paragraph{Expressions}

The typing judgment for an expression $E$ is $\ZJtype{P; \Gamma}{E \!}{\! \comptype}$,
where $\Gamma$ is a typing context mapping variables to types.
This judgment states that under program $P$ and context $\Gamma$, $E$ has type $\comptype$.

The typing rules restrict the input dependencies of certain expressions.
For example, \Cref{rule:ziv-typing-frag:t-op} requires each argument $E_i$
to have a type label at most the label of the operator $\ZEop$.
Since operators with $\ZTreal$ arguments must be $\bot$-labeled (\cref{sec:lang:syntax}),
their applications are all evaluated at compile time.
Similarly, \Cref{rule:ziv-typing-frag:t-newtab} requires each table element $E_i$
to have type $\comptype^\bot$, whose labels are all $\bot$.
This ensures that all tables are independent of runtime inputs and precomputed at compile time.

The typing rules also soundly track data dependencies on runtime inputs.
For example, \Cref{rule:ziv-typing-frag:t-op} propagates the argument type labels $\ell_i$ to the result type
by taking the join $\ell_1 \sqcup \cdots \sqcup \ell_n$.
Similarly, \Cref{rule:ziv-typing-frag:t-gettab} propagates the index type label $\ell$ to the result type
by taking the join $T^\bot\sqcup\ell$,
where $T' \sqcup \ell'$ denotes the type obtained by joining each label in $T'$ with $\ell'$.

\paragraph{Statements}

The typing judgment for a statement $S$ is $\ZJtype[\comptype]{P; \Gamma}{S }{ (\Omega,W)}$,
where $\Omega$ is $\ZJret$ or a typing context $\Gamma'$, and $W$ is a set of variables.
The judgment states that under program $P$ and context $\Gamma$, every value $S$ returns has type $\comptype$.
Moreover, it states that
(i) if $(\Omega, W) = (\ZJret, \varnothing)$, $S$ returns on every terminating path;
and (ii) if $(\Omega, W) = (\Gamma', W)$, $S$ may terminate without returning,
in which case $\Gamma'$ gives the variable types after $S$,
and $W$ records the variables $S$ may assign.
For example, \Cref{rule:ziv-typing-frag:t-ret} for $S \equiv (\ZSret{E})$
requires $E$ to have the return type $\comptype$ and yields $(\ZJret,\varnothing)$;
and \Cref{rule:ziv-typing-frag:t-assn} for $S \equiv (\ZSassn{x}{E})$ yields $(\Gamma', \{x\})$,
recording $x$ in $W$.
The latter rule also illustrates flow-sensitive typing:
$\Gamma'$ maps $x$ to the type of $E$,
which allows variable types to change on reassignment.

The typing rules soundly track control dependencies on runtime inputs.
For example, \Cref{rule:ziv-typing-frag:t-if} for $S \equiv (\ZSif{E}{S_1}{S_2})$ does so in two steps.
First, the rule tracks how \emph{variable values} may depend on the condition $E$,
using the function $\Zmerge$ (formally defined in \cref{sec:lang-def-ziv:typing}).
Intuitively, this function combines the branch typing outputs $(\Omega_i, W_i)$ ($i\,{=}\,1,2$)
using the type label $\ell$ of $E$ as follows:
\begin{itemize}%
  \item
        It computes $\Gamma'$ as the ``intersection'' of the typing contexts $\Gamma_i$ among the outcomes $\Omega_i$:
        it keeps only variables with the same unlabeled type in all $\Gamma_i$,
        and joins the corresponding type labels.
  \item
        It computes $W'$ as the union of the corresponding variable sets $W_i$.
  \item
        It updates $\Gamma'$ by joining $\ell$ into the type of each variable in $\Zdom(\Gamma') \cap W'$.
\end{itemize}
Second, the rule tracks how the \emph{return value} may depend on the condition $E$.
Specifically, if either branch contains a return statement (i.e., $\Zhasret(S_1) \lor \Zhasret(S_2)$),
it requires every label in the return type $T$ to be at least the condition label $\ell$,
where $\Zhasret(S')$ indicates whether $S'$ contains a return statement.

For while loops, \Cref{rule:ziv-typing-full:t-while} (\cref{sec:lang-def-ziv:typing})
imposes two restrictions: %
(i) the loop condition must have type $\ZTbool^{\bot}$, making it independent of runtime inputs; and
(ii) the loop body must contain no return statements, excluding early returns from the loop.
These restrictions simplify the \Ziv*{} compiler {(\cref{sec:compiler})} at negligible practical cost,
since existing math libraries very rarely use runtime-dependent or return-containing loops. %

\paragraph{Programs}

The judgment for a program $P$ is $\ZJtype{}{\! P \!}{\! \ZJok}$,
stating that $P$ is well-typed.
\Cref{rule:ziv-typing-frag:t-prog}
requires each function body to return a value of the declared return type whenever it terminates.

\subsection{Semantics and Type Preservation}
\label{sec:lang:sem}

We define the big-step operational semantics of \Ziv*{} using a judgment %
\( \ZJeval{P}{f(\bar v)}{v'}, \)
which states that the function call $f(\bar v)$ in program $P$ evaluates to a value $v'$.
The rules are given in \cref{sec:lang-def-ziv:semantics}.

We show that \Ziv*{} satisfies type preservation:
in a well-typed program, if a function call with arguments of the expected types evaluates to a value,
then the value has the declared return type.

\begin{theorem}[Type Preservation]
  \label{thm:ziv-type-preserve-main}
  Let $\ZP{P}$ be a well-typed \Ziv*{} program.
  Suppose a function $f$ has type $\ZTfun{(\comptype_1, \ldots, \comptype_n)}{\comptype'}$ in $\ZP{P}$,
  i.e., $\ZJtype{P; \varepsilon}{f}{\ZTfun{(\comptype_1, \ldots, \comptype_n)}{\comptype'}}$.
  Then, for all values $v_1, \ldots, v_n, v'$,
  \[
    (\forall i \in [n].\; \ZJtype{P; \varepsilon}{v_i}{\comptype_i})
    \quad {\land} \quad
    \ZJeval{P}{f(v_1, \ldots, v_n)}{v'}
    \quad {\implies} \quad
    \ZJtype{P; \varepsilon}{v'}{\comptype'}.
  \]
\end{theorem}

%% file: figs/fig-ziv-syntax-frag.tex
\begin{figure}
  \begin{figdefs}
    \setlength{\fboxrule}{0.4pt}
    \setlength{\parskip}{0pt}
    \setlength{\abovedisplayskip}{0pt} %
    \setlength{\abovedisplayshortskip}{\abovedisplayskip}
    \setlength{\belowdisplayskip}{\abovedisplayskip}
    \setlength{\belowdisplayshortskip}{\abovedisplayskip}
    \setlength{\MyMathparLineskip}{4pt}
    \def\MathparLineskip{\lineskip=\MyMathparLineskip}
    \vspace{-3.0pt}
    \begin{alignat*}{10}
      \text{Labels}
       &
       &
      {\ell}
       & \ \Coloneqq \
      {\bot \mid \top}
      \\[-2.0pt]
      \text{Primitive types}
       &
       &
      {\primtype^\ell} \!
       & \ \Coloneqq \
      \rlap{$
          \ZTbool^\ell
          \mid \ZTint{n}^\ell
          \mid \ZTuint{n}^\ell
          \mid \ZTfloat{n}^\ell
          \mid {\ZTreal^\bot}
        $}
      \\[-2.0pt]
      \text{Types}
       &
       &
      \comptype
       & \ \Coloneqq \
      \rlap{$
          \primtype^\ell
          \mid {\ZTdyadic{\primtype^\ell \!, \primtype^\ell \!}{n}}
          \mid {\ZTdyadic{\primtype^\ell \!, \primtype^\ell \!}{\primtype^\ell}}
          \mid {\ZTmulwrd{{\bar{\comptype}}}}
        $}
      \\[-2.0pt]
       &
       &
       & \phantom{ {} \ \Coloneqq \ {} \primtype^\ell}
      \rlap{$
          {}
          \mid \ZTtable{{\comptype^{\bot}} \!}{\bar{n}}
          \mid \ZTtuple{\bar\comptype}
          \mid \ZTfun{\bar\comptype}{\comptype}
        $}
      \\[-2.0pt]
      \text{Expressions}
       &
       &
      E
       & \ \Coloneqq \
      \rlap{$
          v
          \mid x
          \mid {\ZEop^{\ell}}{[\ZTfun{\bar\primtype}{\primtype}]}(\bar E)
          \mid E(\bar E)
          \mid \ZEnew{\comptype}{\bar E}
          \mid \ZEget{E, E}
        $}
      \\[-2.0pt]
      \text{Statements}
       &
       &
      S
       & \ \Coloneqq \
      \rlap{$
          \ZSskip
          \mid \ZSassn{x}{E}
          \mid \ZSret{E}
          \mid \ZSseq{S}{S}
          \mid \ZSif{E}{S}{S}
          \mid \ZSwhile{E}{S}
        $}
      \\[-2.0pt]
      \text{Programs}
       &
       &
      P
       & \ \Coloneqq \
      \rlap{$
          \ZPnil \mid \ZPseq{P}{\ZPdef{f}{\bar x:\bar\comptype}{\comptype}{S}}
        $}
      \\[-2.0pt]
      \noalign{%
        \vskip \MyMathparLineskip
        \nointerlineskip
        \hbox to\linewidth{%
          \hskip 6pt
          \rule{\dimexpr\linewidth-12pt\relax}{\fboxrule}\hss}%
        \nointerlineskip
        \vskip \MyMathparLineskip}
      \text{Primitive operators}
       &
       & \quad
      \ZEop^\ell \!
       & \ \Coloneqq \
      \rlap{$
          \ZEop^\ell_{\Zkw{Bool}}
          \mid
          \ZEop^\ell_{\Zkw{Int}}
          \mid
          \ZEop^\ell_{\Zkw{UInt}}
          \mid
          \ZEop^\ell_{\Zkw{Float}}
          \mid
          {\ZEop^\bot_{\Zkw{Real}}}
          \mid
          \ZEop^\ell_{\Zkw{Cast}}
        $}
      \\[-2.0pt]
       &
       &
      \llap{$
          {\ZEop}_{\Zkw{Bool}}
        $}
       & \ \Coloneqq \
      \Zkw{\&} \mid \Zkw{\shortbar} \mid \Zkw{\textasciicircum} \mid \Zkw{$\textsf{!}$}
      \mid \Zkw{$\textsf{==}$} %
      \mid \cdots
       &
      {\ZEop}_{\Zkw{Real}}
       & \ \Coloneqq \
      \Zkw{$+$} \mid \Zkw{$-$} \mid \Zkw{$\times$} \mid \Zkw{$/$}  %
      \mid \cdots
      \\[-2.0pt]
       &
       &
      \llap{$
          {\ZEop}_{\Zkw{Int}},
          {\ZEop}_{\Zkw{UInt}},
          {\ZEop}_{\Zkw{Float}}
        $}
       & \ \Coloneqq \
      \Zkw{$+$} \mid \Zkw{$-$} \mid \Zkw{$\times$} \mid \Zkw{$/$}
      \mid \Zkw{$\ll$} %
      \mid \Zkw{\&} %
      \mid \Zkw{$\leq$} %
      \mid \cdots
       & \quad\;\;
      {\ZEop}_{\Zkw{Cast}}
       & \ \Coloneqq \
      \Zkw{cast} \mid \Zkw{reint}
      \mid \cdots
    \end{alignat*}
  \end{figdefs}
  \caption{Core syntax of \Ziv.}
  \label{fig:ziv-syntax-frag}
  \vspace{-5pt}
\end{figure}

%% file: figs/fig-ziv-typing-frag.tex
\begin{figure}[!tp]
  \begin{figdefs}
    \msmall
    \setlength{\fboxsep}{2pt}
    \setlength{\fboxrule}{0.4pt}
    \setlength{\parskip}{0pt}
    \setlength{\abovedisplayskip}{2pt}
    \setlength{\abovedisplayshortskip}{\abovedisplayskip}
    \setlength{\belowdisplayskip}{0pt} %
    \setlength{\belowdisplayshortskip}{\belowdisplayskip}
    \setlength{\MyMathparLineskip}{6pt}
    \def\MathparLineskip{\lineskip=4pt}
    \vspace{-2pt}
    \textbf{Definitions} (typing contexts, typing outcomes, write sets)\hfill\fbox{$\Gamma$, $\Omega$, $W$}%
    \begin{align*}
      \Gamma \Coloneqq \varepsilon \mid \Gamma,x:\comptype
      \qquad\qquad
      \Omega \Coloneqq {\ZJret} \mid \Gamma
      \qquad\qquad
      W \subseteq \ZVar
    \end{align*}
    \par\vspace{\dimexpr\MyMathparLineskip-\abovedisplayskip\relax}
    \nointerlineskip
    \hbox to\linewidth{%
      \rule{\dimexpr\linewidth\relax}{0pt}\hss}
    \nointerlineskip
    \vspace{\MyMathparLineskip}
    \vspace{-6pt}
    \textbf{Expressions}\hfill\fbox{$\ZJtype{P; \Gamma}{E}{\comptype}$}%
    \vspace{-1pt}
    \begin{mathpar}
      \Zinfer{T-Op}
      {
        \forall i \in [n].\; \ZJtype{P; \Gamma}{E_i}{\smash{(\primtype_i)^{\ell_i}}}
        \\
        {\forall i \in [n].\; \ell_i \sqsubseteq \ell}
        \\
        {\ell_1 \sqcup \cdots \sqcup \ell_n = \ell'}
      }
      {
        \ZJtype{P; \Gamma}
        {\ZEop^{\ell}{[\ZTfun{(\primtype_1,\ldots,\primtype_n)}{\primtype'}]}(E_1, \ldots, E_n)}
        {\smash{(\primtype')^{\ell'}}}
      }
      \label{rule:ziv-typing-frag:t-op}
      \and
      \Zinfer[width=0.2\linewidth]{T-TabNew}
      {
        {\begin{array}[b]{@{}c@{}}
            \comptype' = \ZTtable{{\smash{\comptype^\bot}}}{(m_1, \ldots, m_k)}
            \\[-1pt]
            \forall i \in [n].\;
            \ZJtype{P; \Gamma}{E_i}{{\smash{\comptype^\bot}}}
            \quad
            n = m_1 \times \cdots \times m_k
          \end{array}}
      }
      {\ZJtype{P; \Gamma}{\ZEnew{\comptype'}{E_1, \ldots, E_n}}{\comptype'}}
      \label{rule:ziv-typing-frag:t-newtab}
      \and \!\!\!\!\!\!\!
      \Zinfer[width=0.2\linewidth]{T-TabGet}
      {
        {\begin{array}[b]{@{}c@{}}
            \ZJtype{P; \Gamma}{E_1}{\ZTtable{\smash{\comptype^\bot} \!}{\bar n}}
            \\[-1pt]
            \ZJtype{P; \Gamma}{E_2}{\ZTuint{k}^\ell}
          \end{array}}
      }
      {\ZJtype{P; \Gamma}{\ZEget{E_1, E_2}}{{\smash{\comptype^\bot} \sqcup \ell}}}
      \label{rule:ziv-typing-frag:t-gettab}
    \end{mathpar}
    \par\vspace{\dimexpr\MyMathparLineskip-\abovedisplayskip\relax}
    \nointerlineskip
    \hbox to\linewidth{%
      \rule{\dimexpr\linewidth\relax}{0pt}\hss}
    \nointerlineskip
    \vspace{\MyMathparLineskip}
    \vspace{-6pt}
    \textbf{Statements}\hfill\fbox{$\ZJtype[\comptype]{P; \Gamma}{S}{(\Omega, W)}$}%
    \vspace{-2pt}
    \begin{mathpar}
      \Zinfer{T-Assn}
      {\ZJtype{P; \Gamma}{E}{\comptype'}}
      {
        \ZJtype[\comptype]{P; \Gamma}{\ZSassn{x}{E}}{((\Gamma, {x:\comptype'}), {\{x\}})}
      }
      \label{rule:ziv-typing-frag:t-assn}
      \and
      \Zinfer{T-Ret}
      {\ZJtype{P; \Gamma}{E}{{\comptype}}}
      {
        \ZJtype[\comptype]{P; \Gamma}{\ZSret{E}}{({\ZJret}, \varnothing)}
      }
      \label{rule:ziv-typing-frag:t-ret}
      \and
      \Zinfer{T-If}
      {
        \ZJtype{P; \Gamma}{E}{\ZTbool^\ell}
        \\
        {\begin{array}[b]{@{}c@{}}
            \ZJtype[\comptype]{P; \Gamma}{S_1}{(\Omega_1, W_1)}
            \\[-1pt]
            \ZJtype[\comptype]{P; \Gamma}{S_2}{(\Omega_2, W_2)}
          \end{array}}
        \\
        {\begin{array}[b]{@{}c@{}}
            {
              \Zmerge
              (\ell, (\Omega_1,W_1), (\Omega_2,W_2))
              =
              (\Omega',W')
            }
            \\[-1pt]
            {
              \Zhasret(S_1)\lor\Zhasret(S_2) {\implies} \comptype\sqcup\ell=\comptype
            }
          \end{array}}
      }
      {
        \ZJtype[\comptype]{P; \Gamma}{\ZSif{E}{S_1}{S_2}}{(\Omega', W')}
      }
      \label{rule:ziv-typing-frag:t-if}
    \end{mathpar}
    \par\vspace{\dimexpr\MyMathparLineskip-\abovedisplayskip\relax}
    \nointerlineskip
    \hbox to\linewidth{%
      \rule{\dimexpr\linewidth\relax}{0pt}\hss}
    \nointerlineskip
    \vspace{\MyMathparLineskip}
    \vspace{-6pt}
    \textbf{Programs}\hfill\fbox{$\ZJtype{}{P}{\ZJok}$}%
    \begin{mathpar}
      \Zinfer{T-Prog}
      {
        P = (\ZPdef{f_j}{\bar{x_j}:\bar{\comptype_j}}{\comptype_j'}{S_j})_{j \in [n]}
        \\
        \forall i \in [n].\; \ZJtype[\comptype_i']{P; (\bar{x_i} : \bar{\comptype_i})}{S_i}{(\ZJret, \varnothing)}
      }
      {
        \ZJtype{}{
          P
        }{\ZJok}
      }
      \label{rule:ziv-typing-frag:t-prog}
    \end{mathpar}
    \vspace{-11pt}
  \end{figdefs}
  \caption{%
    Selected typing rules of \Ziv*{}.}
  \label{fig:ziv-typing-frag}
  \vspace{-0pt}
\end{figure}

%% file: sections/sec04-stdlib.tex
\input{figs/fig-dd-mul}

\section{The \texorpdfstring{\Ziv*{}}{Ziv} Standard Library}
\label{sec:stdlib}

The small core of \Ziv\ presented in \cref{sec:lang}
provides a foundation for higher-level abstractions (\ref{dp:abstraction})
that are better expressed at the library level.
In this section,
we discuss the \Ziv\ standard library \ZivStd,
using standard syntactic sugar for readability.
\zcref[S]{tab:stdlib} lists key \ZivStd\ functions,
including arithmetic (\zcref{sec:stdlib:arith}),
implementation step abstraction (\zcref{sec:stdlib:technique}),
and floating-point-related (\zcref{sec:stdlib:fp}) functions.

\subsection{Arithmetic}\label{sec:stdlib:arith}

\paragraph{Real-Valued Functions}\label{sec:stdlib:arith:real}

\ZivStd\ provides functions on $\ZTreal$ values
and constructors that round them to finite precision.
For example, \Std{makeMultiWord} constructs a $\ZTmulwrd{\bar\comptype}$
from a $\ZTreal$:
\begin{equationcode}
  \begin{equationblock*}
    \Zkw{def}\ \StdFn{makeMultiWord}[{\comptype_1,\ldots,\comptype_n}](x:\ZTreal^\bot):
    \ZTmulwrd{\comptype_1,\ldots,\comptype_n}^\bot \\
    \begin{equationblock}
      \ZSassn{w_i}{\Zkw{cast}[{\comptype_i}]\bigl(
        x - \textstyle\sum_{j=1}^{i-1}\Zkw{cast}[{\ZTreal}](w_j)\bigr)}
      \quad (i=1,\ldots,n); \\
      \ZSret{\ZEnew{\ZTmulwrd{\comptype_1,\ldots,\comptype_n}}{w_1,\ldots,w_n}};
    \end{equationblock}
  \end{equationblock*}
\end{equationcode}

\paragraph{Extended-Precision Arithmetic}\label{sec:stdlib:arith:extended}

Arithmetic on extended-precision types (\zcref{sec:overview:how})
admits multiple operator variants with different properties.
To keep \Ziv\ small,
these operators are a part of \ZivStd.

Arithmetic on $\ZTmulwrd{\bar\comptype}$ is implemented with
error-free transformations (\zcref{sec:overview:how:multiword}),
including \Std{2sum}, \Std{fast2sum}, and \Std{2mul}~\cite{2sum,doubleword,fpsurvey}. %
\zcref[S]{fig:dd-mul} shows four variants of \ZivStd's double-word multiplication.
LLVM-libc~\cite{llvmlibc} uses three chained FMAs (\zcref{fig:dd-mul:llvm}),
while CORE-MATH's~\cite{coremath} hyperbolic functions expand the last two FMAs (\cref{fig:dd-mul:cmhyper}).
CORE-MATH's exponential uses both a different operation order
(\zcref{fig:dd-mul:cmexp}) and a version that omits the residual of the high part
(\zcref{fig:dd-mul:cmexph}).

Multiple operator variants for floating- and fixed-point dyadic arithmetic
are also supported, implemented using integer arithmetic.
In particular,
as floating-point dyadics with 128 bits of precision are widely used,
\ZivStd\ includes specialized
implementations for $\ZTdyadic{\ZTbool,\ZTuint{128}}{\ZTint{64}}$.

\subsection{Implementation Step Abstractions}\label{sec:stdlib:technique}

Beyond basic arithmetic,
\ZivStd\ provides high-level abstractions for
range reduction, polynomial approximation, and polynomial evaluation,
supporting the implementation steps of \zcref{sec:overview:where}.

\paragraph{Range Reduction}\label{sec:stdlib:technique:reduction}

\ZivStd\ supports the Cody-Waite~\cite{codywaite} and Payne-Hanek~\cite{paynehanek} range reduction algorithms.
Given an input $x$ and reduction constant $C$,
both algorithms compute an integer $k$  and a reduced argument $r \approx x - kC$,
but target different input ranges.
An implementation of Cody-Waite reduction is shown below,
using triple-word arithmetic and the \Std{makeMultiWord} function of \zcref{sec:stdlib:arith:real}.

\begin{equationcode}
  \begin{equationblock*}
    \Zkw{def}\ \StdFn{codywaite}(x:\ZTfloat{64},\;C:\ZTreal^\bot) :
    \ZTtuple{\ZTmulwrd{\ZTfloat{64},\ZTfloat{64}},\ZTint{64}} \\
    \begin{equationblock}
      \ZSassn{c}{\StdFn{makeMultiWord}\bigl[
          \ZTdyadic{\ZTbool,\ZTuint{32}}{\ZTint{64}},\;
          \ZTdyadic{\ZTbool,\ZTuint{32}}{\ZTint{64}},\;\ZTfloat{64}\bigr](-C)}; \\
      \ZSassn{k_i}{\Zkw{cast}[{\ZTint{64}}]\bigl(x\times
        \Zkw{cast}[{\ZTfloat{64}}](1/C)\bigr)};\quad \ZSassn{k_f}{\Zkw{cast}[{\ZTfloat{64}}](k_i)}; \\
      \ZSret{\bigl(\StdFn{fma}(k_f,c,x),\;k_i\bigr)};
    \end{equationblock}
  \end{equationblock*}
\end{equationcode}

To implement table-based range reduction (\zcref{sec:overview:where:reduction})~\cite{tableexp,tablebased},
$\StdFn{tabulate}(f, \bar{n})$ can be used,
which creates a $\ZTtable{\comptype^\bot}{\bar{n}}$
by mapping the function $f$ over every index of the table at compile time.

\input{figs/tab-stdlib-fns}

\paragraph{Polynomial Approximation and Evaluation}\label{sec:stdlib:technique:poly}

Polynomial approximation is supported through
Taylor polynomials (\Std{taylor})
and
minimax polynomials approximated with the Remez algorithm~\cite{remez}
on an interval (\Std{remez}) or fixed grid (\Std{gridremez}),
or with floating-point coefficients (\Std{fpminimax})~\cite{fpminimax,sollya}.
Approximated polynomials can be evaluated with
Horner's~\cite{taocp} and Estrin's~\cite{estrin} schemes
via \Std{horner} and \Std{estrin},
which use \Ziv's staging to unroll the evaluation into straight-line code.

\subsection{Floating-Point-Related Functions}\label{sec:stdlib:fp}

In addition to the extended-precision numerical abstractions described above,
\ZivStd\ provides lower-level utilities for inspecting and manipulating
machine floating-point representations.

\paragraph{Representation-Related Functions}\label{sec:stdlib:fp:repr}

Utilities related to the bit representation of floating-point values~\cite{ieee-754-2019} include
constant calculation (e.g., \Std{bias}, \Std{signmask}),
classification (e.g., \Std{issubnormal}, \Std{isinf}),
field extraction (e.g., \Std{decompose}),
and value manipulation (e.g., \Std{prevfloat}, \Std{nextfloat}).
\Ziv's staging allows these bit-level functions to be parameterized by format
and specialized at compile time.

\paragraph{Optimization Functions}\label{sec:stdlib:fp:optim}

The optimization functions
\Std{argmin}\ and \Std{argmax} each perform compile-time searches
for the smallest and largest floating-point value
that satisfies a monotone predicate.
Internally,
both functions perform binary searches over floating-point bit patterns.

%% file: figs/fig-dd-mul.tex
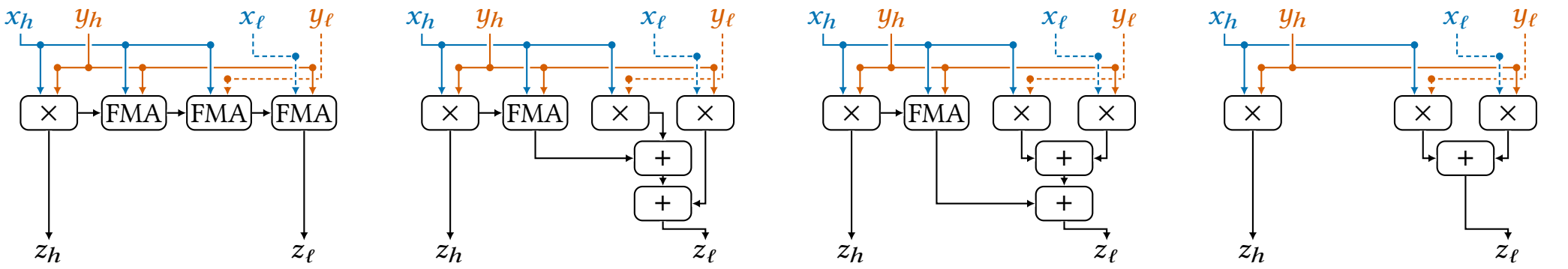
\begin{figure}[t]
  \centering
  \begingroup
  \footnotesize
  \definecolor{ddx}{HTML}{0072B2}
  \definecolor{ddy}{HTML}{D55E00}
  \tikzset{
    dd circuit/.style={x=0.5cm,y=0.5cm,font=\scriptsize,
        every node/.style={inner sep=1pt},
        >={Latex[length=2pt,width=2pt]}},
    dd gate/.style={draw,rounded corners=2pt,minimum width=5mm,
        minimum height=3mm,fill=white},
    dd line/.style={line width=0.4pt,
        preaction={draw=white,line width=1.6pt,arrows=-}},
    dd wire/.style={dd line,->},
    dd low input/.style={dash pattern=on 1.2pt off 0.8pt},
    dd junction/.style={circle,fill,inner sep=0.65pt}
  }
  \newcommand{\ddinputs}{%
    \path[use as bounding box] (-0.25,0.4) rectangle (5.6,-4);
    \draw[dd line,ddx] (0,0) node[above] {$x_h$} -- (0,-0.2) -- (3.35,-0.2);
    \draw[dd line,ddx,dd low input] (4.1,0) node[above] {$x_\ell$} -- (4.1,-0.4) -- (4.85,-0.4);
    \draw[dd line,ddy] (1.2,0) node[above] {$y_h$} -- (1.2,-0.6);
    \draw[dd line,ddy] (0.65,-0.6) -- (5.15,-0.6);
    \node[dd junction,ddy] at (1.2,-0.6) {};
    \draw[dd line,ddy,dd low input] (5.3,0) node[above] {$y_\ell$} -- (5.3,-0.8) -- (3.65,-0.8);
  }
  \newcommand{\ddargs}[4]{%
    \ifdim#3pt=-0.4pt
      \draw[dd wire,ddx,dd low input] (#2-0.15,#3) -- ([xshift=-0.75mm]#1.north);
    \else
      \draw[dd wire,ddx] (#2-0.15,#3) -- ([xshift=-0.75mm]#1.north);
    \fi
    \node[dd junction,ddx] at (#2-0.15,#3) {};
    \ifdim#4pt=-0.8pt
      \draw[dd wire,ddy,dd low input] (#2+0.15,#4) -- ([xshift=0.75mm]#1.north);
    \else
      \draw[dd wire,ddy] (#2+0.15,#4) -- ([xshift=0.75mm]#1.north);
    \fi
    \node[dd junction,ddy] at (#2+0.15,#4) {};
  }
  \newcommand{\ddhigh}{%
    \node[dd gate] (h) at (0.5,-1.4) {$\times$};
    \ddargs{h}{0.5}{-0.2}{-0.6}
    \draw[dd wire] (h.south) -- (0.5,-3.65) node[below] {$z_h$};
  }
  \newcommand{\ddresidual}{%
    \node[dd gate] (e) at (2,-1.4) {FMA};
    \ddargs{e}{2}{-0.2}{-0.6}
    \draw[dd wire] (h.east) -- (e.west);
  }
  \newcommand{\ddcross}{%
    \node[dd gate] (p) at (3.5,-1.4) {$\times$};
    \node[dd gate] (q) at (5,-1.4) {$\times$};
    \ddargs{p}{3.5}{-0.2}{-0.8}
    \ddargs{q}{5}{-0.4}{-0.6}
  }
  \newcommand{\ddlow}{%
    \draw[dd wire] (l.south) -- (4.25,-3.45) -| (5,-3.65)
    node[below] {$z_\ell$};
  }

  \begin{subfigure}[t]{0.237\linewidth}
    \centering
    \begin{tikzpicture}[dd circuit]
      \ddinputs \ddhigh \ddresidual
      \node[dd gate] (s) at (3.5,-1.4) {FMA};
      \node[dd gate] (l) at (5,-1.4) {FMA};
      \ddargs{s}{3.5}{-0.2}{-0.8}
      \ddargs{l}{5}{-0.4}{-0.6}
      \draw[dd wire] (e) -- (s);
      \draw[dd wire] (s) -- (l);
      \draw[dd wire] (l.south) -- (5,-3.65) node[below] {$z_\ell$};
    \end{tikzpicture}
    \caption{$x \,{\times}\, y$ in \Std{LLVMLibc}}\label{fig:dd-mul:llvm}
  \end{subfigure}\hfill
  \begin{subfigure}[t]{0.237\linewidth}
    \centering
    \begin{tikzpicture}[dd circuit]
      \ddinputs \ddhigh \ddresidual \ddcross
      \node[dd gate] (s) at (4.25,-2.2) {$+$};
      \node[dd gate] (l) at (4.25,-3) {$+$};
      \draw[dd wire] (p.east) -| (s.north);
      \draw[dd wire] (e.south) |- (s.west);
      \draw[dd wire] (s) -- (l);
      \draw[dd wire] (q.south) |- (l.east);
      \ddlow
    \end{tikzpicture}
    \caption{$x \,{\times}\, y$ in \Std{CoreMathHyp}}\label{fig:dd-mul:cmhyper} %
  \end{subfigure}\hfill
  \begin{subfigure}[t]{0.237\linewidth}
    \centering
    \begin{tikzpicture}[dd circuit]
      \ddinputs \ddhigh \ddresidual \ddcross
      \node[dd gate] (s) at (4.25,-2.2) {$+$};
      \node[dd gate] (l) at (4.25,-3) {$+$};
      \draw[dd wire] (p.south) |- (s.west);
      \draw[dd wire] (q.south) |- (s.east);
      \draw[dd wire] (s) -- (l);
      \draw[dd wire] (e.south) |- (l.west);
      \ddlow
    \end{tikzpicture}
    \caption{$x \,{\times}\, y$ in \Std{CoreMathExp}}\label{fig:dd-mul:cmexp}
  \end{subfigure}\hfill
  \begin{subfigure}[t]{0.237\linewidth}
    \centering
    \begin{tikzpicture}[dd circuit]
      \ddinputs \ddhigh \ddcross
      \node[dd gate] (l) at (4.25,-2.2) {$+$};
      \draw[dd wire] (p.south) |- (l.west);
      \draw[dd wire] (q.south) |- (l.east);
      \ddlow
    \end{tikzpicture}
    \caption{$x \,{\times_\textit{\textsf{hi}}}\, y$ in \Std{CoreMathExp}}\label{fig:dd-mul:cmexph}
  \end{subfigure}
  \endgroup
  \caption{Selected double-word multiplication variants in \ZivStd{}.
    FMA denotes fused multiply-add~\cite{ieee-754-2019}.}
  \label{fig:dd-mul}
\end{figure}

%% file: figs/tab-stdlib-fns.tex
\begin{table}
  \caption{
    Representative functions of \ZivStd.}\label{tab:stdlib}
  \centering
  \scriptsize
  \begin{booktabs}{
      width=\linewidth,
      colspec={Q[c,m,wd=6pt] Q[l,t,wd=0.155\linewidth]
        Q[l,t,wd=0.340\linewidth] X[l,t]},
      colsep=3pt,
      column{4}={rightsep=0pt},
    }
    \toprule
    & \textbf{Function} & \textbf{Type} & \textbf{Description} \\
    \midrule
    \SetCell[r=15]{c,m}\rotatebox{90}{\textbf{Arithmetic (\zcref{sec:stdlib:arith})}}
    & \SetCell[c=3]{l}\textbf{Real-Valued Functions} & & \\
    & \Std{exp}, \Std{log}, \Std{sin}, ...
      & ${({\ZTreal^n})^\bot}\shortto ({{\ZTreal}^m})^\bot$
      & Real functions \\
    & $\Std{makeMultiWord}[\bar\comptype]$
      & $\ZTreal^\bot \shortto \ZTmulwrd{\bar\comptype}^\bot$
      & Multi-word construction \\
    & $\Std{makeDyadic}[\primtype,\primtype',\alpha]$
      & $\ZTreal^\bot \shortto \ZTdyadic{\primtype,\primtype'}{\alpha}^\bot$
      & Dyadic construction ($\alpha = \primtype''$ for floating, $n$ for fixed) \\
    \cmidrule[l]{2-4}
    & \SetCell[c=3]{l}\textbf{Error-Free Transformations} & & \\
    & \Std{2sum}, \Std{fast2sum}
      & ${\ZTfloat{n}}^2\shortto{\ZTfloat{n}}^2$
      & Rounded sum and its residual~\cite{2sum,taocp,doubleword,fast2sum} \\
    & \Std{2mul}
      & ${\ZTfloat{n}}^2\shortto{\ZTfloat{n}}^2$
      & Rounded product and its residual~\cite{2mul-algo,2mul-proof} \\
    & \Std{2fma}, \Std{fast2fma}
      & ${\ZTfloat{n}}^3\shortto{\ZTfloat{n}}^2$
      & Rounded FMA and its residual~\cite{2fma,2fma-ext,fast2fma-algo,fast2fma-proof-tight} \\ %
    \cmidrule[l]{2-4}
    & \SetCell[c=3]{l}\textbf{Extended-Precision Arithmetic}
      ($\comptype = \ZTmulwrd{\bar{\comptype}}, \ZTdyadic{\primtype,\primtype'}{\alpha}$) & & \\
    & $+, -, \times, /$
      & $\comptype^2\shortto \comptype$
      & Extended-precision arithmetic, default \\
    & $+_\textit{\textsf{hi}}, \times_\textit{\textsf{hi}},+_\textit{\textsf{nm}},\times_\textit{\textsf{nm}}, ...$
      & $\comptype^2\shortto \comptype$
      & Extended-precision arithmetic, variants \\
    & \Std{inv}
      & $\comptype^1\shortto \comptype$
      & Extended-precision reciprocal \\
    & \Std{fma}
      & $\comptype^3\shortto \comptype$
      & Extended-precision FMA \\
    & \Std{ldexp}
      & $(\comptype,\ZTint{n})\shortto\comptype$
      & Scaling by a power of two \\
    & \Std{frexp}
      & $\comptype\shortto \ZTtuple{\comptype, \ZTint{n}}$
      & Decomposition into fractional significand and exponent \\
    \midrule
    \SetCell[r=11]{c,m}\rotatebox{90}{\textbf{Impl. Step Abstractions (\zcref{sec:stdlib:technique})}}
    & \SetCell[c=3]{l}\textbf{Table Operations} & & \\
    & \Std{tabulate}
      & $\bigl(({\ZTuint{64}}^{|\bar{n}|} \shortto \comptype),{\ZTuint{64}}^{|\bar{n}|}\bigr)^\bot
          \shortto\ZTtable{\comptype^\bot}{\bar{n}}$
      & Table construction by mapping over indices (shape $\bar{n}$) \\
    & \Std{getelem}
      & $\bigl(\ZTtable{\comptype^\bot}{\bar{n}}, {\ZTuint{64}}^{\abs{\bar{n}}}\bigr)\shortto \comptype$
      & Multidimensional table access \\
    \cmidrule[l]{2-4}
    & \SetCell[c=3]{l}\textbf{Range Reduction} & & \\
    & \Std{codywaite}
      & $(\ZTfloat{n}, \ZTreal^\bot)\shortto \ZTtuple{T, \ZTint{m}}$
      & Cody-Waite range reduction~\cite{codywaite} \\
    & \Std{paynehanek}
      & $(\ZTfloat{n}, \ZTreal^\bot)\shortto \ZTtuple{T, \ZTint{m}}$
      & Payne-Hanek range reduction~\cite{paynehanek} \\
    \cmidrule[l]{2-4}
    & \SetCell[c=3]{l}\textbf{Polynomial Approximation and Evaluation} & & \\
    & \Std{taylor}
      & $((\ZTreal\shortto\ZTreal),\textit{opts...})^\bot
          \shortto \ZTtuple{\smash[t]{\overline{\smash[t]{\ZTreal}\rule{0pt}{1.4ex}}}}^\bot$
      & Taylor approximation \\
    & \Std{remez}, \Std{gridremez}
      & $((\ZTreal\shortto\ZTreal),\textit{opts...})^\bot
          \shortto \ZTtuple{\smash[t]{\overline{\smash[t]{\ZTreal}\rule{0pt}{1.4ex}}}}^\bot$
      & Minimax approximation, on interval and fixed grid~\cite{remez,sollya} \\
    & \Std{fpminimax}
      & $((\ZTreal\shortto\ZTreal),\textit{opts...})^\bot
          \shortto \ZTtuple{\smash[t]{\overline{\smash[t]{\ZTfloat{n}}\rule{0pt}{1.4ex}}}}^\bot$
      & Minimax approximation, floating-point coefficients~\cite{fpminimax,sollya} \\
    & \Std{horner}, \Std{estrin}
      & $(\comptype,\ZTtuple{\bar{\comptype}}^\bot)\shortto\comptype$
      & Polynomial evaluation, Horner's and Estrin's methods~\cite{taocp,estrin} \\
    \midrule
    \SetCell[r=9]{c,m}\rotatebox{90}{\textbf{Float-Related (\zcref{sec:stdlib:fp})}}
    & \SetCell[c=3]{l}\textbf{Representation-Related Functions} & & \\
    & \Std{decompose}
      & $\ZTfloat{n}\shortto\ZTtuple{\ZTbool,\ZTuint{n},\ZTint{n}}$
      & Decomposition into sign, significand, and exponent \\
    & $\Std{issubnormal}, \Std{isinf}, ...$
      & $\ZTfloat{n}\shortto\ZTbool$
      & Classification of floating-point values \\
    & $\Std{minexponent}, \Std{bias}, ...$
      & $\ZTint{64}\shortto\ZTint{n}$
      & Calculation of IEEE-754-related constants \\
    & \Std{nextfloat}, \Std{prevfloat}
      & $\ZTfloat{n}\shortto\ZTfloat{n}$
      & Next and previous floating-point value \\
    & \Std{argmin}, \Std{argmax}
      & $((\ZTfloat{n}\shortto\ZTbool),{\ZTfloat{n}}^2)\shortto\ZTfloat{n}$
      & Optimization on monotone predicate \\
    \cmidrule[l]{2-4}
    & \SetCell[c=3]{l}\textbf{Testing} & & \\
    & \Std{ulperror}
      & $(\ZTfloat{n},\ZTreal^\bot)\shortto\ZTfloat{64}$
      & Error, in units of the reference value's last place~\cite{fphandbook} \\
    & \Std{randomtest}
      & $((\ZTfloat{n}\shortto\ZTfloat{n}),
          (\ZTreal\shortto\ZTreal)^\bot)\shortto\ZTfloat{64}$
      & Random ulp error testing of a floating-point function \\
    \bottomrule
  \end{booktabs}
\end{table}

%% file: sections/sec05-compiler.tex
\section{The \texorpdfstring{\Ziv*}{Ziv}~Compiler}\label{sec:compiler}

\input{figs/fig-ir-syntax-frag.tex}

We now formally introduce the \Ziv*{} compiler,
which translates programs written in  \Ziv*{}~(\cref{sec:lang}) with \ZivStd{}~(\cref{sec:stdlib})
into \Czero{}, a simplified fragment of C.
The compiler supports staged computation (\ref{dp:staging}) via a \emph{tracing} step,
which combines \emph{partial evaluation} and \emph{symbolic tracing}:
the former evaluates computations at compile time when their input values are known,
while the latter records the remaining computations as intermediate code for runtime execution.
For well-typed programs, the compiler ensures that
expressions of $\bot$-labeled types (\cref{sec:lang}) are evaluated at compile time. %

The compiler proceeds in three steps:
normalization~(\cref{sec:compiler:norm}), tracing~(\cref{sec:compiler:trace}), and code emission~(\cref{sec:compiler:emit}).
First, \emph{normalization} eliminates early \textsf{return} statements
by translating \Ziv*{} into \Znorm*{}, a subset of \Ziv*{} in which each function has a single \textsf{return} at the end of its body.
Second, \emph{tracing} translates \Znorm*{} into the intermediate language \IR*{} using partial evaluation and symbolic tracing.
Third, \emph{code emission} performs a syntax-directed translation from \IR*{} into \Czero*{}.
In~\cref{sec:compiler:soundness}, we prove that end-to-end compilation preserves the semantics of well-typed programs. %
To distinguish the languages, we use \Zhl{yellow} backgrounds for \Ziv*{} and \Znorm*{}, \Ihl{blue} for \IR*{}, and white for \Czero*{}.

\subsection{Normalization (\texorpdfstring{$\Ziv* \to \Znorm*$}{Ziv → ZivNorm})}
\label{sec:compiler:norm}

The normalization step rewrites each function to have a single \textsf{return} at the end of its body.
This simplifies tracing~(\cref{sec:compiler:trace}) since all branches can then be traced in the same way,
whether they originally returned early or continued execution.
We define a syntax-directed translation $\Ztrans$ from \Ziv*{} to \Znorm*{},
using a helper $\smash{\Ztrans^{{\renewcommand{\Zhlpadding}{0pt}\ZE{y}}}(\ZS{S}, \ZS{S'})}$
to normalize statement $\ZP{S}$
with $\ZE{y}$ as a result variable, and
$\ZP{S'}$ as normalized statements to run afterward if $\ZP{S}$ does not return.
Selected rules are:
\begin{gather*}
  \Ztrans\big(
  {\ZPdef{f}{\bar x:\bar\comptype}{\comptype'}{S}}
  \big)
  \coloneqq
  \ZPdef{f}{\bar x:\bar\comptype}{\comptype'}{
    \ZSseq{\Ztrans^{\ZE{y}}(\ZS{S},\ZSskip)}{\ZSret{y}}
  }
  \;\; \text{for fresh $\ZE{y}$},
  \\[-3pt]
  \Ztrans^{\ZE{y}}(\ZSret{E},\ZS{S'})
  \coloneqq \ZSassn{y}{E},
  \qquad
  \Ztrans^{\ZE{y}}(\ZSseq{S_1}{S_2}, \ZS{S'})
  \coloneqq \Ztrans^{\ZE{y}}(
  \ZS{S_1}, {\Ztrans^{\ZE{y}}(\ZS{S_2},\ZS{S'})}
  ).
\end{gather*}
The function case starts with $\ZP{S'} = \ZSskip$ and appends $\ZSret{y}$.
The return case stores the return value in $\ZP{y}$ and discards $\ZP{S'}$,
while the sequencing case normalizes $\ZP{S_2}$ before $\ZP{S_1}$.
The full translation is given in \cref{sec:compiler-def:norm},
and the type and semantic preservation theorems are stated in \cref{sec:comp-res:norm}.

\subsection{Tracing (\texorpdfstring{$\Znorm* \to \IR*$}{ZivNorm → ZivIR})}
\label{sec:compiler:trace}

\paragraph{Intermediate Language}

\IR*{} is a statically typed imperative language for expressing the computations left for runtime execution.
\Cref{fig:ir-syntax-frag} shows the core syntax.
It uses static single-assignment form (i.e., each variable has a single definition)
and requires expressions to occur only in assignments.
Unlike \Ziv*{}, \IR*{} supports only boolean, integer, and float types (and their array types),
as computations on other types are evaluated at compile time or lowered to scalar and array operations during tracing.
It lacks function calls and while loops, as they are inlined and unrolled during tracing.
Its conditionals assign branch results to fresh variables, which simplifies code emission~(\cref{sec:compiler:emit}).
\IR*{}'s semantics, type system, and type preservation theorem are given in \zcref{sec:lang-def-irc0,sec:lang-res:ir}.

\paragraph{Overview of Tracing}

\input{figs/fig-trans-norm-ir-all-frag.tex}

The tracing step runs \Znorm*{} programs under a nonstandard semantics
that (i) evaluates computations whose inputs are known at compile time,
and (ii)  generates \IR*{} code for the remaining computations.
We formalize this step using operational semantics based on tracing values and states.
A \emph{tracing value} $\widehat{v}$ is either
a compile-time value $\Tcon{\ZV{v}}$, a runtime variable $\Tref{\IT{\primtype}}{\IT{x}}$,
or a composite $\Tcomp{{\theta}}{\widehat v_1, \ldots, \widehat v_n}$.
Here, $\Tcon{\ZV{v}}$ denotes a \Znorm*{} value $\ZV{v}$ known at compile time,
$\Tref{\IT{\primtype}}{\IT{x}}$ denotes a \IR*{} variable $\IT{x}$
that will store a value of type $\IT{\primtype}$ at runtime,
and $\Tcomp{{\theta}}{\widehat v_1, \ldots, \widehat v_n}$ denotes a composite
of $\widehat{v}_1, \ldots, \widehat{v}_n$ having unlabeled type $\theta$.
A \emph{tracing state} $\widehat{\sigma}$ maps \Znorm*{} variables to tracing values.
The semantics updates this state and generates \IR*{} code as tracing proceeds.
\Cref{fig:trans-norm-ir-expr-frag} shows selected rules;
the full definition is given in \cref{sec:compiler-def:trace}.

\paragraph{Tracing Expressions}

The tracing judgment for an expression $\ZP{E}$ is
$\ZJcomp{\ZP{P}; \widehat\sigma}{E}{(\IS{S},\widehat v)}$.
It states that tracing $\ZP{E}$ under program $\ZP{P}$ and state $\widehat{\sigma}$
produces a \IR*{} statement $\IP{S}$ and a tracing value $\widehat{v}$.
For example, the rules for operators
choose between partial evaluation and symbolic tracing,
according to the traced values of the arguments.
If all arguments trace to compile-time values,
\Cref{rule:trans-norm-ir-expr-frag:c-staticop}
evaluates the operator and returns the result $\ZP{v'}$ as a compile-time value,
where $\Zsem{\cdot}$ denotes operator evaluation.
Otherwise,
\Cref{rule:trans-norm-ir-expr-frag:c-dynop}
emits \IR*{} code for runtime operator evaluation
and returns a fresh \IR*{} variable $\IP{y'}$ that will hold the result.
In the rule, $\Tmat(\widehat{v}')$ returns $(\IP{S'},\allowbreak \Tref{\primtype}{x'})$,
where $\IP{x'}$ will store the value represented by $\widehat{v}'$ after $\IP{S'}$ runs;
the definition is in \cref{sec:compiler-def:trace}.

\paragraph{Tracing Statements}

The tracing judgment for a statement $\ZP{S}$ is
$\smash{\ZJcomp{\ZP{P}; \widehat\sigma}{S}{(\IS{S'},\widehat\sigma')}}$.
It states that tracing $\smash{\ZP{S}}$ under program $\smash{\ZP{P}}$ and state $\smash{\widehat{\sigma}}$
produces a \IR*{} statement $\smash{\IP{S'}}$ and an updated state $\smash{\widehat{\sigma}'}$.
For example, the rules for conditionals choose between partial evaluation and symbolic tracing,
according to the traced value of the condition $\ZP{E}$.
If $\ZP{E}$ traces to $\Tcon{\ZEtrue}$,
\Cref{rule:trans-norm-ir-stmt-frag:c-iftrue} traces only the true branch;
the $\Tcon{\ZEfalse}$ case is analogous.
If $\ZP{E}$ traces to $\Tref{\ITbool}{x}$, %
\Cref{rule:trans-norm-ir-stmt-frag:c-dynif} traces \emph{both} branches from the same starting state,
and \emph{merges} the generated \IR*{} code and tracing states
into a single \IR*{} conditional $\smash{\IP{S''}}$ and a merged state $\smash{\widehat\sigma''}$ for subsequent tracing.
The rule then returns $\smash{\IP{S_0';S''}}$ and $\smash{\widehat\sigma''}$,
where $\smash{\IP{S_0'}}$ is the \IR*{} code generated when tracing the condition $\ZP{E}$.
Defining $\Tjoin(\cdot)$ requires care in
combining compile-time and runtime values across branches and
representing composite values using primitive \IR*{} variables;
its full definition is in \cref{sec:compiler-def:trace}.

\enlargethispage*{1.5\baselineskip} %

\paragraph{Tracing Programs}

The tracing judgment for a function $\smash{\ZP{f}}$ is
$\smash{\ZP{P}\vdash\ZE{f}\leadsto\IP{P'}}$.
It states that tracing an entry function $\smash{\ZP{f}}$ of program $\ZP{P}$ produces a \IR*{} program $\IP{P'}$.
\Cref{rule:trans-norm-ir-stmt-frag:c-prog}
starts tracing from an initial state
that maps \Znorm*{} parameters of $\smash{\ZP{f}}$ to the corresponding \IR*{} parameters of $\smash{\IP{f}}$.

\paragraph{Staging Guarantee and Soundness}

We show that tracing respects \Ziv*{}'s input-dependency labels:
tracing an expression of $\bot$-labeled primitive type produces a compile-time value.
This ensures that computations marked as input-independent by \Ziv*{}'s type system are evaluated at compile time.

\begin{theorem}[$\bot \,{\Rightarrow}\, \Tcon{}$]
  \label{thm:trace-bot-static-main}
  Let {$\ZP{P}$ be a well-typed \Znorm*{} program}.
  Suppose an expression $\ZP{E}$ has primitive type $\ZP{\primtype^\bot}$ under context $\ZP{\Gamma}$, %
  i.e., $\ZJtype{\ZP{P}; \Gamma}{E}{\primtype^\bot}$.
  For all tracing states $\widehat{\sigma}$ ``compatible'' with $\ZP{\Gamma}$,
  \begin{align*}
    \ZJcomp{\ZP{P}; \widehat{\sigma}}{E}{(\IS{S}, \widehat{v})}
    \quad {\implies} \quad
    \exists \ZP{v}.\;
    \widehat{v} = \Tcon{\ZP{v}}
    \land
    \ZJtype{\ZP{P}; \varepsilon}{v}{\primtype^\bot}.
  \end{align*}
\end{theorem}

\noindent
The general version of the theorem (with the formal definition of ``compatibility'')
as well as type and semantic preservation theorems for tracing are given in \Cref{sec:comp-res:trace}.

\subsection{Code Emission (\texorpdfstring{$\IR* \to \Czero*$}{ZivIR → C₀})}
\label{sec:compiler:emit}

The code emission step translates \IR*{} programs into \Czero*{}, a simplified fragment of C.
The syntax, semantics, type system, and type preservation theorem of \Czero*{}
are given in \zcref{sec:lang-def-irc0,sec:lang-res:c0}.
Since the tracing step has already evaluated compile-time computations and lowered high-level constructs,
this step is a simple syntax-directed translation, which we denote by $\Ctrans$.
For example,
\begin{align*}
  \Ctrans\bigl(
  \ISif{(\bar x:\bar\primtype)}{y}
  {S_1;(\bar z_1)}{S_2;(\bar z_2)}\bigr)
  \coloneqq
  \CSseq{\CSdecl{\bar\primtype}{\bar x}}{
    \CSif{y}
    {\CSseq{\Ctrans(\IS{S_1})}{\CSassn{\bar x}{\bar z_1}}}
    {\CSseq{\Ctrans(\IS{S_2})}{\CSassn{\bar x}{\bar z_2}}}
  }
\end{align*}
The rule declares $\bar{x}$ before the conditional and assigns them in each branch.
The full translation is given in \cref{sec:compiler-def:emit},
and its type and semantic preservation theorems are stated in \cref{sec:comp-res:emit}.

\subsection{Soundness of Compilation}
\label{sec:compiler:soundness}

We establish a soundness theorem for end-to-end compilation from \Ziv*{} to \Czero*{}.
It states that
(i) successful compilation of a function $\smash{\ZP{f}}$ in a well-typed \Ziv*{} program $\ZP{P}$
produces a well-typed \Czero*{} program $R$;
and (ii) for each well-typed input, the compiled function $f$ in $R$ evaluates to a value
if and only if the original function $\smash{\ZP{f}}$ evaluates to the same value.
The proof combines the type and semantic preservation theorems for the three compiler steps
(see \cref{sec:comp-res:all}).

\begin{theorem}[Compiler Soundness]
  \label{thm:compiler-sound-main}
  Let {$\ZP{P}$ be a well-typed \Ziv*{} program}
  containing a function $\smash{\ZP{f}}$ of type $\ZTfun{(\comptype_1, \ldots, \comptype_n)}{\comptype'}$.
  Suppose compiling $\smash{\ZP{f}}$ produces $\CP{R}$,
  i.e., $\smash{\ZJcomp{{\Ztrans(\ZP{P})}}{f}{\IP{Q}}}$ and $\Ctrans(\IP{Q}) = \CP{R}$.
  Then, {$\CP{R}$ is a well-typed \Czero*{} program},
  and for all values $v_1, \ldots, v_n, v'$ with each $v_i$ of type $\ZP{\comptype_i}$,
  \begin{align*}
    \ZJeval{P}{f(v_1, \ldots, v_n)}{v'}
    \quad {\iff} \quad
    \CJeval{R}{f(v_1, \ldots, v_n)}{v'}.
  \end{align*}
\end{theorem}

\noindent
Our Lean formalization of all theorems is in the supplementary material (see \cref{sec:lang-res,sec:comp-res}).

%% file: figs/fig-ir-syntax-frag.tex
\begin{figure}
  \begin{figdefs}
    \setlength{\fboxrule}{0.4pt}
    \setlength{\parskip}{0pt}
    \setlength{\abovedisplayskip}{0pt} %
    \setlength{\abovedisplayshortskip}{\abovedisplayskip}
    \setlength{\belowdisplayskip}{\abovedisplayskip}
    \setlength{\belowdisplayshortskip}{\abovedisplayskip}
    \setlength{\MyMathparLineskip}{6pt}
    \def\MathparLineskip{\lineskip=\MyMathparLineskip}
    \vspace{-2pt}
    \begin{alignat*}{10}
      \text{Primitive types}
       &
       &
      \Ihl{\, \primtype \,\vphantom{f^8_8}}
       & \ \Coloneqq \
      \Ihl{\,
        \ITbool
        \mid \ITint{n}
        \mid \ITuint{n}
        \mid \ITfloat{n}
        \,\vphantom{f^8_8}
      }
      \\[-2.0pt]
      \text{Primitive operators}
       &
       & \quad
      \Ihl{\, \IEop \,\vphantom{f^8_8}}
       & \ \Coloneqq \
      \Ihl{\,
        \IEop_{\Ikw{Bool}}
        \mid \IEop_{\Ikw{Int}}
        \mid \IEop_{\Ikw{UInt}}
        \mid \IEop_{\Ikw{Float}}
        \mid \IEop_{\Ikw{Cast}}
        \,\vphantom{f^8_8}
      }
      \\[-2.0pt]
      \text{Expressions}
       &
       &
      \Ihl{\, E \,\vphantom{f^8_8}}
       & \ \Coloneqq \
      \Ihl{\,
      \IEval{v}
      \mid \IEvar{x}
      \mid \IEapp{\IEop[{\ITfun{\bar\primtype}{\primtype}]}}{\IEvar{\bar x}}
      \mid \IEget{\IEvar{x},\IEvar{x}}
      \,\vphantom{f^8_8}
      }
      \\[-2.0pt]
      \text{Statements}
       &
       &
      \Ihl{\, S \,\vphantom{f^8_8}}
       & \ \Coloneqq \
      \Ihl{\,
        \ISskip
        \mid \ISassn{\IEvar{x}:\primtype}{E}
        \mid \ISassn{\IEvar{x}:\IT{\primtype[n]}}{\IEnew{\IEval{\bar v}}}
        \mid \ISseq{S}{S}
        \mid \ISif{(\IEvar{\bar{x}}:\bar\primtype)}{\IEvar{x}}{S; \IEvar{\bar{x}}}{S; \IEvar{\bar{x}}}
        \,\vphantom{f^8_8}
      }
      \\[-2.0pt]
      \text{Programs}
       &
       &
      \IP{\, P \,\vphantom{f^8_8}}
       & \ \Coloneqq \
      \Ihl{\,
        \IPdef{f}{\IEvar{\bar x}:\bar\primtype}{\primtype}{S; \ISret{\IEvar{x}}}
        \,\vphantom{f^8_8}
      }
    \end{alignat*}
  \end{figdefs}
  \caption{Core syntax of \IR*{}.}
  \label{fig:ir-syntax-frag}
\end{figure}

%% file: figs/fig-trans-norm-ir-all-frag.tex
\begin{figure}[!t]
  \begin{figdefs}
    \msmall
    \setlength{\fboxsep}{2pt}
    \setlength{\fboxrule}{0.4pt}
    \setlength{\parskip}{0pt}
    \setlength{\abovedisplayskip}{2pt} %
    \setlength{\abovedisplayshortskip}{\abovedisplayskip}
    \setlength{\belowdisplayskip}{0pt} %
    \setlength{\belowdisplayshortskip}{\abovedisplayskip}
    \setlength{\MyMathparLineskip}{6pt}
    \def\MathparLineskip{\lineskip=6pt}
    \vspace{-2pt}
    \textbf{Definitions} (tracing values, tracing states)\hfill\fbox{$\widehat v$, $\widehat\sigma$}%
    \begin{align*}
      \widehat v     \Coloneqq \Tcon{\ZV{v}} \mid \Tref{\IT{\primtype}}{\IEvar{x}}
      \mid \Tcomp{{\theta}}{\widehat v_1,\ldots,\widehat v_n}
      \qquad\qquad
      \widehat\sigma \Coloneqq \varepsilon
      \mid \widehat\sigma, \ZE{x}\mapsto\widehat v
    \end{align*}
    \par\vspace{\dimexpr\MyMathparLineskip-\abovedisplayskip\relax}
    \nointerlineskip
    \hbox to\linewidth{%
      \rule{\dimexpr\linewidth\relax}{0pt}\hss}
    \nointerlineskip
    \vspace{\MyMathparLineskip}
    \vspace{-6pt}
    \textbf{Expressions}\hfill\fbox{$\ZJcomp{\ZP{P}; \widehat\sigma}{E}{(\IS{S},\widehat v)}$}%
    \begin{mathpar}
      \Tinfer{C-Const}
      { }
      {\ZJcomp{\ZP{P}; \widehat\sigma}{v}{(\ISskip, \Tcon{\ZV{v}})}}
      \label{rule:trans-norm-ir-expr-frag:c-const}
      \and
      \Tinfer{C-Var}
      {\widehat\sigma(\ZE{x})=\widehat v}
      {\ZJcomp{\ZP{P}; \widehat\sigma}{x}{(\ISskip,\widehat v)}}
      \label{rule:trans-norm-ir-expr-frag:c-var}
      \and
      \Tinfer{C-OpCt}
      {
        \forall i\in[n].\
        \ZJcomp{\ZP{P}; \widehat\sigma}{E_i}{(\IS{S_i},\Tcon{\ZV{v_i}})}
        \\
        {\begin{array}[b]{@{}c@{}}
            \Zsem{\Zhl{\ZEop[{\ZTfun{(\primtype_1, \ldots, \primtype_n)}{\primtype'}}]}}
            (\Zhl{\ZV{v_1}, \ldots, \ZV{v_n}})=\ZV{v'}
          \end{array}}
      }
      {\ZJcomp{\ZP{P}; \widehat\sigma}{\ZEop^\ell[{\ZTfun{(\primtype_1, \ldots, \primtype_n)}{\primtype'}}](E_1, \ldots, E_n)}
        {(\IS{S_1;\cdots;S_n}, {\Tcon{\Zhl{v'}}})}}
      \label{rule:trans-norm-ir-expr-frag:c-staticop}
      \and
      \Tinfer{C-OpRt}
      {
        \forall i\in[n].\
        \ZJcomp{\ZP{P}; \widehat\sigma}{E_i}{(\IS{S_i},\widehat v_i)}
        \qquad
        \exists j \in[n].\ \tcr{\forall \ZV{v}.\ \widehat v_j \neq \Tcon{\ZV{v}}}
        \\
        \forall i\in[n].\ \tcr{\Tmat(\widehat v_i)} = (\IS{S_{i+n}},\Tref{\IT{\primtype_i}}{\IEvar{x_i}})
        \\
        \IS{
          S' \defas (
          S_1;\ \cdots;\ S_{2n};\
          \tcr{
            \ISassn{
              \IEvar{y'} \!: \primtype'
            }{
              \IEapp{\IEop[{\ITfun{(\primtype_1, \ldots, \primtype_n)}{\primtype'}}]}{\IEvar{x_1}, \ldots, \IEvar{x_n}}
            }
          }
          )
        }
        \text{ for fresh $\IE{y'}$}
      }
      {
        \ZJcomp{\ZP{P}; \widehat\sigma}{\ZEop^\ell[{\ZTfun{(\primtype_1, \ldots, \primtype_n)}{\primtype'}}](E_1, \ldots, E_n)}
        {(\IS{S'},
          \tcr{\Tref{\IT{\primtype'}}{\IEvar{y'} \!}})}
      }
      \label{rule:trans-norm-ir-expr-frag:c-dynop}
    \end{mathpar}
    \par\vspace{\dimexpr\MyMathparLineskip-\abovedisplayskip\relax}
    \nointerlineskip
    \hbox to\linewidth{%
      \rule{\dimexpr\linewidth\relax}{0pt}\hss}
    \nointerlineskip
    \vspace{\MyMathparLineskip}
    \vspace{-6pt}
    \textbf{Statements}\hfill\fbox{$\ZJcomp{\ZP{P}; \widehat\sigma}{S}{(\IS{S'},\widehat\sigma')}$}%
    \begin{mathpar}
      \Tinfer{C-Assn}
      {\ZJcomp{\ZP{P}; \widehat\sigma}{E}{(\IS{S},\widehat v)}}
      {
        \ZJcomp{\ZP{P}; \widehat\sigma}{\ZSassn{x}{E}}
        {(\IS{S},(\widehat\sigma, \Zhl{x}\mapsto \widehat v))}
      }
      \label{rule:trans-norm-ir-stmt-frag:c-assn}
      \and\!\!\!\!\!\!
      \Tinfer[width=0.2\linewidth]{C-IfTrue}
      {\ZJcomp{\ZP{P}; \widehat\sigma}{E}
        {(\IS{S'_0},\Tcon{\ZEtrue})} %
        \\ \ZJcomp{\ZP{P}; \widehat\sigma}{S_1}{(\IS{S'_1},\widehat\sigma'_1)}}
      {\ZJcomp{\ZP{P}; \widehat\sigma}{\ZSif{E}{S_1}{S_2}}
        {(\IS{S'_0;S'_1},\widehat\sigma'_1)}}
      \label{rule:trans-norm-ir-stmt-frag:c-iftrue}
      \and
      \Tinfer{C-IfRt}
      {
        \ZJcomp{
          \ZP{P}; \widehat\sigma
        }{E}{
          ( \IS{S'_0}, \tcr{\Tref{\ITbool}{\IEvar{x}}} )
        }
        \\
        \forall i \in \{1,2\}.\
        \ZJcomp{\ZP{P}; \widehat\sigma}{S_i}{(\IS{S'_i},\widehat\sigma'_i)}
        \\
        \tcr{
          \Tjoin(
          \IEvar{x},
          (\IS{S'_1},\widehat\sigma'_1),
          (\IS{S'_2},\widehat\sigma'_2)
          )
        }
        =(\IS{S''},\widehat\sigma'')
      }
      {\ZJcomp{\ZP{P}; \widehat\sigma}{\ZSif{E}{S_1}{S_2}}
        {(\IS{S'_0;S''},\widehat\sigma'')}}
      \label{rule:trans-norm-ir-stmt-frag:c-dynif}
    \end{mathpar}
    \par\vspace{\dimexpr\MyMathparLineskip-\abovedisplayskip\relax}
    \nointerlineskip
    \hbox to\linewidth{%
      \rule{\dimexpr\linewidth\relax}{0pt}\hss}
    \nointerlineskip
    \vspace{\MyMathparLineskip}
    \vspace{-6pt}
    \textbf{Programs}\hfill\fbox{$\ZJcomp{\ZP{P}}{\ZE{f}}{\IP{P'}}$}%
    \begin{mathpar}
      \Tinfer{C-Prog}
      {
        {\begin{array}[b]{@{}c@{}}
            \smash{\Zhl{\ZP{P}
                =(\ZPdef{f_j}{\bar{x_j}:\bar{\comptype_j}}{\comptype_j'}
                {\ZSseq{S_j}{\ZSret{E_j}}})_{j \in [n]}}}
            \\
            i \in [n]
            \land
            \smash{\Zhl{
                \bar{\comptype_i} = \tcr{(\bar{\primtype_i})^\top}
                \land
                \comptype_i' = \tcr{(\primtype_i')^\top}
              }}
            \\
            \widehat\sigma
            \defas
            (
            \ZV{\bar{x_i}} \mapsto
            \tcr{ \Tref{\IT{\bar{\primtype_i}}}{\IEvar{\bar{y_i}}} }
            )
            \text{ for fresh } \IEvar{\bar{y_i}}
          \end{array}}
        \\
        {\begin{array}[b]{@{}c@{}}
            \ZJcomp{\ZP{P}; \widehat\sigma}{S_i}{(\IS{S'_1},\widehat\sigma')}
            \\
            \ZJcomp{\ZP{P}; \widehat\sigma'}{E_i}{(\IS{S'_2},\widehat v')}
            \\
            \tcr{\Tmat(\widehat v')}=(\IS{S'_3},\Tref{\IT{\primtype_i'}}{\IEvar{z'}})
          \end{array}}
      }
      {
        \ZJcomp{\ZP{P}}{\ZE{f_i}}{
          \IPdef{f_i}{\IEvar{\bar{y_i}}:\bar{\primtype_i}}{\primtype_i'}
          {S'_1;S'_2;S'_3;\ISret{\IEvar{z'}}}
        }
      }
      \label{rule:trans-norm-ir-stmt-frag:c-prog}
    \end{mathpar}
    \vspace{-7pt}
  \end{figdefs}
  \caption[]{%
    Selected tracing rules ($\Znorm* \to \IR*$).
  } %
  \label{fig:trans-norm-ir-expr-frag}
\end{figure}

%% file: sections/sec06-implementation.tex
\section{Implementation}
\label{sec:impl}

We implemented \Ziv\ as an embedded language in Julia~\cite{julia},
exploiting its multiple dispatch, type system, and metaprogramming.
We now discuss the implementation of \Ziv\ and its compiler.

\paragraph{Language}

Both \Ziv’s primitive operators and user-defined functions
are implemented as Julia methods,
which may be overloaded.
Julia’s \emph{multiple dispatch} selects which method to invoke based on the types of all arguments.
This allows integer and floating-point addition to share the syntax \texttt{+},
as assumed in \zcref{sec:lang:syntax}.
The user-defined functions in
\ZivStd\ use the same mechanism to extend \texttt{+} to addition between
two multi-word values or between a multi-word value and a floating-point value.

Most type checking is delegated to Julia's type system.
\Ziv's primitive types are represented as native Julia types,
and composite types as Julia \texttt{struct}s.
While our discussion of \Ziv\ omitted type inference,
it is implemented through Julia's type system,
including inference for labels.

\paragraph{Compiler}

The normalization step of the compiler (\zcref{sec:compiler:norm})
requires syntactic transformations,
which are implemented with Julia's metaprogramming.
The \texttt{@ziv} macro applies these transformations.

During tracing (\zcref{sec:compiler:trace}),
a single function supports
both partial evaluation and symbolic tracing through multiple dispatch.
A static value has a Julia type \texttt{T}
representing a \IR\ primitive type
(e.g., \texttt{Float64}), while a dynamic
value has the parameterized type \texttt{Tracer\{T\}}.
Multiple dispatch allows the same function to be overloaded
on both \texttt{T} and \texttt{Tracer\{T\}}.
For example, a function call
\texttt{f(x)} is partially evaluated when \texttt{x} has type \texttt{T},
and is symbolically traced when \texttt{x} has type \texttt{Tracer\{T\}}.

%% file: sections/sec07-evaluation.tex
\section{Evaluation}\label{sec:eval}
\input{figs/fig-eval-existing-libraries}

We evaluate whether \Ziv\ makes correctly rounded math libraries
easier to write, understand, and modify, addressing the
challenges identified in \zcref{sec:intro:existing-approach}.
We consider four research questions:

\begin{enumerate}[label=\textbf{RQ\arabic*.},ref=RQ\arabic*,leftmargin=*,wide=0pt]
  \item\label{rq:eval:perf} {\itshape Can \Ziv{} express existing implementations
          while achieving competitive performance?}
  \item\label{rq:eval:concise} {\itshape Can \Ziv{} express existing implementations more concisely?}
  \item\label{rq:eval:novel} {\itshape Is \Ziv{} effective for writing new implementations?}
  \item\label{rq:eval:variants} {\itshape  Is \Ziv{} effective for exploring implementation variants?}
\end{enumerate}

We investigate these questions through three case studies:
reimplementing existing libraries (\zcref{sec:eval:existing}),
writing new implementations (\zcref{sec:eval:novel}), and
exploring implementation variants (\zcref{sec:eval:variants}).

\paragraph{Experimental Setup}

We use the Julia-embedded implementation of \Ziv\ (\zcref{sec:impl}) to compile
\Ziv\ programs to C.
We compile both \Ziv-emitted C code and baseline C/C++ code
with Clang 21.1 using flags \texttt{-O3 -march=native}.
All experiments are run on an AMD EPYC\texttrademark\ 9575F processor.

We compare \Ziv\ implementations with
baseline libraries in terms of accuracy and performance.
Accuracy is measured by the error between exact and computed values,
expressed in \emph{ulp} (units in the last place)~\cite{fphandbook}.
Broadly, error expressed in ulp counts the number of floating-point steps between
the exact and computed values.
Under the round-to-nearest mode,
correct rounding implies a maximum error of \SI{0.5}{\ulp},
attained when the exact result lies halfway between two adjacent floats.
Performance is measured by reciprocal throughput,
expressed in CPU cycles per function call.

We focus our performance and accuracy tests on 64-bit floating-point inputs,
as support for larger formats is still nascent~\cite{crsurvey,llvmlibc}.
Accuracy tests use $10^9$ bit-random inputs,
in addition to available hard-to-round inputs~\cite{worstelem,worsttrig}.
Performance tests report the minimum cycles per call over 10 trials.
Each trial is performed on 200,000 inputs
randomly sampled from function-specific intervals~\cite{coremath}.

\subsection{Reimplementing Existing Libraries in \Ziv{} (\ref{rq:eval:perf}, \ref{rq:eval:concise})}
\label{sec:eval:existing}
\input{figs/tab-eval-existing-libraries-loc}

We reimplement existing correctly rounded math libraries in \Ziv\ to
evaluate our system's
expressivity and performance (\ref{rq:eval:perf}),
as well as conciseness (\ref{rq:eval:concise}).
We choose CORE-MATH~\cite{coremath}
and LLVM-libc~\cite{llvmlibc} as baseline libraries for reimplementation,
as they represent the state of the art in math libraries
claiming correct rounding for 64-bit inputs.
We reimplement 11 functions from CORE-MATH commit \texttt{07cf01e1}
and 8 functions from LLVM-libc commit \texttt{efe31d6d}.

The \Ziv\ reimplementations matched
CORE-MATH and LLVM-libc bit for bit on all tested inputs,
demonstrating \Ziv’s ability to express correctly rounded functions (\ref{rq:eval:perf}).
Reimplementing the baselines in \Ziv\ also exposed
numerical bugs that, to our knowledge,
were previously undocumented.
LLVM-libc's \texttt{cos} was not correctly rounded,
and its floating-point dyadic addition
could return incorrect exponents.
Julia's \texttt{ldexp} also was not correctly rounded.
Furthermore, we recovered the mathematical definitions
for some of CORE-MATH's undocumented polynomials and tables.

We compare the performance of our \Ziv\ reimplementations
with the original baselines in \zcref{fig:eval-existing-libraries},
demonstrating that \Ziv\ can achieve
competitive performance (\ref{rq:eval:perf}).
Relative to CORE-MATH, \Ziv\ delivers
$0.83\times$ performance on average (geometric mean),
ranging from $0.65\times$ to $1.03\times$.
Relative to LLVM-libc, \Ziv\
delivers $0.82\times$ performance on average,
ranging from $0.74\times$ to $0.96\times$.
Using \Ziv's high-level abstractions (\ref{dp:abstraction}) instead
of the function-specific code of existing implementations
incurs modest performance overhead but substantially reduces code size,
as we quantify next.

We report line counts of \Ziv\ and the original baselines
in \cref{tab:eval-existing-libraries-loc},
showing that \Ziv\ can express existing libraries more concisely (\ref{rq:eval:concise}).
Our \Ziv\ reimplementation gives a
line count ratio of {2.17:1} (53.8\% reduction) against CORE-MATH's
original implementation, and {3.59:1} (72.1\% reduction) against LLVM-libc.
Our reimplementations show that \Ziv\ can express
existing correctly rounded libraries and achieve competitive
performance while providing more concise implementations.
In addition, \Ziv\ code can be more readable than existing implementations
since it expresses high-level mathematical intent more directly,
as compared in \zcref{sec:intro,sec:overview}
(\e.g., \zcref{fig:intro-code,fig:overview-exp}; \zcref{sec:overview:how,sec:overview:where}).

\subsection{Writing New Implementations in \Ziv{} (\ref{rq:eval:novel})}\label{sec:eval:novel}

\input{figs/tab-eval-novel-functions}

We write new implementations in \Ziv\ to examine its
effectiveness in developing new math library implementations (\ref{rq:eval:novel}).
We focus on three functions that, to our knowledge,
currently lack correctly rounded implementations in existing fixed-precision math libraries:
the dilogarithm~\cite{dilog}, GELU~\cite{gelu}, and bivariate softmax~\cite{softmax}.
For each function, we evaluate \Ziv\ and baseline implementations
targeting correct rounding as well as relaxed accuracy.
We assess correct rounding empirically.

We report the accuracy and performance of
the new \Ziv\ implementations and their baselines in
\zcref{tab:eval-novel-functions}.
For targeting correct rounding ($\leq \SI{0.5}{\ulp}$),
\Ziv\ matches the maximum observed errors of baselines using
the MPFR~\cite{mpfr} arbitrary-precision library.
We use MPFR baselines as we found no correctly rounded
fixed-precision alternatives.
Avoiding the overhead of arbitrary precision,
hand-written \Ziv\ achieves
from $9.52\times$ to $359.7\times$ the performance of the baselines.
Searched \Ziv\ implementation variants (\cref{sec:eval:variants})
yield from $1.57\times$ to $2.35\times$ the performance of hand-written \Ziv.

For the relaxed accuracy target ($> \SI{0.5}{\ulp}$) in \zcref{tab:eval-novel-functions},
searched \Ziv\ implementations achieve lower maximum observed errors than the
GNU Scientific Library (GSL)~\cite{gsl}
and C baseline implementations~\cite{pytorch} for all three functions.
The \Ziv\ implementations achieve $6.17\times$ the performance of the GSL dilogarithm
and $0.96\times$ and $1.24\times$ the performance of
the C baselines for GELU and softmax, respectively.
Our new implementations show that \Ziv\ supports developing
math library functions with different accuracy goals,
with comparable or better performance than existing libraries.

\subsection{Exploring Implementation Variants with \Ziv{} (\ref{rq:eval:variants})}
\label{sec:eval:variants}
\input{figs/fig-eval-variants}

We parameterize, generate, and test \Ziv\ implementations to assess
whether \Ziv\ supports systematic exploration of
implementation variants (\ref{rq:eval:variants}).
For the three functions in \zcref{sec:eval:novel},
we parameterize their \Ziv\ code
by working precisions, table sizes, and polynomial degrees
and automatically generate variants by varying these parameters.
To keep exploration manageable,
we test each variant's accuracy on $10^7$ random inputs and
hard-to-round inputs found in the experiments of \zcref{sec:eval:novel}.

We explore the accuracy--performance trade-offs
of the generated \Ziv\ implementation variants in \zcref{fig:eval-variants}.
We use combinations of working precisions, table sizes, and polynomial degrees
to generate 72,000 plotted variants for the dilogarithm,
734,400 for GELU,
and 17,856 for softmax.
Variant counts and accuracy--performance trends differ as
each function is parameterized differently,
with parameters having distinct effects on accuracy and performance.
Horizontal plateaus in the plots appear
when increasing the polynomial degree
no longer improves accuracy due to other
factors, such as insufficient working precision.
Within the explored parameter ranges,
no variant using only 64-bit floats is correctly rounded,
demonstrating the need for \Ziv's extended-precision types.

During our experiments,
we found that \Ziv's staging (\ref{dp:staging}) make it well-suited
for generating and exploring implementation variants through parameterization.
For example,
our parameterized GELU implementation has the simplified label signature of
$\texttt{gelu}(\texttt{x}:\top,\ \texttt{DEGREES}:\bot,\ \texttt{TABLES}:\bot)$.
The \Ziv\ compiler specializes the implementation
to the compile-time values of \texttt{DEGREES} and \texttt{TABLES},
leaving \texttt{x} as the only input at runtime.
Sweeping over different values of these parameters automatically
generates implementation variants with different polynomial degrees and table sizes.

%% file: figs/fig-eval-existing-libraries.tex
\begin{figure}[t]
  \centering
  \includegraphics[width=\linewidth]{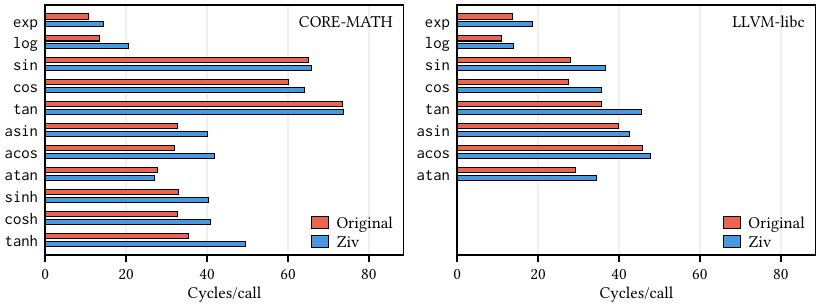}
  \caption{Cycles per call for original CORE-MATH and LLVM-libc
    functions and their \Ziv{} reimplementations.}
  \label{fig:eval-existing-libraries}
\end{figure}

%% file: figs/tab-eval-existing-libraries-loc.tex
\begin{table}[t]
  \caption{%
    Source lines of code (SLOC) in original CORE-MATH and LLVM-libc
    implementations and their \Ziv\ reimplementations.
    For fairness,
    we exclude numerical support code, and
    SLOC for individual functions only include function-specific code;
    code shareable within a function family is counted only in its \emph{Shared} row.
  }
  \label{tab:eval-existing-libraries-loc}
  \centering
  \begingroup
  \small
  \setlength{\aboverulesep}{0.3ex}
  \setlength{\belowrulesep}{0.5ex}
  \begin{booktabs}{
      width=\linewidth,
      colspec={l X[l] r r r r r r},
      rowsep=0.3pt,stretch=0.5,
      column{2}={rightsep=13pt}, column{5}={rightsep=13pt},
      row{Z}={font=\bfseries},
    }
    \toprule
    &  & \SetCell[c=3]{c}\textbf{CORE-MATH} & & & \SetCell[c=3]{c}\textbf{LLVM-libc} & & \\
    \cmidrule[lr]{3-5}\cmidrule[l]{6-8}
    Function Family & Function
    & Original & \Ziv{} & Original\hspace{1pt}:\hspace{1pt}\Ziv
    & Original & \Ziv{} & Original\hspace{1pt}:\hspace{1pt}\Ziv \\
    \midrule
    Exponential & \texttt{exp} & 210 & 136 & 1.54 : 1 & 466 & 123 & 3.79 : 1 \\
    \midrule
    Logarithmic & \texttt{log} & 1,051 & 155 & 6.78 : 1 & 941 & 125 & 7.53 : 1 \\
    \midrule
    Trigonometric & \texttt{sin} & 140 & 12 & 11.67 : 1 & 112 & 69 & 1.62 : 1 \\
    & \texttt{cos} & 138 & 12 & 11.50 : 1 & 111 & 67 & 1.66 : 1 \\
    & \texttt{tan} & 195 & 109 & 1.79 : 1 & 138 & 87 & 1.59 : 1 \\
    & Shared & 1,012 & 565 & 1.79 : 1 & 588 & 119 & 4.94 : 1 \\
    \cmidrule[l]{2-8}
    & Subtotal & 1,485 & 698 & 2.13 : 1 & 949 & 342 & 2.77 : 1 \\
    \midrule
    Inverse Trig. & \texttt{asin} & 133 & 76 & 1.75 : 1 & 89 & 59 & 1.51 : 1 \\
    & \texttt{acos} & 108 & 63 & 1.71 : 1 & 99 & 79 & 1.25 : 1 \\
    & \texttt{atan} & 192 & 133 & 1.44 : 1 & 163 & 58 & 2.81 : 1 \\
    & Shared & 91 & 95 & 0.96 : 1 & 406 & 82 & 4.95 : 1 \\
    \cmidrule[l]{2-8}
    & Subtotal & 524 & 367 & 1.43 : 1 & 757 & 278 & 2.72 : 1 \\
    \midrule
    Hyperbolic & \texttt{sinh} & 230 & 115 & 2.00 : 1 &
    \SetCell[r=5,c=3]{c,m,font=\footnotesize}
    {Not implemented in\\LLVM-libc commit \texttt{efe31d6d}.} & & \\
    & \texttt{cosh} & 193 & 86 & 2.24 : 1 & & & \\
    & \texttt{tanh} & 181 & 116 & 1.56 : 1 & & & \\
    & Shared & 69 & 147 & 0.47 : 1 & & & \\
    \cmidrule[lr]{2-5}
    & Subtotal & 673 & 464 & 1.45 : 1 & & & \\
    \midrule
    & \textbf{Total} & 3,943 & 1,820 & 2.17 : 1 & 3,113 & 868 & 3.59 : 1 \\
    \bottomrule
  \end{booktabs}
  \endgroup
\end{table}

%% file: figs/tab-eval-novel-functions.tex
\begin{table}[t]
  \caption{
    Maximum observed ulp errors and cycles per call for
    new functions implemented in \Ziv\ and their baselines.
    MPFR working precisions (in parentheses) are the minimum found sufficient in testing.
    C baselines reimplement PyTorch's numerical algorithms.
    \textsuperscript{\textdagger}Parentheses report the maximum over finite ulp errors.}
  \label{tab:eval-novel-functions}
  \centering
  \begingroup
  \small
  \setlength{\aboverulesep}{0.3ex}
  \setlength{\belowrulesep}{0.5ex}
  \sisetup{
    round-mode=places,
    round-precision=1,
    round-pad=true,
    group-digits=false
  }
  \begin{booktabs}{
      width=\linewidth,
      colspec={X[0.8,l] X[l] X[1.3,l] X[1.2,r] X[r]},
      rowsep=0.3pt,stretch=0.5,
    }
    \toprule
    Function & Target Accuracy & Implementation & Max Error (ulp) & {Cycles per Call} \\
    \midrule
    Dilogarithm & $\le \SI{0.5}{\ulp}$ & MPFR (53-bit) & \num{0.4999999986} & 44,197.6 \\
    & & \Ziv{} (hand-written) & \num{0.4999999986} & 1,209.8 \\
    & & \Ziv{} (searched) & \num{0.4999999986} & 768.0 \\
    \cmidrule[l]{2-5}
    & $> \SI{0.5}{\ulp}$ & GSL & $\infty$ ({3,060,232.6}\rlap{)\textsuperscript{\textdagger}} & 277.7 \\
    & & \Ziv{} (searched) & {2,364,539.2} & 45.0 \\
    \midrule
    GELU & $\le \SI{0.5}{\ulp}$ & MPFR (80-bit) & \num{0.5} & 267,155.9 \\
    & & \Ziv{} (hand-written) & \num{0.5} & 742.8 \\
    & & \Ziv{} (searched) & \num{0.5} & 383.4 \\
    \cmidrule[l]{2-5}
    & $> \SI{0.5}{\ulp}$ & C & \num[exponent-mode=scientific]{9.0072e15} & 36.2 \\
    & & \Ziv{} (searched) & \num[exponent-mode=scientific]{1.3433e13} & 37.7 \\
    \midrule
    Softmax & $\le \SI{0.5}{\ulp}$ & MPFR (79-bit) & \num{0.4999999959} & 4,137.0 \\
    (Bivariate) & & \Ziv{} (hand-written) & \num{0.4999999959} & 434.4 \\
    & & \Ziv{} (searched) & \num{0.4999999959} & 184.8 \\
    \cmidrule[l]{2-5}
    & $> \SI{0.5}{\ulp}$ & C & \num{511.525} & 38.8 \\
    & & \Ziv{} (searched) & \num{449.5433720517246} & 31.2 \\
    \bottomrule
  \end{booktabs}
  \endgroup
\end{table}

%% file: figs/fig-eval-variants.tex
\begin{figure}[t]
  \centering
  \includegraphics[width=\linewidth]{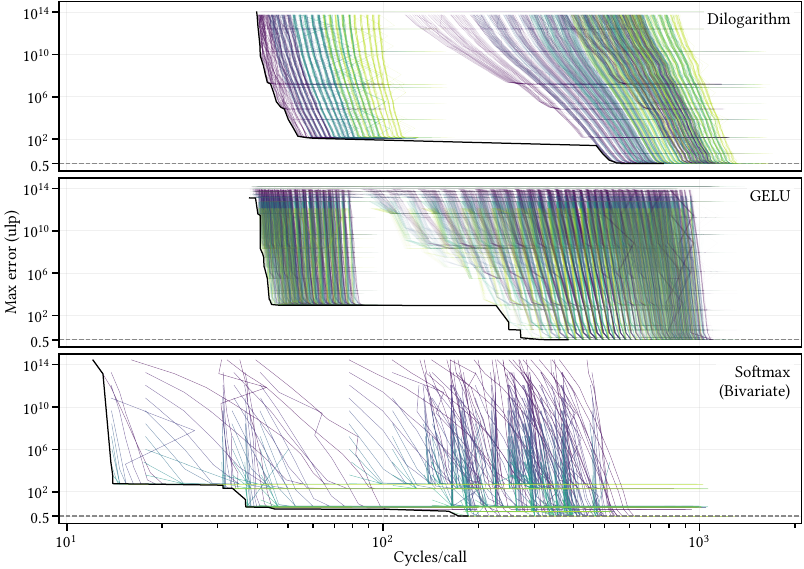}
  \caption{Maximum observed ulp errors versus cycles per call for
    implementation variants generated by \Ziv.
    Each colored curve varies polynomial degree while keeping all other parameters constant;
    curves are colored sequentially to indicate table size
    from \textcolor[HTML]{754F84}{small} to \textcolor[HTML]{5A9F6E}{large}.
    Black curves represent Pareto frontiers.}
  \label{fig:eval-variants}
\end{figure}

%% file: sections/sec08-related.tex
\section{Related Work}\label{sec:related}

\paragraph{Correctly Rounded Math Libraries}

Correctly rounded math libraries have garnered sustained interest
in both research and practice, as recently surveyed by \citet{crsurvey}.
\Ziv\ provides abstractions for established techniques in
floating-point arithmetic~\cite{fphandbook,fpsurvey}
and math library implementation~\cite{elementaryfns,fnhandbook,ia64}.
Several libraries target correct rounding,
including CR-LIBM~\cite{crlibm}, RLibm~\cite{rlibm},
CORE-MATH~\cite{coremath}, and LLVM-libc~\cite{llvmlibc}.
We evaluate \Ziv\ against the last two in \zcref{sec:eval:existing},
as they largely define the state of the art
in 64-bit math libraries.
Beyond fixed-precision math libraries,
libraries such as MPFR~\cite{mpfr} and Arb~\cite{arb}
support arbitrary precision,
which is outside our scope.
Despite advances in the verification
of math libraries~\cite{verify-pldi16,verify-popl18,verify-itp24,verify-arith23}, %
proving the correct rounding of these implementations %
still requires extensive manual effort~\cite{atanhcr,powcr}
and is also outside our scope.

\paragraph{Languages for Math Libraries}

Prior work has proposed languages for implementing math libraries,
but correct rounding has not been a primary line of research.
Metalibm~\cite{metalibm-icms}
aims to generate math functions from high-level specifications,
prioritizing automated program synthesis.
By comparison, \Ziv\ supports explicit control over numerical algorithms
in addition to high-level abstractions.
A subsequent version of Metalibm~\cite{metalibm-arith}
introduces a graph-based language for math library code generation,
but focuses on graph-level control of assembly code generation
rather than language design.
In contrast, \Ziv\ provides high-level abstractions tailored for
correctly rounded math libraries.

More recently, MegaLibm~\cite{megalibm} introduced
a high-level language for math libraries, albeit not targeting correct rounding.
Beyond their accuracy goals,
MegaLibm and \Ziv\ greatly differ in their design principles.
MegaLibm programs express \emph{intent} with real-valued expressions,
leaving floating-point details to the compiler.
By contrast, \Ziv\ programs specify floating-point \emph{implementations},
using an imperative language
with staging and abstractions for correct rounding.
Consequently,
\Ziv\ requires more manual programming but offers greater expressivity.
\Ziv\ supports techniques beyond MegaLibm's intent-oriented constructs,
including Ziv's method, tabulation, dyadics, and bitwise operations,
allowing it to reimplement state-of-the-art
correctly rounded libraries (\zcref{sec:eval:existing}).

\paragraph{Approaches to Staging}

Staging in programming languages has been extensively studied,
with broader context available in existing overviews~\cite{metasurvey,partialeval};
we focus here on approaches most relevant to \Ziv.
Lightweight Modular Staging (LMS)~\cite{lms} uses types to
distinguish static values from staged expressions
and to construct graphs of dynamic computations.
\Ziv\ similarly uses type labels and tracing,
but leaves final staging decisions to symbolic tracing rather than
the type system (\zcref{sec:compiler:trace}), unlike LMS.
Similarly, machine learning frameworks use
symbolic tracing to record programs~\cite{jax,tfeager}.

More classically, \citet{autobta} present a method for inferring
static and dynamic binding-time annotations for a typed $\lambda$-calculus.
\Ziv\ instead uses labels to soundly track which expressions
are guaranteed to statically evaluate (\zcref{sec:compiler:trace}).
Multi-stage programming~\cite{metaml} performs staging with
explicitly quasiquoted expressions,
and Terra~\cite{terra} uses similar expression-level annotations
to implement a Lua-embedded staged language.
By contrast, \Ziv's staging is performed implicitly using labels,
allowing unobtrusive programming without explicitly quoted expressions.

\section{Conclusion}

We introduced \Ziv{}, a programming system for developing correctly rounded math libraries.
\Ziv{} supports
high-level numerical abstractions~(\ref{dp:abstraction})
via extended-precision types and a rich standard library,
and staged computation~(\ref{dp:staging})
via compile-time evaluation of input-independent computations.
These features enable high-level code that unifies precomputation and runtime computation.
We demonstrated that \Ziv{} is effective for expressing existing implementations concisely,
writing practical implementations for new functions, and automatically exploring variants.

%% file: sections/sec09-ai-data.tex
\paragraph{AI Use}

The authors fully designed the \Ziv\ language, standard library,
and compiler.
The authors performed the vast majority of the implementation and evaluation,
with minor assistance from OpenAI Codex (GPT-6 Astra).
The only direct use of Codex during implementation
was to assist in porting polynomial approximation routines
and identifying undocumented constants in existing libraries;
all resulting code was independently tested and verified by the authors.
Codex was also used to construct portions of the Lean artifact,
with all proofs checked by the Lean kernel.

\begin{acks}
  We thank Hyeonjung Kim of POSTECH for his assistance in
  reimplementing CORE-MATH and LLVM-libc.
  We also thank Paul Zimmermann and Alexei Sibidanov of the CORE-MATH project,
  and Tue Ly of the LLVM-libc project, for their discussion and guidance on
  the design of correctly rounded math libraries.
  Additionally, we thank Guillaume Melquiond,
  Vincent Lefèvre,
  Claude-Pierre Jeannerod, and
  Tom Hubrecht of the Pascaline team at ENS Lyon for their valuable feedback.

  This work was supported by the National Research Foundation of Korea (NRF) under Grant RS-2026-25470496,
  funded by the Korea Ministry of Science and ICT (MSIT).
\end{acks}

%% file: appendices/app-lang-def-ziv.tex
\section{Complete Definition of the \texorpdfstring{\Ziv*{}}{Ziv} Language}
\label{sec:lang-def-ziv}

In this section, we define the syntax, type system, and operational semantics of the \Ziv*{} language.

\subsection{Syntax}
\label{sec:lang-def-ziv:syntax}

The abstract syntax of \Ziv{} is shown in \Cref{fig:ziv-syntax-full}.

\paragraph{Notation.}
We use the following notation.
\begin{itemize}
  \item
        $n \in \mathbb{Z}$ is an integer.
  \item
        $\ell \in \mathit{Lab}$ is a label,
        where $\mathit{Lab} \defas \{\bot, \top\}$ is ordered by
        $\bot \sqsubseteq \bot$, $\bot \sqsubseteq \top$, and $\top \sqsubseteq \top$.
  \item
        $x \in \mathit{Var}$ is a variable, where $\mathit{Var}$ is a countable set of variable names.
  \item
        $f \in \mathit{Fun}$ is a function, where $\mathit{Fun}$ is a countable set of function names.
  \item
        $v \in \ZVal_{\Ziv}$ is a value, where $\ZVal_{\Ziv}$ is a set of values in \Ziv{}.
        The set $\ZVal_{\Ziv}$ consists of all function names and all non-function values:
        \begin{align*}
          \ZVal_{\Ziv} \defas
          \mathit{Fun}
          \cup
          {
            \bigcup_{\substack{
                \comptype \in \Ziv
                \\
                \ZT{\comptype} \text{ is not a function type}
              }}
            \ZVal_{\Zerase(\comptype)}
          }.
        \end{align*}
        Here, $\Zerase(\ZT{\comptype})$ is the erased type of $\ZT{\comptype}$
        (i.e., the type without labels),
        and $\ZVal_\theta$ is the set of values of erased type ${\theta} = \Zerase(\comptype)$
        defined as follows:
        \begin{align*}
          \ZVal_{\ZTbool}
           & \defas \{\ZEval{\ZTbool}{c} \mid c \in \{\ZEtrue, \ZEfalse\}\},
          \\
          \ZVal_{\ZTint{n}}
           & \defas \{\ZEval{\ZTint{n}}{c} \mid c \in \mathbb{Z} \land c \in [-2^{n-1}, 2^{n-1}) \},
          \\
          \ZVal_{\ZTuint{n}}
           & \defas \{\ZEval{\ZTuint{n}}{c} \mid c \in \mathbb{Z} \land c \in [0, 2^n) \},
          \\
          \ZVal_{\ZTfloat{n}}
           & \defas \{\ZEval{\ZTfloat{n}}{c} \mid c \in \{0,1\}^n\}
          \text{ is the set of $n$-bit IEEE-754 floats},
          \\
          \ZVal_{\ZTreal}
           & \defas \{\ZEval{\ZTreal}{c} \mid c \in \mathbb{R} \},
          \\
          \ZVal_{\ZTdyadic{\primtype_1, \primtype_2}{n}}
           & \defas \{
          \ZEval{\ZTdyadic{\primtype_1, \primtype_2}{n}}{c}
          \mid c \in \ZVal_{\primtype_1} \times \ZVal_{\primtype_2}
          \},
          \\
          \ZVal_{\ZTdyadic{\primtype_1, \primtype_2}{\primtype_3}}
           & \defas \{
          \ZEval{\ZTdyadic{\primtype_1, \primtype_2}{\primtype_3}}{c}
          \mid c \in \ZVal_{\primtype_1} \times \ZVal_{\primtype_2} \times \ZVal_{\primtype_3}
          \},
          \\
          \ZVal_{\ZTmulwrd{\theta_1, \ldots, \theta_n}}
           & \defas \{
          \ZEval{\ZTmulwrd{\theta_1, \ldots, \theta_n}}{c}
          \mid c \in \ZVal_{\theta_1} \times \cdots \times \ZVal_{\theta_n}
          \},
          \\
          \ZVal_{\ZTtuple{\theta_1, \ldots, \theta_n}}
           & \defas \{
          \ZEval{\ZTtuple{\theta_1, \ldots, \theta_n}}{c}
          \mid c \in \ZVal_{\theta_1} \times \cdots \times \ZVal_{\theta_n}
          \},
          \\
          \ZVal_{\ZTtable{\theta}{(n_1, \ldots, n_k)}}
           & \defas \{ \textstyle
          \ZEval{\ZTtable{\theta}{(n_1, \ldots, n_k)}}{c}
          \mid c \in (\ZVal_{\theta})^{n_1 \times \cdots \times n_k}
          \},
          \\
          \ZVal_{\ZTfun{\bar{\theta}}{\theta'}}
           & \defas \varnothing.
        \end{align*}
        As shown above, each non-function value $\ZEval{\theta}{c}$ is tagged with its erased type $\theta$.
  \item
        $\comptype^\bot$ is a type in which all labels are $\bot$.
  \item
        A bar is a finite sequence, which can be empty. For example:
        \begin{itemize}
          \item $\bar{E}$ is $E_1, \ldots, E_n$ for some $n \ge 0$.
          \item $\bar{x} : \bar{\comptype}$ is $x_1 : \comptype_1, \ldots, x_n : \comptype_n$ for some $n \ge 0$.
        \end{itemize}
\end{itemize}

\paragraph{Syntactic Restrictions.}
We make the following restrictions on the syntax of \Ziv{}.
\begin{itemize}
  \item
        Types $\ZTint{n}$, $\ZTuint{n}$, and $\ZTfloat{n}$: $n \geq 1$.
  \item
        Type $\ZTdyadic{\primtype_1^{\ell_1},\! \primtype_2^{\ell_2}\!}{n}$:
        $\primtype_1$ is $\ZTbool$, $\primtype_2$ is $\ZTuint{k}$, and $n \in \mathbb{Z}$.
  \item
        Type $\ZTdyadic{\primtype_1^{\ell_1},\! \primtype_2^{\ell_2}\!}{\primtype_3^{\ell_3}}$:
        $\primtype_1$ is $\ZTbool$,
        $\primtype_2$ is $\ZTuint{k}$, and
        $\primtype_3$ is $\ZTint{k'}$.
  \item
        Type $\ZTmulwrd{\bar{\comptype}}$:
        $|\bar{\comptype}| \geq 1$ and
        $\comptype_i$ is either $\ZT{\ZTfloat{n}^\ell}$,
        $\ZTdyadic{\primtype_1^{\ell_1}\!, \primtype_2^{\ell_2}\!}{n}$, or
        $\ZTdyadic{\primtype_1^{\ell_1}\!, \primtype_2^{\ell_2}\!}{\primtype_3^{\ell_3}}$.
  \item
        Type $\ZTtuple{\bar{\comptype}}$:
        $|\bar{\comptype}| \geq 1$ and
        $\comptype_i$ is either $\ZT{\primtype^\ell}$,
        $\ZTdyadic{\primtype_1^{\ell_1}\!, \primtype_2^{\ell_2}\!}{n}$,
        $\ZTdyadic{\primtype_1^{\ell_1}\!, \primtype_2^{\ell_2}\!}{\primtype_3^{\ell_3}}$, or
        $\ZTmulwrd{\bar{\comptype}'}$.
  \item
        Type $\ZTtable{\comptype^\bot}{(n_1, \ldots, n_k)}$:
        $k \geq 1$, $n_1, \ldots, n_k \geq 1$, and
        $\comptype$ is either $\ZT{\primtype^\ell}$,
        $\ZTdyadic{\primtype_1^{\ell_1}\!, \primtype_2^{\ell_2}\!}{n}$,
        $\ZTdyadic{\primtype_1^{\ell_1}\!, \primtype_2^{\ell_2}\!}{\primtype_3^{\ell_3}}$,
        $\ZTmulwrd{\bar{\comptype}'}$, or
        $\ZTtuple{\bar{\comptype}'}$.
  \item
        Statement $\ZSwhile{E}{S}$:
        $S$ has no return statement.
  \item
        Program $(\ZPdef{f_i}{\bar{x_i} \,{:}\, \bar{\comptype_i}}{\comptype_i'}{S_i})_{i \in [n]}$:
        $f_1, \ldots, f_n$ are distinct,
        and $x_{i,1}, \ldots, x_{i,|\bar{x_i}|}$ are distinct. %
  \item
        Type-annotated operator $\ZEop^\ell[\ZTfun{\bar{\primtype}}{\primtype'}]$:
        $\ZEop[\ZTfun{\bar{\primtype}}{\primtype'}] \in \ZOp_{\Ziv}$
        and $(\ell = \bot) \;{\iff}\; (\exists i.\; \primtype_i \notin \IR)$,
        where $\ZOp_{\Ziv}$ is a countable set of operators in \Ziv{}
        and the types in $\IR$ are defined in \Cref{sec:lang-def-irc0:syntax}.
        An example of $\ZOp_{\Ziv}$ is as follows:
        \begin{align*}
          \ZOp_{\Ziv} = \{
           &
          \Zkw{\&}[\ZTfun{(\ZTbool, \ZTbool)}{\ZTbool}],
          \Zkw{==}[\ZTfun{(\ZTbool, \ZTbool)}{\ZTbool}], \ldots,
          \\ &
          {+}[\ZTfun{(\ZTint{n}, \ZTint{n})}{\ZTint{n}}],
          \Zkw{$\leq$}[\ZTfun{(\ZTint{n}, \ZTint{n})}{\ZTbool}], \ldots,
          \\ &
          {+}[\ZTfun{(\ZTuint{n}, \ZTuint{n})}{\ZTuint{n}}],
          \Zkw{$\leq$}[\ZTfun{(\ZTuint{n}, \ZTuint{n})}{\ZTbool}], \ldots,
          \\ &
          {+}[\ZTfun{(\ZTfloat{n}, \ZTfloat{n})}{\ZTfloat{n}}],
          \Zkw{$\leq$}[\ZTfun{(\ZTfloat{n}, \ZTfloat{n})}{\ZTbool}], \ldots,
          \\ &
          {+}[\ZTfun{(\ZTreal, \ZTreal)}{\ZTreal}],
          {-}[\ZTfun{(\ZTreal, \ZTreal)}{\ZTreal}], \ldots,
          \\ &
          \Zkw{cast}[\ZTfun{\ZTuint{n}}{\ZTint{n}}],
          \Zkw{cast}[\ZTfun{\ZTfloat{n}}{\ZTreal}], \ldots,
          \\ &
          \Zkw{reint}[\ZTfun{\ZTfloat{n}}{\ZTuint{n}}],
          \Zkw{reint}[\ZTfun{\ZTuint{n}}{\ZTfloat{n}}], \ldots
          \}.
        \end{align*}
\end{itemize}

\subsection{Type System}
\label{sec:lang-def-ziv:typing}

The type system of \Ziv{} is shown in \Cref{fig:ziv-typing-full}.

\paragraph{Notation.}
We use the following notation.
\begin{itemize}
  \item For $n \in \mathbb{Z}$,
        $[n]$ and $[n]_0$ are the sets $\{1, \ldots, n\}$ and $\{0, \ldots, n-1\}$, respectively.
  \item For an erased type $\theta$,
        $\theta^\bot$ is the type obtained by annotating $\theta$ only with $\bot$.
  \item For $\ell, \ell' \in \{\bot, \top\}$,
        $\ell \sqcup \ell'$ is the least upper bound (or join) of $\ell$ and $\ell'$.
  \item For a type $\comptype$ and $\ell \in \mathit{Lab}$,
        $\comptype \sqcup \ell$ is the type obtained by replacing each label $\ell'$ in $\comptype$
        with $\ell' \sqcup \ell$.
        It is undefined if the resulting type is invalid (e.g., $\ZTreal^\top$),
        or if $\comptype$ is a function type and $\ell = \top$.
  \item For types $\comptype$ and $\comptype'$,
        $\comptype \sqcup \comptype'$ is the type
        obtained by joining the corresponding labels in $\comptype$ and $\comptype'$
        if $\Zerase(\comptype) = \Zerase(\comptype')$.
        It is undefined if $\Zerase(\comptype) \neq \Zerase(\comptype')$,
        or if either $\comptype$ or $\comptype'$ is a function type and $\comptype \neq \comptype'$.
  \item For types $\comptype$ and $\comptype'$,
        the subtyping relation $\comptype \sqsubseteq \comptype'$ is defined as follows:
        \begin{align*}
          \comptype \sqsubseteq \comptype'
          \iff
          \Zerase(\comptype) = \Zerase(\comptype')
           & {} \land
          (\ell \sqsubseteq \ell' \text{ for all corresponding labels $\ell$ in $\comptype$ and $\ell'$ in $\comptype'$})
          \\
           & {} \land
          (\text{$T$ or $T'$ is a function type} \implies T = T').
        \end{align*}
  \item For typing contexts ${\Gamma}$ and ${\Gamma'}$,
        the refinement relation ${\Gamma \preceq \Gamma'}$ is defined as follows:
        \begin{align*}
          {\Gamma\preceq\Gamma'}
          \iff
           & \Zdom(\Gamma) \supseteq \Zdom(\Gamma')
          \land (\Gamma(x) \sqsubseteq \Gamma'(x) \text{ for all } x \in \Zdom(\Gamma')).
        \end{align*}
\end{itemize}

\paragraph{Auxiliary Functions.}
We define the following auxiliary functions used in \Cref{fig:ziv-typing-full}.
\begin{itemize}
  \item
        For a statement $S$, the function $\Zhasret(S)$ is defined as follows:
        \begin{align*}
          \Zhasret(S) \defas
          \begin{cases}
            \mathit{true}  & \text{if $S$ has a return statement}
            \\
            \mathit{false} & \text{otherwise.}
          \end{cases}
        \end{align*}
  \item
        For a type $\comptype$, the function $\Zfields(\comptype)$ is defined as follows:
        \begin{align*}
          \Zfields(\ZT{\primtype^\ell})
           & \defas \text{undefined},
           &
          \Zfields(\ZTfun{\bar\comptype}{\comptype'})
           & \defas \text{undefined},                                                                                          %
          \\
          \Zfields(\ZTdyadic{\primtype_1^{\ell_1}, \primtype_2^{\ell_2}}{n})
           & \defas (\ZT{\primtype_1^{\ell_1}, \primtype_2^{\ell_2}}),
           &
          \Zfields(\ZTtuple{\bar\comptype})
           & \defas (\ZT{\bar\comptype}),
          \\
          \Zfields(\ZTdyadic{\primtype_1^{\ell_1}, \primtype_2^{\ell_2}}{\primtype_3^{\ell_3}})
           & \defas (\ZT{\primtype_1^{\ell_1}, \primtype_2^{\ell_2}, \primtype_3^{\ell_3}}),
           &
          \Zfields(\ZTtable{\comptype^\bot}{\bar{n}})
           & \defas \smash{(
                      \underbracecpt{\ZT{\comptype^\bot, \ldots, \comptype^\bot}}_{\;\; n_1 \times \cdots \times n_{|\bar{n}|}}
                      )},
          \\
          \Zfields(\ZTmulwrd{\bar\comptype})
           & \defas (\ZT{\bar\comptype}).
        \end{align*}
  \item
        For typing outcomes $\Omega, \Omega'$, sets $W, W' \subseteq \ZVar$, and $\ell \in \mathit{Lab}$,
        the function $\Zmerge(\ell, (\Omega,W), \allowbreak (\Omega',W'))$ is defined as follows:
        \begin{align*}
          \Zmerge
          \bigl(
          \ell,
          (\Omega,W),
          (\Omega',W')
          \bigr)
           & \defas
          \begin{cases}
            \big(
            (\Gamma\cap\Gamma') \sqcup_{W\cup W'} \ell,
            \, W \cup W'
            \big)
             &
            \text{if } (\Omega, \Omega') = (\Gamma, \Gamma')
            \\
            \left(
            \Gamma' \sqcup_{W'} \ell,
            \,
            W'
            \right)
             &
            \text{if } (\Omega, \Omega') = (\ZJret, \Gamma')
            \\
            \left(
            \Gamma \sqcup_{W} \ell,
            \,
            W
            \right)
             &
            \text{if } (\Omega, \Omega') = (\Gamma, \ZJret)
            \\
            (\ZJret,\varnothing)
             &
            \text{if } (\Omega, \Omega') = (\ZJret, \ZJret).
          \end{cases}
        \end{align*}
        Here, the context intersection $\ZCtx{\Gamma\cap\Gamma'}$ and the context update $\ZCtx{\Gamma \sqcup_W \ell}$
        are defined as follows:
        \begin{align*}
          (\Gamma\cap\Gamma')(x)
           & \defas
          \begin{cases}
            \Gamma(x) \sqcup \Gamma'(x)
             & \text{if }
            x \in \Zdom(\Gamma) \cap \Zdom(\Gamma')
            \\
            \text{undefined}
             & \text{otherwise,}
          \end{cases}
          \\
          (\Gamma \sqcup_W \ell)(x)
           & \defas
          \begin{cases}
            \big(x \in W\ ?\ \Gamma(x) \sqcup \ell : \Gamma(x) \big)
             & \text{if } x \in \Zdom(\Gamma)
            \\
            \text{undefined}
             & \text{otherwise.}
          \end{cases}
        \end{align*}
\end{itemize}

\subsection{Operational Semantics}
\label{sec:lang-def-ziv:semantics}

The operational semantics of \Ziv{} is shown in \Cref{fig:ziv-semantics-full}.

We use the following notation.
\begin{itemize}
  \item
        For a finite sequence $\bar{a}$, $|\bar{a}|$ is the length of $\bar{a}$.
  \item
        For $\ZEop[\ZTfun{(\primtype_1, \ldots, \primtype_n)}{\primtype'}] \in \ZOp_{\Ziv}$,
        we denote its semantics by the total function
        \[
          \Zsem{\ZEop[\ZTfun{(\primtype_1, \ldots, \primtype_n)}{\primtype'}]}
          : \ZVal_{\primtype_1} \times \cdots \times \ZVal_{\primtype_n} \to \ZVal_{\primtype'}.
        \]
\end{itemize}

\input{figs/fig-ziv-syntax-full}
\input{figs/fig-ir-syntax-full}
\input{figs/fig-c0-syntax-full}
\afterpage{\clearpage}

\input{figs/fig-ziv-typing-full}
\afterpage{\clearpage}

\input{figs/fig-ziv-semantics-full}
\afterpage{\clearpage}

%% file: figs/fig-ziv-syntax-full.tex
\begin{figure}[!p]
  \languagecoloringfalse
  \begin{figdefs}
    \setlength{\parskip}{0pt}
    \setlength{\abovedisplayskip}{0pt} %
    \setlength{\abovedisplayshortskip}{\abovedisplayskip}
    \setlength{\belowdisplayskip}{\abovedisplayskip}
    \setlength{\belowdisplayshortskip}{\abovedisplayskip}
    \setlength{\MyMathparLineskip}{6pt}
    \def\MathparLineskip{\lineskip=\MyMathparLineskip}
    \begin{alignat*}{10}
      \text{Labels}
       &
       &
      \tcr{\ell}
       & \ \Coloneqq \
      \tcr{\bot \mid \top}
      \\[-1pt]
      \text{Primitive types}
       &
       &
      \tcr{\primtype^\ell} \!
       & \ \Coloneqq \
      \rlap{$
          \ZTbool^\ell
          \mid \ZTint{n}^\ell
          \mid \ZTuint{n}^\ell
          \mid \ZTfloat{n}^\ell
          \mid \tcr{\ZTreal^\bot}
        $}
      \\[-1pt]
      \text{Composite types}
       &
       &
      \comptype
       & \ \Coloneqq \
      \rlap{$
          \primtype^\ell
          \mid \tcr{\ZTdyadic{\primtype^\ell \!, \primtype^\ell \!}{n}}
          \mid \tcr{\ZTdyadic{\primtype^\ell \!, \primtype^\ell \!}{\primtype^\ell}}
          \mid \tcr{\ZTmulwrd{{\bar{\comptype}}}}
        $}
      \\[-1pt]
       &
       &
       & \phantom{ {} \ \Coloneqq \ {} \primtype^\ell}
      \rlap{$
          {}
          \mid \ZTtable{\tcr{\comptype^{\bot}} \!}{\bar{n}}
          \mid \ZTtuple{\bar\comptype}
          \mid \ZTfun{\bar\comptype}{\comptype}
        $}
      \\[-1pt]
      \text{Expressions}
       &
       &
      E
       & \ \Coloneqq \
      \rlap{$
          v
          \mid x
          \mid \tcr{\ZEop^{\ell}}{[\ZTfun{\bar\primtype}{\primtype}]}(\bar E)
          \mid E(\bar E)
          \mid \ZEnew{\comptype}{\bar E}
          \mid \ZEget{E, E}
        $}
      \\[-1pt]
      \text{Statements}
       &
       &
      S
       & \ \Coloneqq \
      \rlap{$
          \ZSskip
          \mid \ZSassn{x}{E}
          \mid \ZSret{E}
          \mid \ZSseq{S}{S}
          \mid \ZSif{E}{S}{S}
          \mid \ZSwhile{E}{S}
        $}
      \\[-1pt]
      \text{Programs}
       &
       &
      P
       & \ \Coloneqq \
      \rlap{$
          \ZPnil \mid \ZPseq{P}{\ZPdef{f}{\bar x:\bar\comptype}{\comptype}{S}}
        $}
      \\[-1pt]
      \noalign{%
        \vskip \MyMathparLineskip
        \nointerlineskip
        \hbox to\linewidth{%
          \hskip 6pt
          \rule{\dimexpr\linewidth-12pt\relax}{\fboxrule}\hss}%
        \nointerlineskip
        \vskip \MyMathparLineskip}
      \text{Primitive operators}
       &
       & \quad
      \ZEop^\ell \!
       & \ \Coloneqq \
      \rlap{$
          \ZEop^\ell_{\Zkw{Bool}}
          \mid
          \ZEop^\ell_{\Zkw{Int}}
          \mid
          \ZEop^\ell_{\Zkw{UInt}}
          \mid
          \ZEop^\ell_{\Zkw{Float}}
          \mid
          \tcr{\ZEop^\bot_{\Zkw{Real}}}
          \mid
          \ZEop^\ell_{\Zkw{Cast}}
        $}
      \\[-1pt]
       &
       &
      \llap{$
          {\ZEop}_{\Zkw{Bool}}
        $}
       & \ \Coloneqq \
      \Zkw{\&} \mid \Zkw{\shortbar} \mid \Zkw{\textasciicircum} \mid \Zkw{$\textsf{!}$}
      \mid \Zkw{$\textsf{==}$} %
      \mid \cdots
       &
      {\ZEop}_{\Zkw{Real}}
       & \ \Coloneqq \
      \Zkw{$+$} \mid \Zkw{$-$} \mid \Zkw{$\times$} \mid \Zkw{$/$}  %
      \mid \cdots
      \\[-1pt]
       &
       &
      \llap{$
          {\ZEop}_{\Zkw{Int}},
          {\ZEop}_{\Zkw{UInt}},
          {\ZEop}_{\Zkw{Float}}
        $}
       & \ \Coloneqq \
      \Zkw{$+$} \mid \Zkw{$-$} \mid \Zkw{$\times$} \mid \Zkw{$/$}
      \mid \Zkw{$\ll$} %
      \mid \Zkw{\&} %
      \mid \Zkw{$\leq$} %
      \mid \cdots
       & \quad\;\;
      {\ZEop}_{\Zkw{Cast}}
       & \ \Coloneqq \
      \Zkw{cast} \mid \Zkw{reint}
      \mid \cdots
    \end{alignat*}
  \end{figdefs}
  \caption{Core syntax of \Ziv.}
  \label{fig:ziv-syntax-full}
  \languagecoloringtrue
  \bigskip
\end{figure}

%% file: figs/fig-ir-syntax-full.tex
\begin{figure}[!t]
  \languagecoloringfalse
  \begin{figdefs}
    \setlength{\parskip}{0pt}
    \setlength{\abovedisplayskip}{0pt} %
    \setlength{\abovedisplayshortskip}{\abovedisplayskip}
    \setlength{\belowdisplayskip}{\abovedisplayskip}
    \setlength{\belowdisplayshortskip}{\abovedisplayskip}
    \setlength{\MyMathparLineskip}{6pt}
    \def\MathparLineskip{\lineskip=\MyMathparLineskip}
    \begin{alignat*}{10}
      \text{Primitive types}
       &
       &
      \IT{\primtype}
       & \ \Coloneqq \
      \ITbool
      \mid \ITint{n}
      \mid \ITuint{n}
      \mid \ITfloat{n}
      \\[-1pt]
      \text{Primitive operators}
       &
       & \quad
      \IEop
       & \ \Coloneqq \
      \IEop_{\Ikw{Bool}}
      \mid \IEop_{\Ikw{Int}}
      \mid \IEop_{\Ikw{UInt}}
      \mid \IEop_{\Ikw{Float}}
      \mid \IEop_{\Ikw{Cast}}
      \\[-1pt]
      \text{Expressions}
       &
       &
      \IE{E}
       & \ \Coloneqq \
      \IEval{v}
      \mid \IEvar{x}
      \mid \IEapp{\IEop[{\ITfun{\bar\primtype}{\primtype}]}}{\IEvar{\bar x}}
      \mid \IEget{\IEvar{x},\IEvar{x}}
      \\[-1pt]
      \text{Statements}
       &
       &
      \IS{S}
       & \ \Coloneqq \
      \ISskip
      \mid \ISassn{\IEvar{x}:\primtype}{E}
      \mid \ISassn{\IEvar{x}:\IT{\primtype[n]}}{\IEnew{\IEval{\bar v}}}
      \mid \ISseq{S}{S}
      \mid \ISif{(\IEvar{\bar{x}}:\bar\primtype)}{\IEvar{x}}{S; \IEvar{\bar{x}}}{S; \IEvar{\bar{x}}}
      \\[-1pt]
      \text{Programs}
       &
       &
      \IP{P}
       & \ \Coloneqq \IPdef{f}{\IEvar{\bar x}:\bar\primtype}{\primtype}{S; \ISret{\IEvar{x}}}
    \end{alignat*}
  \end{figdefs}
  \caption{Core syntax of \IR{}.}
  \label{fig:ir-syntax-full}
  \languagecoloringtrue
  \bigskip
\end{figure}

%% file: figs/fig-c0-syntax-full.tex
\begin{figure}[!t]
  \languagecoloringfalse
  \begin{figdefs}
    \setlength{\parskip}{0pt}
    \setlength{\abovedisplayskip}{0pt} %
    \setlength{\abovedisplayshortskip}{\abovedisplayskip}
    \setlength{\belowdisplayskip}{\abovedisplayskip}
    \setlength{\belowdisplayshortskip}{\abovedisplayskip}
    \setlength{\MyMathparLineskip}{6pt}
    \def\MathparLineskip{\lineskip=\MyMathparLineskip}
    \begin{alignat*}{10}
      \text{Primitive types}
       &
       &
      \primtype
       & \ \Coloneqq \
      \CTbool
      \mid \CTint{n}
      \mid \CTuint{n}
      \mid \CTfloat{n}
      \\[-1pt]
      \text{Primitive operators}
       &
       & \quad
      \CEop
       & \ \Coloneqq \
      \CEop_{\Ckw{Bool}}
      \mid \CEop_{\Ckw{Int}}
      \mid \CEop_{\Ckw{UInt}}
      \mid \CEop_{\Ckw{Float}}
      \mid \CEop_{\Ckw{Cast}}
      \\[-1pt]
      \text{Expressions}
       &
       &
      E
       & \ \Coloneqq \
      \CE{v}\mid \CEvar{x}
      \mid\CEapp{\CEop[{\CTfun{\bar\primtype}{\primtype}}]}{\CEvar{\bar x}}
      \mid \CEget{\CEvar{x}}{\CEvar{x}}
      \\[-1pt]
      \text{Statements}
       &
       &
      S
       & \ \Coloneqq \
      \CSskip
      \mid \CSdecl{\primtype}{x}
      \mid \CSnew{\primtype}{x}{n}{\bar{v}}
      \mid \CSassn{x}{E}
      \mid \CSseq{S}{S}
      \mid \CSif{x}{S}{S}
      \\[-1pt]
      \text{Programs}
       &
       &
      P
       & \ \Coloneqq \
      \CPdef{f}{\bar{\primtype}\ \bar{x}}{\primtype'}{S\ \CSret{x}}
    \end{alignat*}
  \end{figdefs}
  \caption{Core syntax of \Czero{}.}
  \label{fig:c0-syntax-full}
  \languagecoloringtrue
\end{figure}

%% file: figs/fig-ziv-typing-full.tex
\begin{figure}[!p]
  \begin{figdefs}[5pt]
    \setlength{\fboxsep}{2pt}
    \textbf{Definitions} (typing contexts, typing outcomes, write sets)
    \hfill\fbox{$\Gamma$, $\Omega$, $W$}
    \begin{Zgather*}
      \Gamma \Coloneqq \varepsilon \mid \Gamma,x:\comptype
      \qquad\qquad
      \Omega \Coloneqq \Gamma \mid \tcr{\ZJret}
      \qquad\qquad
      W \subseteq \ZVar
    \end{Zgather*}

    \textbf{Expressions}
    \hfill\smash{\fbox{$\ZJtype{P; \Gamma}{E}{\comptype}$}}
    \begin{Zmathpar}
      \Zinfer{T-Const}
      {
        \tcr{v \in \ZVal_{\theta}}
      }
      {\ZJtype{P; \Gamma}{v}{\tcr{\theta^\bot}}}
      \and %
      \Zinfer[width=0.2\linewidth]{T-Fun}
      {
        \tcr{v \in \ZFun}
        \\
        (\ZPdef{v}{\bar{x}:\bar{\comptype}}{\comptype'}{S}) \in P
      }
      {\ZJtype{P; \Gamma}{v}{\tcr{\ZTfun{\bar{\comptype}}{\comptype'}}}}
      \and %
      \Zinfer{T-Var}
      {~}
      {\ZJtype{P; \Gamma}{x}{\Gamma(x)}}
      \and %
      \Zinfer[width=0.2\linewidth]{T-Op}
      {
        \forall i \in [n].\; \ZJtype{P; \Gamma}{E_i}{\primtype_i^{\ell_i}}
        \\
        \tcr{\forall i \in [n].\; \ell_i \sqsubseteq \ell}
        \qquad
        \tcr{\ell_1 \sqcup \cdots \sqcup \ell_n = \ell'}
      }
      {
        \ZJtype{P; \Gamma}
        {\ZEop^{\ell}{[\ZTfun{(\primtype_1,\ldots,\primtype_n)}{\primtype'}]}(E_1, \ldots, E_n)}
        {(\primtype')^{\ell'}}
      }
      \and \!\!\!\!
      \Zinfer[width=0.2\linewidth]{T-Call}
      {
        \ZJtype{P; \Gamma}{E_0}{\ZTfun{(\comptype_1, \ldots, \comptype_n)}{\comptype'}}
        \\
        \forall i \in [n].\; \ZJtype{P; \Gamma}{E_i}{\comptype_i}
      }
      {
        \ZJtype{P; \Gamma}{E_0(E_1, \ldots, E_n)}{\comptype'}
      }
      \and
      \Zinfer[width=0.2\linewidth]{T-New}
      {
        \Zfields(\comptype')=(\comptype_1, \ldots, \comptype_n)
        \land n \geq 1
        \\
        \forall i \in [n].\;
        \ZJtype{P; \Gamma}{E_i}{\comptype_i}
      }
      {\ZJtype{P; \Gamma}{\ZEnew{\comptype'}{E_1, \ldots, E_n}}{\comptype'}}
      \and \!\!\!\!\!\!\!
      \Zinfer[width=0.2\linewidth]{T-GetConsInd}
      {
        \ZJtype{P; \Gamma}{E}{\comptype'}
        \\
        \Zfields(\comptype') = (\comptype_1, \ldots, \comptype_n)
        \land
        i \in [n]_0
      }
      {\ZJtype{P; \Gamma}{\ZEget{E,\ZEval{\ZTuint{k}}{i}}}{\comptype_{i+1}}}
      \and
      \Zinfer[width=0.2\linewidth]{T-GetExprInd}
      {
        \ZJtype{P; \Gamma}{E_1}{\ZTtable{\comptype}{\bar n}}
        \\
        \ZJtype{P; \Gamma}{E_2}{\ZTuint{k}^\ell}
      }
      {
        \ZJtype{P; \Gamma}{\ZEget{E_1, E_2}}{\tcr{\comptype \sqcup \ell}}
      }
      \and
      \Zinfer[width=0.5\linewidth]{T-Sub}
      {
        \ZJtype{P; \Gamma}{E}{\comptype}
        \\
        \comptype\sqsubseteq\comptype'
      }
      {\ZJtype{P; \Gamma}{E}{\comptype'}}
    \end{Zmathpar}

    \textbf{Statements}
    \hfill\smash{\fbox{$\ZJtype[\comptype]{P; \Gamma}{S}{(\Omega, W)}$}}
    \begin{Zmathpar}
      \Zinfer{T-Skip}
      { }
      {
        \ZJtype[\comptype]{P; \Gamma}{\ZSskip}{(\Gamma, \varnothing)}
      }
      \and
      \Zinfer{T-Assn}
      {\ZJtype{P; \Gamma}{E}{\comptype'}}
      {
        \ZJtype[\comptype]{P; \Gamma}{\ZSassn{x}{E}}{((\Gamma, x:\comptype'), \{x\})}
      }
      \and
      \Zinfer{T-Ret}
      {\ZJtype{P; \Gamma}{E}{\tcr{\comptype}}}
      {
        \ZJtype[\comptype]{P; \Gamma}{\ZSret{E}}{(\tcr{\ZJret}, \varnothing)}
      }
      \and
      \Zinfer{T-Seq1}
      {
        \ZJtype[\comptype]{P; \Gamma}{S_1}{(\ZJret, \varnothing)}
      }
      {
        \ZJtype[\comptype]{P; \Gamma}{\ZSseq{S_1}{S_2}}{(\ZJret, \varnothing)}
      }
      \and
      \Zinfer{T-Seq2}
      {
        \ZJtype[\comptype]{P; \Gamma}{S_1}{(\Gamma_1, W_1)}
        \\
        \ZJtype[\comptype]{P; \Gamma_1}{S_2}{(\Omega_2, W_2)}
      }
      {
        \ZJtype[\comptype]{P; \Gamma}{\ZSseq{S_1}{S_2}}{
          (\Omega_2, (\Omega_2 = \ZJret\ ?\ \varnothing : W_1 \cup W_2))
        }
      }
      \and
      \Zinfer{T-If}
      {
        \ZJtype{P; \Gamma}{E}{\ZTbool^\ell}
        \\
        {\begin{array}[b]{@{}c@{}}
            \ZJtype[\comptype]{P; \Gamma}{S_1}{(\Omega_1, W_1)}
            \\
            \ZJtype[\comptype]{P; \Gamma}{S_2}{(\Omega_2, W_2)}
          \end{array}}
        \\
        {\begin{array}[b]{@{}c@{}}
            \tcr{
              \Zmerge
              (\ell, (\Omega_1,W_1), (\Omega_2,W_2))
              =
              (\Omega',W')
            }
            \\
            \tcr{
              \Zhasret(S_1)\lor\Zhasret(S_2) {\implies} \comptype\sqcup\ell=\comptype
            }
          \end{array}}
      }
      {
        \ZJtype[\comptype]{P; \Gamma}{\ZSif{E}{S_1}{S_2}}{(\Omega', W')}
      }
      \and
      \Zinfer{T-While}
      {
        \tcr{\neg\Zhasret(S)}
        \\
        {\begin{array}[b]{@{}c@{}}
            \ZJtype{P; \Gamma_I}{E}{\tcr{\ZTbool^{\bot}}}
            \\
            \ZJtype[\comptype]{P; \Gamma_I}{S}{(\Gamma', W')}
          \end{array}}
        \\
        {\begin{array}[b]{@{}c@{}}
            \tcr{\Gamma\preceq\Gamma_I}
            \\
            \tcr{\Gamma'\preceq\Gamma_I}
          \end{array}}
      }
      {
        \ZJtype[\comptype]{P; \Gamma}{\ZSwhile{E}{S}}{(\Gamma_I, W')}
      }
      \label{rule:ziv-typing-full:t-while}
    \end{Zmathpar}

    \textbf{Programs}
    \hfill\smash{\fbox{$\ZJtype{}{P}{\ZJok}$}}
    \begin{Zmathpar}
      \Zinfer{T-Prog}
      {
        P = (\ZPdef{f_j}{\bar{x_j}:\bar{\comptype_j}}{\comptype_j'}{S_j})_{j \in [n]}
        \\
        \forall i \in [n].\; \ZJtype[\comptype_i']{P; (\bar{x_i} : \bar{\comptype_i})}{S_i}{(\ZJret, \varnothing)}
      }
      {
        \ZJtype{}{
          P
        }{\ZJok}
      }
    \end{Zmathpar}
  \end{figdefs}
  \caption{Type system of \Ziv{}.}
  \label{fig:ziv-typing-full}
\end{figure}

%% file: figs/fig-ziv-semantics-full.tex
\begin{figure}[!p]
  \begin{figdefs}[5pt]
    \setlength{\fboxsep}{2pt}
    \textbf{Definitions} (states, evaluation outcomes)
    \hfill\fbox{$\sigma$, $\omega$}
    \begin{Zgather*}
      \sigma \Coloneqq \varepsilon \mid \sigma,x\mapsto v
      \qquad\qquad
      \omega \Coloneqq \sigma \mid \tcr{\ZJret(v)}
    \end{Zgather*}

    \textbf{Expressions}
    \hfill\smash{\fbox{$\ZJeval{P; \sigma}{E}{v}$}}
    \begin{Zmathpar}
      \Zinfer{E-Const}
      { \tcr{v \in \ZVal_{\Ziv}} }
      {\ZJeval{P; \sigma}{v}{v}}
      \and \!\!\!\!\!\!\!\!
      \Zinfer{E-Var}
      {\sigma(x)=v}
      {\ZJeval{P; \sigma}{x}{v}}
      \and \!\!\!\!\!\!\!\!
      \Zinfer[width=0.2\linewidth]{E-Op}
      {
        \forall i \in [n].\; \ZJeval{P; \sigma}{E_i}{v_i}
        \\
        \Zsem{\ZEop_{\ZTfun{(\primtype_1, \ldots, \primtype_n)}{\primtype'}}}(v_1, \ldots, v_n)=v'
      }
      {
        \ZJeval{P; \sigma}
        {\ZEop^\ell[{\ZTfun{(\primtype_1, \ldots, \primtype_n)}{\primtype'}}](E_1, \ldots, E_n)}{v'}
      }
      \and
      \Zinfer{E-Call} %
      {
        {\begin{array}[b]{@{}c@{}}
            \ZJeval{P; \sigma}{E_0}{f}
            \\
            \forall i \in [n].\; \ZJeval{P; \sigma}{E_i}{v_i}
          \end{array}}
        \\
        {\begin{array}[b]{@{}c@{}}
            (\ZPdef{f}{x_1:\comptype_1, \ldots, x_n:\comptype_n}{\comptype'}{S}) \in P
            \\
            \ZJeval{P; (x_1 \mapsto v_1, \ldots, x_n \mapsto v_n)}{S}{\ZJret(v')}
          \end{array}}
      }
      {
        \ZJeval{P; \sigma}{E_0(E_1, \ldots, E_n)}{v'}
      }
      \and
      \Zinfer[width=0.2\linewidth]{E-New}
      {
        \Zfields(\comptype)=(\comptype_1, \ldots, \comptype_n)
        \land
        \Zerase(\comptype) = \theta
        \\
        \forall i \in [n].\; \ZJeval{P; \sigma}{E_i}{v_i}
        \land
        \tcr{v_i \in \ZVal_{\Zerase(\comptype_i)}}
      }
      {\ZJeval{P; \sigma}{\ZEnew{\comptype}{E_1, \ldots, E_n}}
        {\ZEval{\theta}{v_1, \ldots, v_n}}}
      \and \!\!\!\!\!\!\!\!\!
      \Zinfer[width=0.5\linewidth]{E-Get}
      {
        \ZJeval{P; \sigma}{E_1}{\ZEval{\theta}{v_1, \ldots, v_n}}
        \\
        \ZJeval{P; \sigma}{E_2}{\ZEval{\ZTuint{k}}{i}}
        \land
        i \in [n]_0
      }
      {\ZJeval{P; \sigma}{\ZEget{E_1, E_2}}{v_{i+1}}}
    \end{Zmathpar}

    \textbf{Statements}
    \hfill\smash{\fbox{$\ZJeval{P; \sigma}{S}{\omega}$}}
    \begin{Zmathpar}
      \Zinfer{E-Skip}
      { }
      {\ZJeval{P; \sigma}{\ZSskip}{\sigma}}
      \and\!\!\!\!
      \Zinfer{E-Assn}
      {\ZJeval{P; \sigma}{E}{v}}
      {
        \ZJeval{P; \sigma}{\ZSassn{x}{E}}{(\sigma, x\mapsto v)}
      }
      \and\!\!\!\!
      \Zinfer{E-Ret}
      {\ZJeval{P; \sigma}{E}{v}}
      {
        \ZJeval{P; \sigma}{\ZSret{E}}{\tcr{\ZJret(v)}}
      }
      \and
      \Zinfer{E-Seq1}
      {\ZJeval{P; \sigma}{S_1}{\ZJret(v)}}
      {
        \ZJeval{P; \sigma}{\ZSseq{S_1}{S_2}}{\ZJret(v)}
      }
      \and
      \Zinfer{E-Seq2}
      {
        \ZJeval{P; \sigma}{S_1}{\sigma_1}
        \\
        \ZJeval{P; \sigma_1}{S_2}{\omega_2}
      }
      {
        \ZJeval{P; \sigma}{\ZSseq{S_1}{S_2}}{\omega_2}
      }
      \and
      \Zinfer[width=0.2\linewidth]{E-IfTrue}
      {
        \ZJeval{P; \sigma}{E}{\ZEval{\ZTbool}{\ZEtrue}}
        \\
        \ZJeval{P; \sigma}{S_1}{\omega_1}
      }
      {
        \ZJeval{P; \sigma}{\ZSif{E}{S_1}{S_2}}{\omega_1}
      }
      \and
      \Zinfer[width=0.2\linewidth]{E-IfFalse}
      {
        \ZJeval{P; \sigma}{E}{\ZEval{\ZTbool}{\ZEfalse}}
        \\
        \ZJeval{P; \sigma}{S_2}{\omega_2}
      }
      {
        \ZJeval{P; \sigma}{\ZSif{E}{S_1}{S_2}}{\omega_2}
      }
      \and
      \Zinfer{E-WhileFalse}
      {
        \ZJeval{P; \sigma}{E}{\ZEval{\ZTbool}{\ZEfalse}}
      }
      {
        \ZJeval{P; \sigma}{\ZSwhile{E}{S}}{\sigma}
      }
      \and \!\!\!\!\!\!\!\!
      \Zinfer[width=0.2\linewidth]{E-WhileTrue}
      {
        \ZJeval{P; \sigma}{E}{\ZEval{\ZTbool}{\ZEtrue}}
        \\
        \ZJeval{P; \sigma}{S}{\sigma'}
        \quad
        \ZJeval{P; \sigma'}{\ZSwhile{E}{S}}{\sigma''}
      }
      {
        \ZJeval{P; \sigma}{\ZSwhile{E}{S}}{\sigma''}
      }
    \end{Zmathpar}

    \textbf{Programs}
    \hfill\smash{\fbox{$\ZJeval{P}{f(\bar v)}{v'}$}}
    \begin{Zmathpar}
      \Zinfer{E-Prog}
      {
        P = (\ZPdef{f_j}{\bar{x_j}:\bar{\comptype_j}}{\comptype_j'}{S_j})_{j \in [n]}
        \\
        i\in[n] \land |\bar{x_i}| = |\bar{T_i}| = |\bar{v_i}|
        \\
        \ZJeval{P; (\bar{x_i} \mapsto \bar{v_i})}{S_i}{\ZJret(v')}
      }
      {
        \ZJeval{
          P
        }{f_i(\bar{v_i})}{v'}
      }
    \end{Zmathpar}
  \end{figdefs}
  \caption{Operational semantics of \Ziv{}.}
  \label{fig:ziv-semantics-full}
\end{figure}

%% file: appendices/app-lang-def-irc0.tex
\section{Complete Definition of the Intermediate Languages}
\label{sec:lang-def-irc0}

In this section, we define the syntax, type system, and operational semantics of
the three intermediate languages: \Znorm*{}, \IR*{}, and \Czero*{}.

\subsection{Syntax}
\label{sec:lang-def-irc0:syntax}

The abstract syntax of \Znorm{} is the same as that of \Ziv{} (\Cref{fig:ziv-syntax-full}),
except that every function in \Znorm{} must have exactly one return statement
and the statement must be at the end of the function body.
The abstract syntax of \IR{} and \Czero{}
is shown in \Cref{fig:ir-syntax-full,fig:c0-syntax-full}, respectively.

\paragraph{Notation.}
We use the same notation as in \Cref{sec:lang-def-ziv:syntax} with the following differences.
\begin{itemize}
  \item
        In \IR{},
        $v \in \IVal_{\IR}$ where $\IVal_{\IR}$ is a set of values in \IR{}.
        The set $\IVal_{\IR}$ consists of all values of primitive types and array types:
        \begin{align*}
          \IVal_{\IR} \defas
          \bigcup_{{\primtype} \in \IR} \ZVal_\primtype \cup
          \bigcup_{{\primtype} \in \IR,\, n \geq 1} \ZVal_{\primtype[n]}.
        \end{align*}
        $\IVal_\primtype$ was defined in \Cref{sec:lang-def-ziv:syntax},
        and $\IVal_{\primtype[n]}$ is defined as follows:
        \begin{align*}
          \IVal_{\primtype[n]} \defas
          \{\ZEval{\primtype[n]}{v_1, \ldots, v_n} \mid v_1, \ldots, v_n \in \ZVal_\primtype\}.
        \end{align*}
  \item
        In \Czero{},
        $v \in \CVal_{\Czero}$ where $\CVal_{\Czero}$ is a set of values in \Czero{}.
        The set $\CVal_{\Czero}$ consists of all values of primitive types and array types:
        \begin{align*}
          \CVal_{\Czero} \defas
          \bigcup_{{\primtype} \in \Czero} \ZVal_\primtype \cup
          \bigcup_{{\primtype} \in \Czero,\, n \geq 1} \ZVal_{\primtype[n]}.
        \end{align*}
        $\CVal_\primtype, \CVal_{\primtype[n]}$ were defined above.
        Since \IR{} and \Czero{} have the same types, $\CVal_{\Czero} = \IVal_{\IR}$.
\end{itemize}

\paragraph{Syntactic Restrictions.}
We make the following restrictions on the syntax of \IR{}:
\begin{itemize}
  \item
        Types $\ITint{n}$, $\ITuint{n}$, and $\ITfloat{n}$:
        $n \geq 1$.
  \item
        Statement
        ($\ISassn{\IEvar{x}:\IT{\primtype[n]}}{\IEnew{\IEval{\bar v}}}$):
        $n \geq 1$.
  \item
        Statement
        ($
          \ISif{
            (\bar{x} : \bar{\primtype})
          }{\IEvar{y}}{
            S_1; \IEvar{\bar{z_1}}
          }{
            S_2; \IEvar{\bar{z_2}}
          }
        $):
        $|\bar{x}| = |\bar{\primtype}| = |\bar{z_1}| = |\bar{z_2}|$.
  \item
        Program
        ($\IPdef{f}{x_1:\primtype_1, \ldots, x_n:\primtype_n}{\primtype'}{S; \ISret{y}}$):
        the variables $x_1, \ldots, x_n$ and all variables appearing on the left-hand side of assignment statements in $S$
        are pairwise distinct.
  \item
        Type-annotated operator $\IEop[\ITfun{\bar\primtype}{\primtype'}]$:
        $\IEop[\ITfun{\bar\primtype}{\primtype'}] \in \mathit{Op}_{\IR}$,
        where $\mathit{Op}_{\IR} \subseteq \mathit{Op}_{\Ziv}$ is defined by
        \begin{align*}
          \mathit{Op}_{\IR} \defas \{
          \IEop[\ITfun{(\primtype_1, \ldots, \primtype_n)}{\primtype_{n+1}}] \in \mathit{Op}_{\Ziv}
          \mid
          \primtype_1, \ldots, \primtype_{n+1} \in \IR
          \}.
        \end{align*}
\end{itemize}

We make the following restrictions on the syntax of \Czero{}:
\begin{itemize}
  \item
        Types $\CTint{n}$, $\CTuint{n}$, and $\CTfloat{n}$:
        $n \geq 1$.
  \item
        Statement
        ($\CSnew{\primtype}{x}{n}{\bar{v}}$):
        $n \geq 1$.
  \item
        Program
        ($\CPdef{f}{x_1:\primtype_1, \ldots, x_n:\primtype_n}{\primtype'}{S; \CSret{y}}$):
        $x_1, \ldots, x_n$ are distinct.
  \item
        Type-annotated operator $\CEop[\CTfun{\bar\primtype}{\primtype'}]$:
        $\CEop[\CTfun{\bar\primtype}{\primtype'}] \in \mathit{Op}_{\Czero}$,
        where $\mathit{Op}_{\Czero} \subseteq \mathit{Op}_{\IR}$ is defined by
        \begin{align*}
          \mathit{Op}_{\Czero} \defas \{
          \CEop[\CTfun{(\primtype_1, \ldots, \primtype_n)}{\primtype_{n+1}}] \in \mathit{Op}_{\IR}
          \mid
          \primtype_1, \ldots, \primtype_{n+1} \in \Czero
          \}.
        \end{align*}
        Since \IR{} and \Czero{} have the same types,
        $\mathit{Op}_{\Czero} = \mathit{Op}_{\IR}$.
\end{itemize}

\subsection{Type System}
\label{sec:lang-def-irc0:typing}

The type system of \Znorm{} is the same as that of \Ziv{} (\Cref{fig:ziv-typing-full}).
The type system of \IR{} and \Czero{} is shown in \Cref{fig:ir-typing-full,fig:c0-typing-full}, respectively.

\subsection{Operational Semantics}
\label{sec:lang-def-irc0:semantics}

The operational semantics of \Znorm{} is the same as that of \Ziv{} (\Cref{fig:ziv-semantics-full}).
The operational semantics of \IR{} and \Czero{} is shown in \Cref{fig:ir-semantics-full,fig:c0-semantics-full}, respectively.

We use the same notation as in \Cref{sec:lang-def-ziv:semantics}, together with the following notation.
\begin{itemize}
  \item
        For $f : X \to Y$ and $X' \subseteq X$,
        $f|_{X'} : X' \to Y$ is the restriction of $f$ to $X'$.
  \item
        For $\IEop[\ITfun{\bar{\primtype}}{\primtype'}]
          \in \mathit{Op}_{\IR} \cup \mathit{Op}_{\Czero}$,
        its semantics is the $\Zsem{\IEop[\ITfun{\bar{\primtype}}{\primtype'}]}$
        defined in \Cref{sec:lang-def-ziv:semantics}.
\end{itemize}

\input{figs/fig-ir-typing-full}
\afterpage{\clearpage}
\input{figs/fig-c0-typing-full}
\afterpage{\clearpage}

\input{figs/fig-ir-semantics-full}
\afterpage{\clearpage}
\input{figs/fig-c0-semantics-full}
\afterpage{\clearpage}

%% file: figs/fig-ir-typing-full.tex
\begin{figure}[!t]
  \begin{figdefs}[5pt]
    \setlength{\fboxsep}{2pt}
    \textbf{Definitions} (typing contexts)
    \hfill\fbox{$\IJctx{\Gamma}$}
    \begin{Igather*}
      \IJctx{\Gamma} \Coloneqq \IJctx{\varepsilon} \mid
      \IJctx{\Gamma,\IEvar{x}:\primtype}\mid
      \IJctx{\Gamma,\IEvar{x}:\IT{\primtype[n]}}
    \end{Igather*}

    \textbf{Expressions}
    \hfill\smash{\fbox{$\IJtype{\IJctx{\Gamma}}{E}{\primtype}$}}
    \begin{Imathpar}
      \Iinfer{T-Const}
      {\IEval{v}\in\IVal[\primtype]}
      {\IJtype{\IJctx{\Gamma}}{\IEval{v}}{\primtype}}
      \and
      \Iinfer{T-Var}
      { \Gamma(x) = \primtype }
      {\IJtype{\IJctx{\Gamma}}{\IEvar{x}}{\primtype}}
      \and
      \Iinfer[width=0.2\linewidth]{T-Op}
      {
        \forall i \in [n].\;
        \Gamma({x_i}) = \primtype_i
      }
      {
        \IJtype{\IJctx{\Gamma}}{\IEapp{\IEop[{\ITfun{(\primtype_1, \ldots, \primtype_n)}{\primtype'}}]}{\IEvar{x_1, \ldots, x_n}}}{\primtype'}
      }
      \and
      \Iinfer{T-Get}
      {
        \Gamma({x}) = \primtype[n]
        \\
        \Gamma(y) = \ITuint{k}
      }
      {
        \IJtype{\IJctx{\Gamma}}{\IEget{\IEvar{x}, \IEvar{y}}}{\primtype}
      }
    \end{Imathpar}

    \textbf{Statements}
    \hfill\smash{\fbox{$\IJtype{\IJctx{\Gamma}}{S}{\IJctx{\Gamma'}}$}}
    \begin{Imathpar}
      \Iinfer{T-Skip}
      { }
      {\IJtype{\IJctx{\Gamma}}{\ISskip}{\IJctx{\Gamma}}}
      \and
      \Iinfer{T-Assn}
      {\IJtype{\IJctx{\Gamma}}{E}{\primtype}}
      {
        \IJtype{\IJctx{\Gamma}}{(\ISassn{\IEvar{x}:\primtype}{E})}{(\IJctx{\Gamma},\IEvar{x}:\primtype)}
      }
      \and
      \Iinfer{T-New}
      {
        n\geq1
        \\
        \forall i\in[n].\ \IEval{v_i}\in\IVal[\primtype]}
      {
        \IJtype{\IJctx{\Gamma}}
        {(\ISassn{\IEvar{x}:\IT{\primtype[n]}}
          {\IEnew{\IEval{v_1},\ldots,\IEval{v_n}}})}
        {(\IJctx{\Gamma},\IEvar{x}:\IT{\primtype[n]})}
      }
      \and
      \Iinfer{T-Seq}
      {
        \IJtype{\IJctx{\Gamma}}{S_1}{\IJctx{\Gamma_1}}
        \\
        \IJtype{\IJctx{\Gamma_1}}{S_2}{\IJctx{\Gamma_2}}
      }
      {
        \IJtype{\IJctx{\Gamma}}{\ISseq{S_1}{S_2}}{\IJctx{\Gamma_2}}
      }
      \and
      \Iinfer{T-If}
      {
        \IJtype{\IJctx{\Gamma}}{\IEvar{y}}{\ITbool}
        \\
        {\begin{array}[b]{@{}c@{\qquad}c@{}}
            \IJtype{\IJctx{\Gamma}}{S_1}{\IJctx{\Gamma_1}}
             &
            \forall i \in [n].\; \IJtype{\Gamma_1}{z_{1,i}}{ \primtype_i }
            \\
            \IJtype{\IJctx{\Gamma}}{S_2}{\IJctx{\Gamma_2}}
             &
            \forall i \in [n].\; \IJtype{\Gamma_2}{z_{2,i}}{ \primtype_i }
          \end{array}}
      }
      {
        \IJtype{\IJctx{\Gamma}}{
          (\ISif{
            (\IEvar{x_1}:\primtype_1, \ldots, \IEvar{x_n}:\primtype_n)
          }{\IEvar{y}}{
            S_1; \IEvar{\bar{z_1}}
          }{
            S_2; \IEvar{\bar{z_2}}
          })
        }{(\IJctx{\Gamma}, \IEvar{x_1}:\primtype_1, \ldots, \IEvar{x_n}:\primtype_n)}
      }
    \end{Imathpar}

    \textbf{Programs}
    \hfill\smash{\fbox{$\IJtype{}{P}{\IJok}$}}
    \begin{Imathpar}
      \Iinfer{T-Prog}
      {
        \IJtype{
          (\IEvar{x_1}:\IT{\primtype_1}, \ldots, \IEvar{x_n}:\IT{\primtype_n})
        }{S}{\IJctx{\Gamma}}
        \\
        \IJtype{\Gamma}{y}{ \primtype' }
      }
      {
        \IJtype{}{
          (\IPdef{f}{
            \IEvar{x_1}:\IT{\primtype_1}, \ldots, \IEvar{x_n}:\IT{\primtype_n}
          }{\primtype'}{
            S;\ISret{\IEvar{y}}
          })
        }{\IJok}
      }
    \end{Imathpar}
  \end{figdefs}
  \caption{Type system of \IR{}.}
  \label{fig:ir-typing-full}
\end{figure}

%% file: figs/fig-c0-typing-full.tex
\begin{figure}[!p]
  \begin{figdefs}[5pt]
    \setlength{\fboxsep}{2pt}
    \textbf{Definitions} (typing contexts)
    \hfill\fbox{$\CJctx{\Gamma}$}
    \begin{Cgather*}
      \CJctx{\Gamma} \Coloneqq \varepsilon\mid
      \CJctx{\Gamma},\CEvar{x}:\primtype\mid
      \CJctx{\Gamma},\CEvar{x}:\primtype[n]
    \end{Cgather*}

    \textbf{Expressions}
    \hfill\smash{\fbox{$\CJtype{\CJctx{\Gamma}}{E}{\primtype}$}}
    \begin{Cmathpar}
      \Cinfer{T-Const}
      {v\in\CVal_\primtype}
      {\CJtype{\CJctx{\Gamma}}{\CE{v}}{\primtype}}
      \and
      \Cinfer{T-Var}
      { \Gamma(x) = \primtype }
      {\CJtype{\CJctx{\Gamma}}{\CEvar{x}}{\primtype}}
      \and
      \Cinfer{T-Op}
      {
        \forall i\in[n].\
        \Gamma({x_i}) = \primtype_i
      }
      {\CJtype{\CJctx{\Gamma}}
        {\CEapp{\CEop[{\CTfun{(\primtype_1, \ldots, \primtype_n)}{\primtype'}}]}
          {\CEvar{x_1, \ldots, x_n}}}{\primtype'}}
      \and
      \Cinfer{T-Get}
      {
        \Gamma(x) = \primtype[n]
        \\
        \Gamma(y) = \CTuint{k}
      }
      {\CJtype{\CJctx{\Gamma}}{\CEget{\CEvar{x}}{\CEvar{y}}}{\primtype}}
    \end{Cmathpar}

    \textbf{Statements}
    \hfill\smash{\fbox{$\CJtype{\CJctx{\Gamma}}{S}{\CJctx{\Gamma'}}$}}
    \begin{Cmathpar}
      \Cinfer{T-Skip}{ }
      {\CJtype{\CJctx{\Gamma}}{\CSskip}{\CJctx{\Gamma}}}
      \and
      \Cinfer{T-Decl}
      {
        \tcr{x \notin \Zdom(\Gamma)}
      }
      {\CJtype{\CJctx{\Gamma}}{(\CSdecl{\primtype}{x})}
        {(\CJctx{\Gamma},\CEvar{x}:\primtype)}}
      \and
      \Cinfer{T-Assn}
      {
        \tcr{\CJtype{\Gamma}{x}{\primtype} }
        \\
        \CJtype{\CJctx{\Gamma}}{E}{\primtype}
      }
      {\CJtype{\CJctx{\Gamma}}{(\CSassn{x}{E})}{\CJctx{\Gamma}}}
      \and
      \Cinfer{T-New}
      {
        \tcr{x \notin \Zdom(\Gamma)}
        \\
        n\geq1
        \\
        \forall i\in[n].\ v_i\in\CVal_\primtype
      }
      {\CJtype{\CJctx{\Gamma}}
        {(\CSnew{\primtype}{x}{n}{v_1,\ldots,v_n})}
        {(\CJctx{\Gamma},\CEvar{x}:\primtype[n])}}
      \and
      \Cinfer{T-Seq}
      {\CJtype{\CJctx{\Gamma}}{S_1}{\CJctx{\Gamma_1}}
        \\ \CJtype{\CJctx{\Gamma_1}}{S_2}{\CJctx{\Gamma_2}}}
      {\CJtype{\CJctx{\Gamma}}{\CSseq{S_1}{S_2}}{\CJctx{\Gamma_2}}}
      \and
      \Cinfer{T-If}
      {
        \CJtype{\Gamma}{x}{ \CTbool }
        \\
        \CJtype{\CJctx{\Gamma}}{S_1}{\CJctx{\Gamma_1}}
        \\
        \CJtype{\CJctx{\Gamma}}{S_2}{\CJctx{\Gamma_2}}
      }
      {\CJtype{\CJctx{\Gamma}}{\CSif{x}{S_1}{S_2}}{\CJctx{\Gamma}}}
    \end{Cmathpar}

    \textbf{Programs}
    \hfill\smash{\fbox{$\CJtype{}{P}{\CJok}$}}
    \begin{Cmathpar}
      \Cinfer{T-Prog}
      {
        \CJtype{(\CEvar{x_1}:\primtype_1, \ldots, \CEvar{x_n}:\primtype_n)}{S}{\CJctx{\Gamma}}
        \\
        \CJtype{\Gamma}{y}{ \primtype' }
      }
      {\CJtype{}
        {\CPdef{f}{\primtype_1\ x_1, \ldots, \primtype_n\ x_n}{\primtype'}{S\ \CSret{y}}}
        {\CJok}}
    \end{Cmathpar}
  \end{figdefs}
  \caption{Type system of \Czero{}.}
  \label{fig:c0-typing-full}
\end{figure}

%% file: figs/fig-ir-semantics-full.tex
\begin{figure}[!p]
  \begin{figdefs}[5pt]
    \setlength{\fboxsep}{2pt}
    \textbf{Definitions} (states)
    \hfill\fbox{$\IJctx{\sigma}$}
    \begin{Igather*}
      \IJctx{\sigma} \Coloneqq \IJctx{\varepsilon} \mid
      \IJctx{\sigma,\IEvar{x}\mapsto \IEval{v}}
    \end{Igather*}

    \textbf{Expressions}
    \hfill\smash{\fbox{$\IJeval{\sigma}{E}{\IEval{v}}$}}
    \begin{Imathpar}
      \Iinfer{E-Const}
      { \tcr{v \in \IVal_{\primtype}} }
      {\IJeval{\sigma}{\IEval{v}}{\IEval{v}}}
      \and
      \Iinfer{E-Var}
      {\sigma(\IEvar{x})=\IEval{v}}
      {\IJeval{\sigma}{\IEvar{x}}{\IEval{v}}}
      \and
      \Iinfer[width=0.2\linewidth]{E-Op}
      {
        \forall i \in [n].\;
        \IJeval{\sigma}{\IEvar{x_i}}{\IEval{v_i}}
        \\
        \Isem{\IEop[{\ITfun{(\primtype_1, \ldots, \primtype_n)}{\primtype'}}]}(\IEval{v_1, \ldots, v_n})=\IEval{v'}
      }
      {
        \IJeval{\sigma}{\IEapp{\IEop[{\ITfun{(\primtype_1, \ldots, \primtype_n)}{\primtype'}}]}{\IEvar{x_1, \ldots, x_n}}}{\IEval{v'}}
      }
      \and
      \Iinfer{E-Get}
      {
        \IJeval{\sigma}{\IEvar{x}}{\IEvall{\IT{\primtype[n]}}{\IEval{v_1}, \ldots, \IEval{v_n}}}
        \\
        \IJeval{\sigma}{\IEvar{y}}{\IEvall{\ITuint{k}}{i}}
        \\
        i \in [n]_0
      }
      {
        \IJeval{\sigma}{\IEget{\IEvar{x}, \IEvar{y}}}{\IEval{v_{i+1}}}
      }
    \end{Imathpar}

    \textbf{Statements}
    \hfill\smash{\fbox{$\IJeval{\sigma}{S}{\sigma'}$}}
    \begin{Imathpar}
      \Iinfer{E-Skip}
      { }
      {\IJeval{\sigma}{\ISskip}{\sigma}}
      \and
      \Iinfer{E-Assn}
      {\IJeval{\sigma}{E}{\IEval{v}}}
      {
        \IJeval{\sigma}{(\ISassn{\IEvar{x}:\primtype}{E})}{(\sigma,\IEvar{x}\mapsto \IEval{v})}
      }
      \and
      \Iinfer{E-New}
      {
        n\geq1
        \\
        \tcr{\forall i\in[n].\ \IEval{v_i}\in\IVal[\primtype]}
      }
      {
        \IJeval{\sigma}
        {(\ISassn{\IEvar{x}:\IT{\primtype[n]}}
          {\IEnew{\IEval{v_1},\ldots,\IEval{v_n}}})}
        {(\sigma,\IEvar{x}\mapsto
          \IEvall{\IT{\primtype[n]}}{\IEval{v_1},\ldots,\IEval{v_n}})}
      }
      \and \!\!\!\!\!\!\!\!\!\!
      \Iinfer[width=0.2\linewidth]{E-Seq}
      {
        \IJeval{\sigma}{S_1}{\sigma_1}
        \\
        \IJeval{\sigma_1}{S_2}{\sigma_2}
      }
      {
        \IJeval{\sigma}{\ISseq{S_1}{S_2}}{\sigma_2}
      }
      \and
      \Iinfer{E-IfTrue} %
      {
        \IJeval{\sigma}{\IEvar{y}}{\IEvall{\ITbool}{\IEtrue}}
        \\
        \IJeval{\sigma}{S_1}{\sigma_1}
        \\
        \forall i \in [n].\;
        \IJeval{\sigma_1}{\IEvar{z_{1,i}}}{\IEval{v_i}}
      }
      {
        \IJeval{\sigma}{
          (\ISif{
            (\IEvar{x_1}:\primtype_1, \ldots, \IEvar{x_n}:\primtype_n)
          }{\IEvar{y}}{
            S_1; \IEvar{\bar{z_1}}
          }{
            S_2; \IEvar{\bar{z_2}}
          })
        }{
          (\sigma,\IEvar{x_1}\mapsto\IEval{v_1}, \ldots, \IEvar{x_n}\mapsto\IEval{v_n})
        }
      }
      \and
      \Iinfer{E-IfFalse} %
      {
        \IJeval{\sigma}{\IEvar{y}}{\IEvall{\ITbool}{\IEfalse}}
        \\
        \IJeval{\sigma}{S_2}{\sigma_2}
        \\
        \forall i \in [n].\;
        \IJeval{\sigma_2}{\IEvar{z_{2,i}}}{\IEval{v_i}}
      }
      {
        \IJeval{\sigma}{
          (\ISif{
            (\IEvar{x_1}:\primtype_1, \ldots, \IEvar{x_n}:\primtype_n)
          }{\IEvar{y}}{
            S_1; \IEvar{\bar{z_1}}
          }{
            S_2; \IEvar{\bar{z_2}}
          })
        }{
          (\sigma,\IEvar{x_1}\mapsto\IEval{v_1}, \ldots, \IEvar{x_n}\mapsto\IEval{v_n})
        }
      }
    \end{Imathpar}

    \textbf{Programs}
    \hfill\smash{\fbox{$\IJeval{P}{f(\IEval{\bar v})}{\IEval{v'}}$}}
    \begin{Imathpar}
      \Iinfer{E-Prog}
      {
        \IJeval{(\IEvar{x_1}\mapsto\IEval{v_1}, \ldots, \IEvar{x_n}\mapsto\IEval{v_n})}{S}{\sigma}
        \\
        \IJeval{\sigma}{\IEvar{y}}{\IEval{v'}}
      }
      {
        \IJeval{
          (\IPdef{f}{
            \IEvar{x_1}:\IT{\primtype_1}, \ldots, \IEvar{x_n}:\IT{\primtype_n}
          }{\primtype'}{
            S;\ISret{\IEvar{y}}
          })
        }{f(\IEval{v_1}, \ldots, \IEval{v_n})}{\IEval{v'}}
      }
    \end{Imathpar}
  \end{figdefs}
  \caption{Operational semantics of \IR{}.}
  \label{fig:ir-semantics-full}
\end{figure}

%% file: figs/fig-c0-semantics-full.tex
\begin{figure}[!p]
  \begin{figdefs}[5pt]
    \setlength{\fboxsep}{2pt}
    \textbf{Definitions} (states)
    \hfill\fbox{$\sigma$}
    \begin{Cgather*}
      \sigma \Coloneqq \varepsilon
      \mid \sigma, x\mapsto v
      \mid \sigma, x\mapsto \tcr{\Cuninit_\primtype}
    \end{Cgather*}

    \textbf{Expressions}
    \hfill\smash{\fbox{$\CJeval{\sigma}{E}{v}$}}
    \begin{Cmathpar}
      \Cinfer{E-Const}
      { \tcr{v \in \CVal_{\primtype}} }
      {\CJeval{\sigma}{\CE{v}}{v}}
      \and
      \Cinfer[width=0.2\linewidth]{E-Var}{
        \sigma(x)=v
      }
      {\CJeval{\sigma}{\CEvar{x}}{v}}
      \and
      \Cinfer[width=0.2\linewidth]{E-Op}{
        \forall i\in[n].\ \CJeval{\sigma}{\CEvar{x_i}}{v_i}
        \\
        \Csem{\CEop[{\CTfun{(\primtype_1, \ldots, \primtype_n)}{\primtype'}}]}(v_1, \ldots, v_n)=v'
      }{\CJeval{\sigma}{\CEapp{\CEop[{\CTfun{(\primtype_1, \ldots, \primtype_n)}{\primtype'}}]}
          {\CEvar{x_1, \ldots, x_n}}}{v'}}
      \and
      \Cinfer{E-Get}{
        \CJeval{\sigma}{\CEvar{x}}{\CEval{\primtype[n]}{v_1,\ldots,v_n}}
        \\
        \CJeval{\sigma}{\CEvar{y}}{\CEval{\CTuint{k}}{i}}
        \\
        i \in [n]_0
      }{\CJeval{\sigma}{\CEget{\CEvar{x}}{\CEvar{y}}}{v_{i+1}}}
    \end{Cmathpar}

    \textbf{Statements}
    \hfill\smash{\fbox{$\CJeval{\sigma}{S}{\sigma'}$}}
    \begin{Cmathpar}
      \Cinfer{E-Skip}{ }{\CJeval{\sigma}{\CSskip}{\sigma}}
      \and\!\!\!
      \Cinfer{E-Decl}{ }
      {
        \CJeval{\sigma}{\CSdecl{\primtype}{x}}{(\sigma,x\mapsto \tcr{\Cuninit_\primtype})}
      }
      \and\!\!\!
      \Cinfer{E-Assn}{
        \CJeval{\sigma}{E}{v}
      }{
        \CJeval{\sigma}{\CSassn{x}{E}}{(\sigma,x\mapsto v)}
      }
      \and
      \Cinfer{E-New}{
        n\geq1
        \\
        \tcr{\forall i\in[n].\ v_i\in\CVal_\primtype}
      }{\CJeval{\sigma}
        {\CSnew{\primtype}{x}{n}{v_1,\ldots,v_n}}
        {(\sigma,x\mapsto\CEval{\primtype[n]}{v_1,\ldots,v_n})}}
      \and
      \Cinfer[width=0.2\linewidth]{E-Seq}{
        \CJeval{\sigma}{S_1}{\sigma_1}
        \\
        \CJeval{\sigma_1}{S_2}{\sigma_2}
      }{\CJeval{\sigma}{\CSseq{S_1}{S_2}}{\sigma_2}}
      \and
      \Cinfer{E-IfTrue}{
        \CJeval{\sigma}{\CEvar{x}}{\CEval{\CTbool}{\CEtrue}}
        \quad
        \CJeval{\sigma}{S_1}{\sigma_1}
      }{
        \CJeval{\sigma}
        {\CSif{x}{S_1}{S_2}}
        { \tcr{\sigma_1|_{\Zdom(\sigma)}} }
      }
      \and \!\!\!\!\!\!\!\!
      \Cinfer{E-IfFalse}{
        \CJeval{\sigma}{\CEvar{x}}{\CEval{\CTbool}{\CEfalse}}
        \quad
        \CJeval{\sigma}{S_2}{\sigma_2}
      }{
        \CJeval{\sigma}
        {\CSif{x}{S_1}{S_2}}
        { \tcr{\sigma_2|_{\Zdom(\sigma)}} }
      }
    \end{Cmathpar}

    \textbf{Programs}
    \hfill\smash{\fbox{$\CJeval{P}{f(\bar{v})}{v'}$}}
    \begin{Cmathpar}
      \Cinfer{E-Prog}{
        \CJeval{(x_1\mapsto v_1,\ldots,x_n\mapsto v_n)}{S}{\sigma}
        \\
        \CJeval{\sigma}{\CEvar{y}}{v'}
      }{\CJeval{\CPdef{f}{\primtype_1\ x_1,\ldots,\primtype_n\ x_n}{\primtype'}{S\ \CSret{y}}}
        {f(v_1,\ldots,v_n)}{v'}}
    \end{Cmathpar}
  \end{figdefs}
  \caption{Operational semantics of \Czero{}.}
  \label{fig:c0-semantics-full}
\end{figure}

%% file: appendices/app-comp-def.tex
\section{Complete Definition of the \texorpdfstring{\Ziv*{}}{Ziv} Compiler}
\label{sec:compiler-def}

In this section, we define the three steps of the \Ziv*{} compiler:
normalization, tracing, and code emission.
We use background colors to distinguish the languages:
\Zhl{yellow for \Ziv*{} and \Znorm*{}},
\Ihl{blue for \IR*{}}, and
\Chl{white for \Czero*{}}.

\subsection{Normalization (\texorpdfstring{$\Ziv*{}\to\Znorm*{}$}{Ziv → ZivNorm})}
\label{sec:compiler-def:norm}

The normalization step of the \Ziv*{} compiler is defined in \Cref{fig:trans-ziv-norm-full}.

The function $\Ztrans^{\Zhl{y}}$ chooses the first satisfying case when multiple cases are applicable.

\subsection{Tracing (\texorpdfstring{$\Znorm*{}\to\IR*{}$}{ZivNorm → ZivIR})}
\label{sec:compiler-def:trace}

The tracing step of the \Ziv*{} compiler is defined in \Cref{fig:trans-norm-ir-expr-full,fig:trans-norm-ir-stmt-full}.

We use the same notation as in \Cref{sec:lang-def-ziv,sec:lang-def-irc0}.
Note that $\Zhl{\theta}$ denotes an erased type in \Ziv{}
(i.e., $\ZT{\theta} = \Zerase(\ZT{\comptype})$ for some type $\ZT{\comptype}$ in $\Ziv$).

We define three auxiliary functions used in \Cref{fig:trans-norm-ir-expr-full,fig:trans-norm-ir-stmt-full}.

\paragraph{Bind Function.}

We define the bind function $\Tmat(\widehat{v})$, where $\widehat v$ is a tracing value:
\begin{align*}
  \Tmat(\widehat v) \defas
  \begin{cases}
    (\ISskip, \Tref{\IT{\primtype}}{\IEvar{x}})
     & \text{if }\widehat v=\Tref{\IT{\primtype}}{\IEvar{x}}
    \\
    (\ISassn{\IEvar{x}:\primtype}{\IEval{v}}, \Tref{\IT{\primtype}}{\IEvar{x}})
    \text{ for fresh }\IEvar{x}
     & \text{if }\widehat v=\Tcon{\ZV{v}}\text{ and }\ZV{v}\in\IVal[\primtype]
    \\
    \text{undefined}
     & \text{otherwise}.
  \end{cases}
\end{align*}
The bind function receives a tracing value $\widehat v$
and returns a pair $(\IS{S},\Tref{\IT{\primtype}}{\IEvar{x}})$,
where $\IS{S}$ is a \IR{} statement
that binds the value $\widehat v$ to the \IR{} variable $\IEvar{x}$.

\paragraph{Lookup Function.}

First, let $\ZT{\theta}$ be an erased type in $\Ziv$.
We define the set of tracing values of type $\ZT{\theta}$ inductively:
\begin{align*}
  \TVal_{\ZT{\theta}}
   & \defas
  \begin{cases}
    \{\Tcon{\ZE{v}}\mid \ZS{v \in \ZVal_\theta}\}
     &
    \text{if $\ZT{\theta = \primtype} \land \ZT{\theta} \notin \IR$}
    \\
    \{\Tcon{\ZE{v}}\mid \ZS{v \in \ZVal_\theta}\}\cup
    \{\Tref{\IT{\theta}}{\IEvar{x}}\mid \IS{\IE{x} \in \IR}\}
     &
    \text{if $\ZT{\theta = \primtype} \land \ZT{\theta} \in  \IR$}
    \\
    \{
    \Tcon{\ZE{v}}\mid \ZS{v \in \ZVal_\theta}\}\cup
    \{\Tcomp{\ZT{\theta}}{\widehat v_1,\ldots,\widehat v_n}
    \mid
    \widehat v_i \in \TVal_{\ZT{\theta_i}}
    \}
     &
    \text{otherwise},
  \end{cases}
\end{align*}
where $( \ZT{\theta_1,\ldots,\theta_n} ) = \Zfields(\ZT{\theta})$
is the ordered list of field types of $\ZT{\theta}$ in the last case.
Remark that $\TVal_{\ZT{\theta}} = \varnothing$ if $\ZT{\theta}$ is a function type,
since $\ZVal_{\ZT{\theta}} = \varnothing$ and $\Zfields(\ZT{\theta}) = \text{undefined}$.

Next, let $\ZT{\theta}$ be a non-function erased type in \Ziv{},
and let $\Tbfields(\ZT{\theta}) = (\ZT{\primtype_1,\ldots,\primtype_m})$
be the ordered list of primitive types that constitute $\ZT{\theta}$,
i.e., $\Tbfields(\ZT{\theta})$ is the result of recursively flattening
the field types of $\ZT{\theta}$ until only primitive types remain.
We define the unwrap function of $\ZT{\theta}$ as follows:
\begin{align*}
   & \Tenc_{\ZT{\theta}}
  : \TVal_{\ZT{\theta}} \to \prod_{j \in [m]} \TVal_{\ZT{\primtype_j}},
  \\
   & \Tenc_{\ZT{\theta}}(\widehat{v})
  \defas
  \begin{cases}
    (\widehat{v})
     & \text{if } \ZT{\theta = \primtype}
    \\
    \Tenc_{\ZT{\theta_1}}(\Tcon{\ZE{v_1}})
    \concat \cdots \concat
    \Tenc_{\ZT{\theta_n}}(\Tcon{\ZE{v_n}})
     & \text{if } \widehat{v} = \Tcon{\ZEval{\ZT{\theta}}{v_1,\ldots,v_n}}
    \\
    \Tenc_{\ZT{\theta_1}}(\widehat{v}_1)
    \concat \cdots \concat
    \Tenc_{\ZT{\theta_n}}(\widehat{v}_n)
     & \text{if } \widehat{v} = \Tcomp{\ZT{\theta}}{\widehat{v}_1,\ldots,\widehat{v}_n},
  \end{cases}
\end{align*}
where $\concat$ is the list concatenation operator,
and $(\ZT{\theta_1,\ldots,\theta_n}) = \Zfields(\ZT{\theta})$ in the last two cases.
The unwrap function converts a tracing value $\widehat{v}$ of type $\ZT{\theta}$
into tracing values $\widehat{v}_1, \ldots, \widehat{v}_m$ of primitive types $\Tbfields(\ZT{\theta})$.
We define the wrap function of $\ZT{\theta}$ as an inverse operation.
In particular, we define it to have the following signature and satisfy the following property:
\begin{align*}
   & \Tdec_{\ZT{\theta}}
  : \prod_{j \in [m]} \TVal_{\ZT{\primtype_j}} \to \TVal_{\ZT{\theta}},
  \\
   & \Tenc_{\ZT{\theta}}(\Tdec_{\ZT{\theta}}(\widehat v_1, \ldots, \widehat v_m))
  = (\widehat v_1, \ldots, \widehat v_m)
  \quad\text{for all $\textstyle (\widehat v_1, \ldots, \widehat v_m) \in \prod_{j \in [m]} \TVal_{\ZT{\primtype_j}}$}.
\end{align*}

We now define the lookup function $\Tlookup (\ZEval{\theta}{v_1,\ldots,v_n},\IEvar{x} )$ as follows,
where $\ZT{\theta = \TTtable{\theta'}{\bar n'}}$ is an erased type in \Ziv{} for tables,
$\ZT{v_1, \ldots, v_n} \in \ZVal_{\ZT{\theta'}}$ are \Ziv{} values for table entries,
and $\IEvar{x}$ is a \IR{} variable for a table index.
Let $\Tbfields(\ZT{\theta'}) = (\ZT{\primtype_1,\ldots,\primtype_m})$
be the ordered list of primitive types that constitute $\ZT{\theta'}$.
Let $\Tenc_{\ZT{\theta'}}(\Tcon{\ZE{v_i}}) = (\Tcon{\ZE{v_{i,1}}},\ldots,\Tcon{\ZE{v_{i,m}}})$
be the unwrapping of the table entry $\ZT{v_i}$ into primitive values ($i \in [n]$).
Then, the lookup function is defined as follows:
\begin{align*}
   & \Tlookup \big(
  \ZEval{\theta}{v_1,\ldots,v_n},
  \IEvar{x} \big)
  \defas
  \begin{cases}
    (\IS{S'}, \widehat v')
     &
    \text{if $\ZT{\primtype_j} \in \IR$ for all $j \in [m]$}
    \\
    \text{undefined}
     &
    \text{otherwise},
  \end{cases}
\end{align*}
where $\IS{S'}$ and $\widehat v'$ are defined by
\begin{align*}
  \IS{S'}
   & \defas
  \IS{
    {\begin{array}[c]{@{}r@{}l@{}r@{}l@{}}
        \ISassn{{\IEvar{y_1}} & {{} :\IT{\primtype_1[n]}} & {}}{{} & {\IEnew{\IEval{v_{1,1},\ldots,v_{n,1}}}}}
        \\
         &  & \cdots
        \\
        \ISassn{{\IEvar{y_m}} & {} :{\IT{\primtype_m[n]}} & {}}{{} & {\IEnew{\IEval{v_{1,m},\ldots,v_{n,m}}}}}
        \\
        \ISassn{{\IEvar{z_1 }}& {} :{\primtype_1} & {}}{{} & {\IEget{\IEvar{y_1},\IEvar{x}}}}
        \\
         &  & \cdots
        \\
        \ISassn{{\IEvar{z_m}} & {} :{\primtype_m} & {}}{{} & {\IEget{\IEvar{y_m},\IEvar{x}}}}
      \end{array}}
  }
  \text{ for fresh $\IEvar{y_j}$ and $\IEvar{z_j}$},
  \\
  \widehat v'
   & \defas
  \Tdec_{\ZT{\theta'}} (\Tref{\IT{\primtype_1}}{\IEvar{z_1}},\ldots,\Tref{\IT{\primtype_m}}{\IEvar{z_m}}).
\end{align*}

\paragraph{Merge Function.}

Let $\IEvar{b}$ be a \IR{} variable
that stores the condition value of a conditional statement,
and $(\IS{S_k},\widehat\sigma_k)$ ($k=1,2$) be a pair of a \IR{} statement and a tracing state
that results from tracing either the true or false branch of the conditional statement.
We define the merge function $\Tjoin \big( \IEvar{b},(\IS{S_1},\widehat\sigma_1), (\IS{S_2},\widehat\sigma_2) \big)$
as follows.

First, let $X$ be the set of variables defined at the end of both branches: %
\begin{align*}
  X
   & \defas
  \Zdom(\widehat\sigma_1)\cap\Zdom(\widehat\sigma_2).
\end{align*}
We consider two disjoint subsets $X^{\mathrm{val}}$ and $X^{\mathrm{typ}}$ of $X$,
based on whether the two branches assign the same tracing value (or erased type) to a variable $\ZT{x} \in X$:
\begin{align*}
  X^{\mathrm{val}}
   & \defas
  \{ \ZT{x} \in X \mid \widehat\sigma_1(\ZT{x}) = \widehat\sigma_2(\ZT{x}) \}
  \asdef \{\ZT{x^{\mathrm{val}}_1, \ldots, x^{\mathrm{val}}_n}\},
  \\
  X^{\mathrm{typ}}
   & \defas
  \big\{
  \ZT{x} \in X \setminus X^{\mathrm{val}}
  \ \big|\
  \begin{array}[t]{@{}l@{}}
    \mathit{etype}(\widehat\sigma_1(\ZT{x}))
    = \mathit{etype}(\widehat\sigma_2(\ZT{x}))
    \land
    \forall \primtype \in \Tbfields(\mathit{etype}(\widehat\sigma_1(\ZT{x}))).\; \primtype \in \IR
  \end{array}
  \big\}
  \\
   & \asdef \{\ZT{x^{\mathrm{typ}}_1, \ldots, x^{\mathrm{typ}}_m}\},
\end{align*}
where $\ZT{x^{\mathrm{val}}_1, x^{\mathrm{val}}_2,
    \ldots, x^{\mathrm{typ}}_1, x^{\mathrm{typ}}_2, \ldots}$ are distinct,
and $\mathit{etype}(\widehat v)$ is the erased type of $\widehat v$ defined by
\begin{align*}
  \mathit{etype}(\widehat v) \defas
  \begin{cases}
    \ZT{\theta}
     & \text{if $\widehat v = \Tcon{\ZE{v}}$ and $\ZV{v} \in \ZVal_{\ZT{\theta}}$}
    \\
    \ZT{\primtype}
     & \text{if $\widehat v = \Tref{\IT{\primtype}}{\IEvar{x}}$}
    \\
    \ZT{\theta}
     & \text{if $\widehat v = \Tcomp{\ZT{\theta}}{\widehat v_1, \ldots, \widehat v_n}$}
    \\
    \text{undefined}
     & \text{otherwise}.
  \end{cases}
\end{align*}
In other words, $\ZP{x} \in X^{\mathrm{val}}$ if and only if $\ZP{x}$ stores the same tracing value at the end of both branches;
and $\ZP{x} \in X^{\mathrm{typ}}$ if and only if
(i) $\ZP{x}$ stores different tracing values at the end of the two branches,
(ii) $\ZP{x}$ has the same non-function erased type $\ZP{\theta}$ at the end of both branches,
and (iii) $\ZP{\theta}$ consists of primitive types in \IR{}.

We now define the merge function $\Tjoin \big( \IEvar{b},(\IS{S_1},\widehat\sigma_1), (\IS{S_2},\widehat\sigma_2) \big)$.
The high-level idea is to
(i) assign the variables in $X^{\mathrm{val}}$ to their common values and
(ii) assign the variables in $X^{\mathrm{typ}}$ to new values
that are computed at runtime through a conditional expression with $\IEvar{b}$.
Let $\ZT{\theta_i} \defas \mathit{etype}(\widehat\sigma_1(\ZT{x^{\mathrm{typ}}_i}))$
be the common erased type of $\ZT{x^{\mathrm{typ}}_i} \in X^{\mathrm{typ}}$.
For each $k \in [2]$ and $i \in [m]$, define
\begin{gather*}
  \ZT{\bar{\primtype_i}}
  \defas \Tbfields (\ZT{\theta_i}),
  \qquad
  \IT{\bar{z_i}}
  \defas \text{$|\ZT{\bar{\primtype_i}}|$ fresh variables in $\IR$},
  \\
  (\IS{\overline{S_{k,i}}}, \Tref{\IT{\bar{\primtype_i}}}{\IEvar{\overline{y_{k,i}}}})
  \defas
  \Tmat \big( \Tenc_{\ZT{\theta_i}}(\widehat\sigma_k(\ZT{x^{\mathrm{typ}}_i})) \big),
\end{gather*}
where the bind function is extended to a list of tracing values such that
it concatenates the computed \IR{} statements and tracing values.
Then, the merge function is defined as follows:
\begin{align*}
  \Tjoin \big( \IEvar{b},(\IS{S_1},\widehat\sigma_1), (\IS{S_2},\widehat\sigma_2) \big)
   & \defas
  (\IS{S'}, \widehat\sigma'),
\end{align*}
where $\IS{S'}$ and $\widehat\sigma'$ are defined by
\begin{align*}
  \IS{S'}
   & \defas
  \IS{
    \begin{array}[c]{@{}l@{}}
      (\IEvar{\bar{z_1}} \,{:}\, \bar{\primtype_1}, \ldots, \IEvar{\overline{z_m}} \,{:}\, \overline{\primtype_m})
      {} = \Ikw{if}\ \IT{b}\
      \{ S_1; \overline{S_{1,1}}; \cdots; \overline{S_{1,m}}; (\IEvar{\overline{y_{1,1}}}, \ldots, \IEvar{\overline{y_{1,m}}}) \}
      \\
      \phantom{
        (\IEvar{\bar{z_1}} \,{:}\, \bar{\primtype_1}, \ldots, \IEvar{\bar{z_m}} \,{:}\, \bar{\primtype_m})
        {} = \Ikw{if}\ x\
      }
      \{ S_2; \overline{S_{2,1}}; \cdots; \overline{S_{2,m}}; (\IEvar{\overline{y_{2,1}}}, \ldots, \IEvar{\overline{y_{2,m}}}) \}
    \end{array}
  }\,,
  \\
  \widehat\sigma'
   & \defas
  \left(
  \big(
  \ZT{x^{\mathrm{val}}_i} \mapsto \widehat\sigma_1(\ZT{x^{\mathrm{val}}_i})
  \big)_{i \in [n]}
  ,{\,}
  \big(
  \ZT{x^{\mathrm{typ}}_i} \mapsto \Tdec_{\ZT{\theta_i}}(\Tref{\IT{\bar{\primtype_i}}}{\IEvar{\bar{z_i}}})
  \big)_{i \in [m]}
  \right).
\end{align*}

\subsection{Code Emission (\texorpdfstring{$\IR*{}\to\Czero*{}$}{ZivIR → C₀})}
\label{sec:compiler-def:emit}

The code emission step of the \Ziv*{} compiler is defined in \Cref{fig:trans-ir-c0-full}.

\input{figs/fig-trans-ziv-norm-full}
\input{figs/fig-trans-ir-c0-full}
\afterpage{\clearpage}

\input{figs/fig-trans-norm-ir-expr-full}
\afterpage{\clearpage}
\input{figs/fig-trans-norm-ir-stmt-full}
\afterpage{\clearpage}

%% file: figs/fig-trans-ziv-norm-full.tex
\begin{figure}[!p]
  \begin{figdefs}[5pt]
    \setlength{\fboxsep}{2pt}
    \textbf{Statements}
    \hfill\fbox{$\Ztrans^{\smash{\ZE{y}}}(\ZS{S},\ZS{S'})$}
    \begin{align*}
      \Ztrans^{\ZE{y}}(\ZS{S},\ZS{S'})
       & \defas \ZSseq{S}{S'}
      \qquad \text{if } \neg\Zhasret(S),
      \\
      \Ztrans^{\ZE{y}}(\ZSret{E},\ZS{S'})
       & \defas \tcr{\ZSassn{y}{E}},
      \\
      \Ztrans^{\ZE{y}}(\ZSseq{S_1}{S_2}, \ZS{S'})
       & \defas \Ztrans^{\ZE{y}}(
      \ZS{S_1}, {\Ztrans^{\ZE{y}}(\ZS{S_2},\ZS{S'})}
      ),
      \\
      \Ztrans^{\ZE{y}}(\ZSif{E}{S_1}{S_2}, \ZS{S'})
       & \defas \ZSif{E}{\Ztrans^{\ZE{y}}(\ZS{S_1},\ZS{S'})}{\Ztrans^{\ZE{y}}(\ZS{S_2},\ZS{S'})}.
    \end{align*}

    \textbf{Programs}
    \hfill\smash{\fbox{$\Ztrans(\ZP{P})$}}
    \begin{align*}
      \Ztrans(\ZPnil)
       & \defas
      \ZPnil,
      \\
      \Ztrans ( %
      \ZPseq{P}{\ZPdef{f}{\bar x:\bar\comptype}{\comptype'}{S}}
      ) %
       & \defas
      \ZPseq{\Ztrans(\ZP{P})}{
        \ZPdef{f}{\bar x:\bar\comptype}{\comptype'}{
          \ZSseq{\Ztrans^{\ZE{y}}(\ZS{S},\ZSskip)}{\tcr{\ZSret{y}}}}},
    \end{align*}
    where $\ZE{y}$ is a fresh variable.
  \end{figdefs}
  \caption{Translation from \Ziv{} to $\Znorm$.}
  \label{fig:trans-ziv-norm-full}
  \bigskip
\end{figure}

%% file: figs/fig-trans-ir-c0-full.tex
\begin{figure}[!t]
  \begin{figdefs}[5pt]
    \setlength{\fboxsep}{2pt}
    \textbf{Expressions}
    \hfill\fbox{$\Ctrans(\IE{E})$}
    \begin{align*}
      \Ctrans(\IEval{v})
       & \defas \CE{v},
      \\
      \Ctrans(\IEvar{x})
       & \defas \CEvar{x},
      \\
      \Ctrans(\IEapp{\IEop[{\ITfun{\bar\primtype}{\primtype'}}]}{\bar x})
       & \defas \CEapp{\CEop[{\CTfun{\bar\primtype}{\primtype'}}]}{\CEvar{\bar x}},
      \\
      \Ctrans(\IEget{x,y})
       & \defas \CEget{\CEvar{x}}{\CEvar{y}}.
    \end{align*}

    \textbf{Statements}
    \hfill\smash{\fbox{$\Ctrans(\IS{S})$}}
    \begin{align*}
      \Ctrans(\ISskip)
       & \defas\CSskip,
      \\
      \Ctrans(\ISassn{x:\primtype}{E})
       & \defas
      \CSseq{
        \CSdecl{\primtype}{x}
      }{
        \CSassn{x}{\Ctrans(\IE{E})}
      },
      \\
      \Ctrans(\ISassn{x:\IT{\primtype[n]}}{\IEnew{\bar{v}}})
       & \defas
      \CSnew{\primtype}{x}{n}{\bar{v}},
      \\
      \Ctrans(\ISseq{S_1}{S_2})
       & \defas\CSseq{\Ctrans(\IS{S_1})}{\Ctrans(\IS{S_2})},
      \\
      \Ctrans\bigl(
      \ISif{(\bar x:\bar\primtype)}{y}
      {S_1; \bar{z_1}}{S_2;\bar{z_2}}\bigr)
       & \defas
      \CSseq{
        \tcr{ \CSdecl{\bar\primtype}{\bar x} }
      }{
        \CSif{y}
        {\CSseq{\Ctrans(\IS{S_1})}{\tcr{\CSassn{\bar x}{\bar{z_1}}}}}
        {\CSseq{\Ctrans(\IS{S_2})}{\tcr{\CSassn{\bar x}{\bar{z_2}}}}}
      },
    \end{align*}
    where $\CSdecl{\bar\primtype}{\bar x}$ and $\CSassn{\bar x}{\bar{z_i}}$
    use semicolons between declarations and assignments.

    \textbf{Programs}
    \hfill\smash{\fbox{$\Ctrans(\IP{P})$}}
    \begin{align*}
      \Ctrans\bigl(\IPdef{f}{\bar x:\bar\primtype}{\primtype'}{S;\ISret{y}}\bigr)
       & \defas\CPdef{f}{\bar\primtype\ \bar x}{\primtype'}{\Ctrans(\IS{S})\ \CSret{y}}.
    \end{align*}
  \end{figdefs}
  \caption{Translation from \IR{} to \Czero{}.}
  \label{fig:trans-ir-c0-full}
\end{figure}

%% file: figs/fig-trans-norm-ir-expr-full.tex
\begin{figure}[!p]
  \begin{figdefs}[5pt]
    \setlength{\fboxsep}{2pt}
    \textbf{Definitions} (tracing values, tracing states)
    \bgroup\msmall
    \hfill\fbox{$\widehat v$, $\widehat\sigma$}
    \begin{gather*}
      \widehat v
      \Coloneqq
      \tcr{
        \Tcon{\ZV{v}}
        \mid \Tref{\IT{\primtype}}{\IEvar{x}}
        \mid \Tcomp{\ZT{\theta}}{\widehat v_1,\ldots,\widehat v_n}
      }
      \qquad\qquad
      \widehat\sigma
      \Coloneqq \varepsilon
      \mid \widehat\sigma, \ZE{x}\mapsto\widehat v
    \end{gather*}
    \egroup

    \textbf{Expressions}
    \bgroup\msmall
    \hfill\smash{\fbox{$\ZJcomp{\ZP{P}; \widehat\sigma}{E}{(\IS{S},\widehat v)}$}}
    \begin{mathpar}
      \Tinfer{C-Const}
      { }
      {\ZJcomp{\ZP{P}; \widehat\sigma}{v}{(\ISskip, \Tcon{\ZV{v}})}}
      \and
      \Tinfer{C-Var}
      {\widehat\sigma(\ZE{x})=\widehat v}
      {\ZJcomp{\ZP{P}; \widehat\sigma}{x}{(\ISskip,\widehat v)}}
      \and
      \Tinfer{C-OpCt}
      {
        \forall i\in[n].\
        \ZJcomp{\ZP{P}; \widehat\sigma}{E_i}{(\IS{S_i},\Tcon{\ZV{v_i}})}
        \\
        {\begin{array}[b]{@{}c@{}}
            {
              \Zsem{\Zhl{\ZEop[{\ZTfun{(\primtype_1, \ldots, \primtype_n)}{\primtype'}}]}}
              (\Zhl{\ZV{v_1}, \ldots, \ZV{v_n}})=\ZV{v'}
            }%
          \end{array}}
      }
      {\ZJcomp{\ZP{P}; \widehat\sigma}{\ZEop^\ell[{\ZTfun{(\primtype_1, \ldots, \primtype_n)}{\primtype'}}](E_1, \ldots, E_n)}
        {(\IS{S_1;\cdots;S_n}, {\Tcon{\Zhl{v'}}})}}
      \and
      \Tinfer{C-OpRt}
      {
        \forall i\in[n].\
        \ZJcomp{\ZP{P}; \widehat\sigma}{E_i}{(\IS{S_i},\widehat v_i)}
        \qquad
        \exists j \in[n].\ \tcr{\forall \ZV{v}.\ \widehat v_j \neq \Tcon{\ZV{v}}}
        \\
        \forall i\in[n].\ \tcr{\Tmat(\widehat v_i)} = (\IS{S_{i+n}},\Tref{\IT{\primtype_i}}{\IEvar{x_i}})
        \\
        \IS{
          S' \defas (
          S_1;\ \cdots;\ S_{2n};\
          \tcr{
            \ISassn{
              \IEvar{y'} \!: \primtype'
            }{
              \IEapp{\IEop[{\ITfun{(\primtype_1, \ldots, \primtype_n)}{\primtype'}}]}{\IEvar{x_1}, \ldots, \IEvar{x_n}}
            }
          }
          )
        }
        \text{ for fresh $\IE{y'}$}
      }
      {
        \ZJcomp{\ZP{P}; \widehat\sigma}{\ZEop^\ell[{\ZTfun{(\primtype_1, \ldots, \primtype_n)}{\primtype'}}](E_1, \ldots, E_n)}
        {(\IS{S'},
          \tcr{\Tref{\IT{\primtype'}}{\IEvar{y'} \!}})}
      }
      \and
      \Tinfer{C-Call} %
      {
        {\begin{array}[b]{@{}c@{}}
            \ZJcomp{\ZP{P}; \widehat\sigma}{E_0}{(\IS{S_{0}}, \Tcon{\ZV{f}})}
            \\
            \forall i\in[n].\ \ZJcomp{\ZP{P}; \widehat\sigma}{E_i}{(\IS{S_i},\widehat v_i)}
          \end{array}}
        \\
        {\begin{array}[b]{@{}c@{}}
            \Zhl{(\ZPdef{f}{x_1:\comptype_1, \ldots, x_n:\comptype_n}{\comptype'}{\ZSseq{S}{\ZSret{E}}}) \in P}
            \\
            \ZJcomp{\ZP{P}; (\Zhl{x_i}\mapsto\widehat v_i)_{i \in [n]}}{S}{(\IS{S_{n+1}},\widehat\sigma')}
            \\
            \ZJcomp{\ZP{P}; \widehat\sigma'}{E}{(\IS{S_{n+2}},\widehat v')}
          \end{array}}
      }
      {
        \ZJcomp{\ZP{P}; \widehat\sigma}{E_0(E_1, \ldots, E_n)}
        {(\IS{S_0;\cdots;S_{n+2}},\widehat v')}
      }
      \and
      \Tinfer{C-NewCt} %
      {\forall i\in[n].\
        \ZJcomp{\ZP{P}; \widehat\sigma}{E_i}{(\IS{S_i}, \Tcon{\ZE{v_i}})}
      }
      {
        \ZJcomp{
          \ZP{P}; \widehat\sigma
        }{
          \ZEnew{\comptype}{E_1, \ldots, E_n}
        }{
          (\IS{S_1;\cdots;S_n}, \Tcon{\ZEval{\Zerase(\ZT{\comptype})}{v_1,\ldots,v_n}})
        }
      }
      \and %
      \Tinfer{C-NewRt} %
      {\forall i\in[n].\
        \ZJcomp{\ZP{P}; \widehat\sigma}{E_i}{(\IS{S_i},\widehat v_i)}
        \\
        \exists j \in[n].\ \tcr{\forall \ZV{v}.\ \widehat v_j \neq \Tcon{\ZV{v}}}
      }
      {
        \ZJcomp{
          \ZP{P}; \widehat\sigma
        }{
          \ZEnew{\comptype}{E_1, \ldots, E_n}
        }{
          (\IS{S_1;\cdots;S_n}, \tcr{\Tcomp{\Zerase(\ZT{\comptype})}{\widehat v_1,\ldots,\widehat v_n}})
        }
      }
      \and
      \Tinfer{C-GetCtInd1} %
      {
        \ZJcomp{
          \ZP{P}; \widehat\sigma
        }{E_1}{
          (\IS{S_1}, \Tcon{\ZEval{\theta}{v_1,\ldots,v_n}})
        }
        \\
        \ZJcomp{\ZP{P}; \widehat\sigma}{E_2}
        {(\IS{S_2},\Tcon{\ZEval{\ZTuint{k}}{i}})}
        \\
        i\in[n]_0
      }
      {\ZJcomp{\ZP{P}; \widehat\sigma}{\ZEget{E_1,E_2}}
        {(\IS{S_1;S_2},\Tcon{\ZV{v_{i+1}}})}}
      \and\!\!\!\!\!\!\!\!\!
      \Tinfer{C-GetCtInd2} %
      {
        \ZJcomp{
          \ZP{P}; \widehat\sigma
        }{E_1}{
          (\IS{S_1}, \Tcomp{\ZT{\theta}}{\widehat v_1,\ldots,\widehat v_n})
        }
        \\
        \ZJcomp{\ZP{P}; \widehat\sigma}{E_2}
        {(\IS{S_2},\Tcon{\ZEval{\ZTuint{k}}{i}})}
        \\
        i\in[n]_0
      }
      {\ZJcomp{\ZP{P}; \widehat\sigma}{\ZEget{E_1,E_2}}
        {(\IS{S_1;S_2},\widehat v_{i+1})}}
      \and
      \Tinfer{C-GetRtInd} %
      {
        \ZJcomp{\ZP{P}; \widehat\sigma}{E_1}{
          (\IS{S_1},\Tcon{\ZV{
              \ZEval{
                \theta
              }{
                v_1,\ldots,v_n
              }
            }})
        }
        \text{ for }
        \ZT{\theta = \TTtable{\theta'}{\bar{n}'}}
        \\
        \ZJcomp{
          \ZP{P}; \widehat\sigma
        }{E_2}{
          (\IS{S_2}, \tcr{\Tref{\ITuint{k}}{\IEvar{x}}})
        }
        \\
        \tcr{
          \Tlookup(
          \ZEval{\theta}{v_1,\ldots,v_n},
          \IEvar{x}
          )
        }
        =(\IS{S_3},\widehat v)
      }
      {\ZJcomp{\ZP{P}; \widehat\sigma}{\ZEget{E_1,E_2}}
        {(\IS{S_1;S_2;S_3},\widehat v)}}
    \end{mathpar}
    \egroup
  \end{figdefs}
  \caption{Translation from $\Znorm$ to $\IR$ for expressions.}
  \label{fig:trans-norm-ir-expr-full}
\end{figure}

%% file: figs/fig-trans-norm-ir-stmt-full.tex
\begin{figure}[!p]
  \begin{figdefs}[5pt]
    \setlength{\fboxsep}{2pt}
    \textbf{Statements} (return-free)
    \bgroup\msmall
    \hfill\fbox{$\ZJcomp{\ZP{P}; \widehat\sigma}{S}{(\IS{S'},\widehat\sigma')}$}
    \begin{mathpar}
      \Tinfer{C-Skip}
      { }
      {\ZJcomp{\ZP{P}; \widehat\sigma}{\ZSskip}{(\ISskip,\widehat\sigma)}}
      \and
      \Tinfer{C-Assn}
      {\ZJcomp{\ZP{P}; \widehat\sigma}{E}{(\IS{S},\widehat v)}}
      {
        \ZJcomp{\ZP{P}; \widehat\sigma}{\ZSassn{x}{E}}
        {(\IS{S},(\widehat\sigma, \Zhl{x}\mapsto \widehat v))}
      }
      \and
      \Tinfer{C-Seq}
      {\ZJcomp{\ZP{P}; \widehat\sigma}{S_1}{(\IS{S'_1},\widehat\sigma'_1)}
        \\ \ZJcomp{\ZP{P}; \widehat\sigma'_1}{S_2}{(\IS{S'_2},\widehat\sigma'_2)}}
      {\ZJcomp{\ZP{P}; \widehat\sigma}{\ZSseq{S_1}{S_2}}
        {(\IS{S'_1;S'_2},\widehat\sigma'_2)}}
      \and
      \Tinfer{C-IfTrue} %
      {\ZJcomp{\ZP{P}; \widehat\sigma}{E}
        {(\IS{S'_0},\Tcon{\ZEval{\ZTbool}{\ZEtrue}})}
        \\ \ZJcomp{\ZP{P}; \widehat\sigma}{S_1}{(\IS{S'_1},\widehat\sigma'_1)}}
      {\ZJcomp{\ZP{P}; \widehat\sigma}{\ZSif{E}{S_1}{S_2}}
        {(\IS{S'_0;S'_1},\widehat\sigma'_1)}}
      \and \!\!\!\!\!
      \Tinfer{C-IfFalse} %
      {\ZJcomp{\ZP{P}; \widehat\sigma}{E}
        {(\IS{S'_0},\Tcon{\ZEval{\ZTbool}{\ZEfalse}})}
        \\ \ZJcomp{\ZP{P}; \widehat\sigma}{S_2}{(\IS{S'_2},\widehat\sigma'_2)}}
      {\ZJcomp{\ZP{P}; \widehat\sigma}{\ZSif{E}{S_1}{S_2}}
        {(\IS{S'_0;S'_2},\widehat\sigma'_2)}}
      \label{rule:trans-norm-ir-stmt-full:c-iffalse}
      \and
      \Tinfer{C-IfRt}
      {
        \ZJcomp{
          \ZP{P}; \widehat\sigma
        }{E}{
          ( \IS{S'_0}, \tcr{\Tref{\ITbool}{\IEvar{x}}} )
        }
        \\
        \forall i \in \{1,2\}.\
        \ZJcomp{\ZP{P}; \widehat\sigma}{S_i}{(\IS{S'_i},\widehat\sigma'_i)}
        \\
        \tcr{
          \Tjoin(
          \IEvar{x},
          (\IS{S'_1},\widehat\sigma'_1),
          (\IS{S'_2},\widehat\sigma'_2)
          )
        }
        =(\IS{S'_3},\widehat\sigma'_3)
      }
      {\ZJcomp{\ZP{P}; \widehat\sigma}{\ZSif{E}{S_1}{S_2}}
        {(\IS{S'_0;S'_3},\widehat\sigma'_3)}}
      \and
      \Tinfer[width=0.2\linewidth]{C-WhileFalse}
      {\ZJcomp{\ZP{P}; \widehat\sigma}{E}
        {(\IS{S'_0},\Tcon{\ZEval{\ZTbool}{\ZEfalse}})}}
      {\ZJcomp{\ZP{P}; \widehat\sigma}{\ZSwhile{E}{S}}
        {(\IS{S'_0},\widehat\sigma)}}
      \and
      \Tinfer{C-WhileTrue}
      {
        \ZJcomp{\ZP{P}; \widehat\sigma}{E}
        {(\IS{S'_0},\Tcon{\ZEval{\ZTbool}{\ZEtrue}})}
        \\
        {\begin{array}[b]{@{}c@{}}
            \ZJcomp{\ZP{P}; \widehat\sigma}{S}{(\IS{S'_1},\widehat\sigma'_1)}
            \\
            \ZJcomp{\ZP{P}; \widehat\sigma'_1}{\ZSwhile{E}{S}}
            {(\IS{S'_2},\widehat\sigma'_2)}
          \end{array}}
      }
      {\ZJcomp{\ZP{P}; \widehat\sigma}{\ZSwhile{E}{S}}
        {(\IS{S'_0;S'_1;S'_2},\widehat\sigma'_2)}}
    \end{mathpar}
    \egroup

    \textbf{Programs} (entry function)
    \bgroup\msmall
    \hfill\smash{\fbox{$\ZJcomp{\ZP{P}}{\ZP{f}}{\IP{P'}}$}}
    \begin{mathpar}
      \Tinfer{C-Prog}
      {
        {\begin{array}[b]{@{}c@{}}
            \Zhl{\ZP{P}
              =(\ZPdef{f_j}{\bar{x_j}:\bar{\comptype_j}}{\comptype_j'}
              {\ZSseq{S_j}{\ZSret{E_j}}})_{j \in [n]}}
            \\
            i \in [n]
            \land
            \Zhl{
              \bar{\comptype_i} = \tcr{(\bar{\primtype_i})^\top}
              \land
              \comptype_i' = \tcr{(\primtype_i')^\top}
            }
            \\
            \widehat\sigma
            \defas
            (
            \ZV{\bar{x_i}} \mapsto
            \tcr{ \Tref{\IT{\bar{\primtype_i}}}{\IEvar{\bar{y_i}}} }
            )
            \text{ for fresh } \IEvar{\bar{y_i}}
          \end{array}}
        \\
        {\begin{array}[b]{@{}c@{}}
            \ZJcomp{\ZP{P}; \widehat\sigma}{S_i}{(\IS{S'_1},\widehat\sigma')}
            \\
            \ZJcomp{\ZP{P}; \widehat\sigma'}{E_i}{(\IS{S'_2},\widehat v')}
            \\
            \tcr{\Tmat(\widehat v')}=(\IS{S'_3},\Tref{\IT{\primtype_i'}}{\IEvar{z'}})
          \end{array}}
      }
      {
        \ZJcomp{\ZP{P}}{\ZE{f_i}}{
          \IPdef{f_i}{\IEvar{\bar{y_i}}:\bar{\primtype_i}}{\primtype_i'}
          {S'_1;S'_2;S'_3;\ISret{\IEvar{z'}}}
        }
      }
    \end{mathpar}
    \egroup
  \end{figdefs}
  \caption{Translation from $\Znorm$ to $\IR$ for statements and programs.}
  \label{fig:trans-norm-ir-stmt-full}
\end{figure}

%% file: appendices/app-lang-res.tex
\section{Soundness of the Languages}
\label{sec:lang-res}

In this section, we present type preservation results for \Ziv*{}, \IR*{}, and \Czero*{}.
All results are formalized in Lean~4.35.0,
and the formalization is included in the supplementary material.
All file paths below are relative to the \codepath{lean/Ziv/} directory in the supplementary material.

\subsection{\texorpdfstring{$\Ziv*$}{Ziv}}
\label{sec:lang-res:ziv}

\begin{theorem}[Type Preservation; \Cref{thm:ziv-type-preserve-main}]
  \label{thm:ziv-type-preserve}
  Let $\ZP{P}$ be a well-typed \Ziv*{} program
  (i.e., $\ZJtype{}{P}{\ZJok}$).
  Suppose a function $\ZP{f}$ has type $\ZTfun{(\comptype_1, \ldots, \comptype_n)}{\comptype'}$ in $\ZP{P}$
  (i.e., $\ZJtype{\ZP{P}; \varepsilon}{f}{\ZTfun{(\comptype_1, \ldots, \comptype_n)}{\comptype'}}$).
  Then, for all values $v_1, \ldots, v_n, v'$,
  \[
    (\forall i \in [n].\; \ZJtype{\ZP{P}; \varepsilon}{v_i}{\comptype_i})
    \quad {\land} \quad
    \ZJeval{P}{f(v_1, \ldots, v_n)}{v'}
    \quad {\implies} \quad
    \ZJtype{\ZP{P}; \varepsilon}{v'}{\comptype'}.
  \]
\end{theorem}
\begin{proof}
  See \texttt{LangZiv.\allowbreak{}type\_\allowbreak{}preservation} in \codepath{Theorems.lean}.
  The proof is in \codepath{Proofs/LangZiv/TypePreservation.lean}.
\end{proof}

\subsection{\texorpdfstring{$\IR*$}{IR}}
\label{sec:lang-res:ir}

\begin{theorem}[Type Preservation]
  \label{thm:ir-type-preserve}
  Let $\IP{\IP{P} = \IPdef{f}{\IEvar{x_1}:\primtype_1, \ldots, \IEvar{x_n}:\primtype_n}{\primtype'}{S;\ISret{\IEvar{y}}}}$
  be a well-typed \IR*{} program
  (i.e., $\IJtype{}{P}{\IJok}$).
  Then, for all values $v_1, \ldots, v_n, v'$,
  \[
    (\forall i \in [n].\; \IJtype{\varepsilon}{v_i}{\primtype_i})
    \quad {\land} \quad
    \IJeval{P}{f(v_1, \ldots, v_n)}{v'}
    \quad {\implies} \quad
    \IJtype{\varepsilon}{v'}{\primtype'}.
  \]
\end{theorem}
\begin{proof}
  See \texttt{LangIR.\allowbreak{}type\_\allowbreak{}preservation} in \codepath{Theorems.lean}.
  The proof is in \codepath{Proofs/LangIR/TypePreservation.lean}.
\end{proof}

\subsection{\texorpdfstring{$\Czero*$}{C₀}}
\label{sec:lang-res:c0}

\begin{theorem}[Type Preservation]
  \label{thm:c0-type-preserve}
  Let $\CP{\CP{P} = \CPdef{f}{\CEvar{x_1}:\primtype_1, \ldots, \CEvar{x_n}:\primtype_n}{\primtype'}{S;\CSret{\CEvar{y}}}}$
  be a well-typed \Czero*{} program
  (i.e., $\CJtype{}{P}{\CJok}$).
  Then, for all values $v_1, \ldots, v_n, v'$,
  \[
    (\forall i \in [n].\; \CJtype{\varepsilon}{v_i}{\primtype_i})
    \quad {\land} \quad
    \CJeval{P}{f(v_1, \ldots, v_n)}{v'}
    \quad {\implies} \quad
    \CJtype{\varepsilon}{v'}{\primtype'}.
  \]
\end{theorem}
\begin{proof}
  See \texttt{LangC0.\allowbreak{}type\_\allowbreak{}preservation} in \codepath{Theorems.lean}.
  The proof is in \codepath{Proofs/LangC0/TypePreservation.lean}.
\end{proof}

%% file: appendices/app-comp-res.tex
\section{Soundness of the Compiler}
\label{sec:comp-res}

In this section, we present soundness results for each compiler step and for the entire compiler.
All results are formalized in Lean~4.35.0,
and the formalization is included in the supplementary material.
All file paths below are relative to the \codepath{lean/Ziv/} directory in the supplementary material.

\subsection{Normalization (\texorpdfstring{$\Ziv*{}\to\Znorm*{}$}{Ziv → ZivNorm})}
\label{sec:comp-res:norm}

\begin{theorem}[Type Preservation]
  \label{thm:norm-type-preserve}
  Let \Zhl{$\ZP{P}$ be a well-typed \Ziv*{} program}
  (i.e., $\ZJtype{}{P}{\ZJok}$),
  and $\ZP{P'} = \Ztrans(\ZP{P})$ be the result of normalizing $\ZP{P}$. %
  Then, \Zhl{$\ZP{P'}$ is a well-typed \Znorm*{} program}:
  \begin{align*}
    \ZJtype{}{P'}{\ZJok}.
  \end{align*}
\end{theorem}
\begin{proof}
  See \texttt{CompNorm.\allowbreak{}type\_\allowbreak{}preservation} in \codepath{Theorems.lean}.
  The proof is in \codepath{Proofs/CompNorm/TypePreservation.lean}.
\end{proof}

\begin{theorem}[Semantic Preservation]
  \label{thm:norm-semantic-preserve}
  Let \Zhl{$\ZP{P}$, $\ZP{P'}$} be as in \Cref{thm:norm-type-preserve},
  and $\ZP{f}$ be a function of type $\ZTfun{(\comptype_1, \ldots, \comptype_n)}{\comptype'}$ in $\ZP{P}$
  (i.e., $\ZJtype{\ZP{P}; \varepsilon}{f}{\ZTfun{(\comptype_1, \ldots, \comptype_n)}{\comptype'}}$).
  Then, for all values $v_1, \ldots, v_n, v'$ with each $v_i$ of type $\ZP{\comptype_i}$
  (i.e., $\ZJtype{\ZP{P}; \varepsilon}{v_i}{\comptype_i}$),
  \Zhl{$\ZP{f(v_1, \ldots, v_n)}$ evaluates to ${v'}$ in $\ZP{P}$}
  if and only if
  \Zhl{$\ZP{f(v_1, \ldots, v_n)}$} \Zhl{evaluates to ${v'}$ in $\ZP{P'}$}:
  \begin{align*}
    \ZJeval{P}{f(v_1, \ldots, v_n)}{v'}
    \quad {\iff} \quad
    \ZJeval{P'}{f(v_1, \ldots, v_n)}{v'}.
  \end{align*}
\end{theorem}
\begin{proof}
  See \texttt{CompNorm.\allowbreak{}semantic\_\allowbreak{}preservation} in \codepath{Theorems.lean}.
  The proof is in \codepath{Proofs/CompNorm/SemanticPreservation.lean}.
\end{proof}

\subsection{Tracing (\texorpdfstring{$\Znorm*{}\to\IR*{}$}{ZivNorm → ZivIR})}
\label{sec:comp-res:trace}

\begin{theorem}[Type Preservation]
  \label{thm:trace-type-preserve}
  Let \Zhl{$\ZP{P}$ be a well-typed \Znorm*{} program}
  (i.e., $\ZJtype{}{P}{\ZJok}$),
  and $\ZP{f}$ be a function defined in $\ZP{P}$.
  Suppose tracing $\ZP{f}$ produces $\IP{Q}$, %
  i.e., $\ZJcomp{\ZP{P}}{f}{\IP{Q}}$.
  Then, \Ihl{$\IP{Q}$ is a well-typed} \Ihl{\IR*{} program}:
  \begin{align*}
    \IJtype{}{Q}{\IJok}.
  \end{align*}
\end{theorem}
\begin{proof}
  See \texttt{CompTrace.\allowbreak{}type\_\allowbreak{}preservation} in \codepath{Theorems.lean}.
  The proof is in \codepath{Proofs/CompTrace/TypePreservation.lean}.
\end{proof}

\begin{theorem}[Semantic Preservation]
  \label{thm:trace-semantic-preserve}
  Let \Zhl{$\ZP{P}$, $\ZP{f}$}, and $\IP{Q}$ be as in \Cref{thm:trace-type-preserve},
  where $\ZP{f}$ has type $\ZTfun{(\comptype_1, \ldots, \comptype_n)}{\comptype'}$ in $\ZP{P}$
  (i.e., $\ZJtype{\ZP{P}; \varepsilon}{f}{\ZTfun{(\comptype_1, \ldots, \comptype_n)}{\comptype'}}$).
  Then, for all values $v_1, \ldots, v_n, v'$ with each $v_i$ of type $\ZP{\comptype_i}$
  (i.e., $\ZJtype{\ZP{P}; \varepsilon}{v_i}{\comptype_i}$),
  \Zhl{$\ZP{f(v_1, \ldots, v_n)}$ evaluates to $\ZP{v'}$ in $\ZP{P}$}
  if and only if
  \Ihl{$\IP{f(v_1, \ldots, v_n)}$ evaluates} \Ihl{to $\IP{v'}$ in $\IP{Q}$}:
  \begin{align*}
    \ZJeval{P}{f(v_1, \ldots, v_n)}{v'}
    \quad {\iff} \quad
    \IJeval{Q}{f(v_1, \ldots, v_n)}{v'}.
  \end{align*}
\end{theorem}
\begin{proof}
  See \texttt{CompTrace.\allowbreak{}semantic\_\allowbreak{}preservation} in \codepath{Theorems.lean}.
  The proof is in \codepath{Proofs/CompTrace/SemanticPreservation.lean}.
\end{proof}

\paragraph{Type Compatibility}
\label{sec:comp-res:trace:type-compat}

Consider a \Znorm*{} program $\ZP{P}$.
For a value and state $\widehat{v}$ and $\widehat{\sigma}$ in tracing,
and for a type and typing context $\ZP{\comptype}$ and $\ZP{\Gamma}$ in \Znorm*{},
we define the relation $\widehat{v} \models_P \ZP{\comptype}$ and $\widehat{\sigma} \models_P \ZP{\Gamma}$
to denote that $\widehat{v}$ and $\widehat{\sigma}$ are compatible with $\ZP{\comptype}$ and $\ZP{\Gamma}$ in $\ZP{P}$, respectively:
\begin{align*}
   & \widehat{v} \models_P \ZT{\comptype}
  \iff
  \begin{cases}
    \big({
      \widehat v=\Tcon{\ZP{v}}
      \land
      \ZJtype{\ZP{P}; \varepsilon}{\ZP{v}}{\comptype}
    }\big)
     & \text{if } \ZP{\comptype = \ZTfun{\bar{\comptype}'}{\comptype''}}
    \\
    \big({
      \widehat{v}=\Tcon{\ZP{v}}
      \land
      \ZJtype{\ZP{P}; \varepsilon}{\ZP{v}}{\comptype}
    }\big)
    \lor \big( \widehat{v}=\Tref{\primtype}{x} \land \ZP{\ell = \top} \big)
     & \text{if } \ZP{\comptype = \primtype^\ell}
    \\
    \begin{array}[t]{@{}l@{}}
      \big({
        \widehat v=\Tcon{\ZP{v}}
        \land
        \ZJtype{\ZP{P}; \varepsilon}{\ZP{v}}{\comptype}
      }\big)
      \\
      \; {} \lor {}
      \big({ %
        \widehat v = \Tcomp{{\Zerase(\ZP{\comptype})}}{\widehat v_1,\ldots,\widehat v_n}
        \land
        \Zfields(\ZT{\comptype})=(\ZT{\comptype_1,\ldots,\comptype_n})
        \land
        \forall i\in[n].\ \widehat v_i \models_P \ZT{\comptype_i}
      }\big), %
    \end{array}
    \!\!\!\!\!\!\!\!\!\!\!\!\!\!\!\!\!\!\!\!
    \!\!\!\!\!\!\!\!\!\!\!\!\!\!\!\!\!
     &
    \text{otherwise}
  \end{cases}
  \\
   &
  \widehat{\sigma} \models_P \ZCtx{\Gamma}
  \iff
  \big( \Zdom(\widehat\sigma) \supseteq \Zdom(\ZP{\Gamma}) \big)
  \land
  \big( \forall \ZE{x}\in\Zdom(\ZP{\Gamma}).\ \widehat\sigma(\ZE{x}) \models_P \ZCtx{\Gamma(x)} \big).
\end{align*}

\begin{theorem}[Type Compatibility]
  \label{thm:trace-type-compatible}
  Let \Zhl{$\ZP{P}$ be a well-typed \Znorm*{} program}
  (i.e., $\ZJtype{}{P}{\ZJok}$).
  Suppose \Zhl{an expression $\ZP{E}$ has type $\ZP{\comptype}$ under context $\ZP{\Gamma}$}
  (i.e., $\ZJtype{\ZP{P}; \Gamma}{E}{\comptype}$).
  For all tracing states $\widehat{\sigma}$ compatible with $\ZP{\Gamma}$
  (i.e., $\widehat{\sigma} \models_P \ZP{\Gamma}$),
  if {tracing} $\ZP{E}$ under $\widehat{\sigma}$ succeeds,
  \mbox{the resulting value is compatible with $\ZP{\comptype}$:}
  \begin{align*}
    \ZJcomp{\ZP{P}; \widehat{\sigma}}{E}{(\IS{S}, \widehat{v})}
    \quad {\implies} \quad
    \widehat{v} \models_P \ZT{\comptype}.
  \end{align*}
\end{theorem}
\begin{proof}
  See \texttt{CompTrace.\allowbreak{}type\_\allowbreak{}compatibility} in \codepath{Theorems.lean}.
  The proof is in \codepath{Proofs/CompTrace/TypeCompatibility.lean}.
\end{proof}

\begin{theorem}[$\bot \,{\Rightarrow}\, \Tcon{}$; \Cref{thm:trace-bot-static-main}]
  \label{thm:trace-bot-static}
  Let \Zhl{$\ZP{P}$ be a well-typed \Znorm*{} program}
  (i.e., $\ZJtype{}{P}{\ZJok}$).
  Suppose \Zhl{an expression $\ZP{E}$ has primitive type $\ZP{\primtype^\bot}$ under context $\ZP{\Gamma}$} %
  (i.e., $\ZJtype{\ZP{P}; \Gamma}{E}{\primtype^\bot}$).
  For all tracing states $\widehat{\sigma}$ compatible with $\ZP{\Gamma}$
  (i.e., $\widehat{\sigma} \models_P \ZP{\Gamma}$),
  if {tracing} $\ZP{E}$ under $\widehat{\sigma}$ succeeds, the resulting value is $\Tcon{}$:
  \begin{align*}
    \ZJcomp{\ZP{P}; \widehat{\sigma}}{E}{(\IS{S}, \widehat{v})}
    \quad {\implies} \quad
    \exists \ZP{v}.\;
    \widehat{v} = \Tcon{\ZP{v}}
    \land
    \ZJtype{\ZP{P}; \varepsilon}{v}{\primtype^\bot}.
  \end{align*}
\end{theorem}
\begin{proof}
  It follows from \Cref{thm:trace-type-compatible} and the definition of $\models_P$.
  See \texttt{CompTrace.\allowbreak{}ctval\_\allowbreak{}of\_\allowbreak{}bottom} in \codepath{Theorems.lean}.
  The proof is in \codepath{Proofs/CompTrace/TypeCompatibility.lean}.
\end{proof}

\subsection{Code Emission (\texorpdfstring{$\IR*{}\to\Czero*{}$}{ZivIR → C₀})}
\label{sec:comp-res:emit}

\begin{theorem}[Type Preservation]
  \label{thm:emit-type-preserve}
  Let \Ihl{$\IP{Q}$ be a well-typed \IR*{} program}
  (i.e., $\IJtype{}{Q}{\IJok}$),
  and $\CP{R} = \Ctrans(\IP{Q})$ be the result of code emission from $\IP{Q}$. %
  Then, \Chl{$\CP{R}$ is a well-typed \Czero*{} program}:
  \begin{align*}
    \CJtype{}{\CP{R}}{\CJok}.
  \end{align*}
\end{theorem}
\begin{proof}
  See \texttt{CompEmit.\allowbreak{}type\_\allowbreak{}preservation} in \codepath{Theorems.lean}.
  The proof is in \codepath{Proofs/CompEmit/TypePreservation.lean}.
\end{proof}

\begin{theorem}[Semantic Preservation]
  \label{thm:emission-semantic-preserve}
  Let $\IP{Q}$ and $\CP{R}$ be as in \Cref{thm:emit-type-preserve},
  where \Ihl{$Q = \Ikw{def}\ {f}($}
  \Ihl{${\IEvar{x_1}:\primtype_1, \ldots, \IEvar{x_n}:\primtype_n}) : {\primtype'}\ \{{S;\ISret{\IEvar{y}}}\}$}.
  Then, for all values $v_1, \ldots, v_n, v'$ with each $v_i$ of type $\IP{\primtype_i}$
  (i.e., $\IJtype{\varepsilon}{v_i}{\primtype_i}$),
  \Ihl{$\IP{f(v_1, \ldots, v_n)}$ evaluates to $\IEval{v'}$ in $\IP{Q}$}
  if and only if
  \Chl{$\CP{f(v_1, \ldots, v_n)}$ evaluates to $\CP{v'}$ in $\CP{R}$}:
  \begin{align*}
    \IJeval{Q}{f(v_1, \ldots, v_n)}{v'}
    \quad {\iff} \quad
    \CJeval{R}{f(v_1, \ldots, v_n)}{v'}.
  \end{align*}
\end{theorem}
\begin{proof}
  See \texttt{CompEmit.\allowbreak{}semantic\_\allowbreak{}preservation} in \codepath{Theorems.lean}.
  The proof is in \codepath{Proofs/CompEmit/SemanticPreservation.lean}.
\end{proof}

\subsection{Entire Compiler (\texorpdfstring{$\Ziv*{}\to\Czero*{}$}{Ziv → C₀})}
\label{sec:comp-res:all}

\begin{theorem}[Compiler Soundness; \Cref{thm:compiler-sound-main}]
  \label{thm:compiler-sound}
  Let {$\ZP{P}$ be a well-typed \Ziv*{} program}
  (i.e., $\ZJtype{}{P}{\ZJok}$),
  containing a function $\ZP{f}$ of type $\ZTfun{(\comptype_1, \ldots, \comptype_n)}{\comptype'}$
  (i.e., $\ZJtype{\ZP{P}; \varepsilon}{f}{\ZTfun{(\comptype_1, \ldots, \comptype_n)}{\comptype'}}$).
  Suppose compiling $\ZP{f}$ produces $\CP{R}$,
  i.e., $\ZJcomp{{\Ztrans(\ZP{P})}}{f}{\IP{Q}}$ and $\Ctrans(\IP{Q}) = \CP{R}$.
  Then, {$\CP{R}$ is a well-typed \Czero*{} program}
  (i.e., $\CJtype{}{R}{\CJok}$),
  and for all values $v_1, \ldots, v_n, v'$ with each $v_i$ of type $\ZP{\comptype_i}$
  (i.e., $\ZJtype{\ZP{P}; \varepsilon}{v_i}{\comptype_i}$),
  \begin{align*}
    \ZJeval{P}{f(v_1, \ldots, v_n)}{v'}
    \quad {\iff} \quad
    \CJeval{R}{f(v_1, \ldots, v_n)}{v'}.
  \end{align*}
\end{theorem}
\begin{proof}
  It follows immediately from
  \cref{thm:norm-type-preserve,thm:norm-semantic-preserve,thm:trace-type-preserve,thm:trace-semantic-preserve,thm:emit-type-preserve,thm:emission-semantic-preserve}.
  See \texttt{CompAll.\allowbreak{}soundness} in \codepath{Theorems.lean}.
  The proof is in \codepath{Proofs/CompAll/Soundness.lean}.
\end{proof}

%% file: paper.bbl

\begin{thebibliography}{66}


\ifx \showCODEN    \undefined \def \showCODEN     #1{\unskip}     \fi
\ifx \showISBNx    \undefined \def \showISBNx     #1{\unskip}     \fi
\ifx \showISBNxiii \undefined \def \showISBNxiii  #1{\unskip}     \fi
\ifx \showISSN     \undefined \def \showISSN      #1{\unskip}     \fi
\ifx \showLCCN     \undefined \def \showLCCN      #1{\unskip}     \fi
\ifx \shownote     \undefined \def \shownote      #1{#1}          \fi
\ifx \showarticletitle \undefined \def \showarticletitle #1{#1}   \fi
\ifx \showURL      \undefined \def \showURL       {\relax}        \fi
\providecommand\bibfield[2]{#2}
\providecommand\bibinfo[2]{#2}
\providecommand\natexlab[1]{#1}
\providecommand\showeprint[2][]{arXiv:#2}
\makeatletter
\@ifundefined{NAT@parse@date}{}{\let\NAT@parse@date@orig\NAT@parse@date}
\@ifundefined{NAT@parse@date}{}{\def\NAT@parse@date#1#2#3#4#5#6@@{\NAT@parse@date@orig#1#2#3#4#5#6@@\def\NAT@tempyear{0000}\def\NAT@tempexlab{{?}}\ifx\NAT@year\NAT@tempyear\ifx\NAT@exlab\NAT@tempexlab\def\NAT@date{[n.\,d.]}\else\edef\NAT@date{[n.\,d.]\NAT@exlab}\fi\fi}}
\makeatother

\bibitem[Agrawal et~al\mbox{.}(2019)]%
        {tfeager}
\bibfield{author}{\bibinfo{person}{Akshay Agrawal}, \bibinfo{person}{Akshay
  Modi}, \bibinfo{person}{Alexandre Passos}, \bibinfo{person}{Allen Lavoie},
  \bibinfo{person}{Ashish Agarwal}, \bibinfo{person}{Asim Shankar},
  \bibinfo{person}{Igor Ganichev}, \bibinfo{person}{Josh Levenberg},
  \bibinfo{person}{Mingsheng Hong}, \bibinfo{person}{Rajat Monga}, {and}
  \bibinfo{person}{Shanqing Cai}.} \bibinfo{year}{2019}\natexlab{}.
\newblock \showarticletitle{TensorFlow Eager: A Multi-Stage, Python-Embedded
  DSL for Machine Learning}. In \bibinfo{booktitle}{\emph{Proceedings of the
  2nd Conference on Machine Learning and Systems (SysML)}}.
\newblock
\urldef\tempurl%
\url{https://mlsys.org/Conferences/2019/doc/2019/88.pdf}
\showURL{%
\tempurl}


\bibitem[Beebe(2017)]%
        {fnhandbook}
\bibfield{author}{\bibinfo{person}{Nelson H.~F. Beebe}.}
  \bibinfo{year}{2017}\natexlab{}.
\newblock \bibinfo{booktitle}{\emph{The {Mathematical}-{Function} {Computation}
  {Handbook}: {Programming} {Using} the {MathCW} {Portable} {Software}
  {Library}}}.
\newblock \bibinfo{publisher}{Springer}.
\newblock
\showISBNx{978-3-319-64110-2}
\href{https://doi.org/10.1007/978-3-319-64110-2}{doi:\nolinkurl{10.1007/978-3-319-64110-2}}


\bibitem[Bezanson et~al\mbox{.}(2017)]%
        {julia}
\bibfield{author}{\bibinfo{person}{Jeff Bezanson}, \bibinfo{person}{Alan
  Edelman}, \bibinfo{person}{Stefan Karpinski}, {and} \bibinfo{person}{Viral~B.
  Shah}.} \bibinfo{year}{2017}\natexlab{}.
\newblock \showarticletitle{Julia: {A} {Fresh} {Approach} to {Numerical}
  {Computing}}.
\newblock \bibinfo{journal}{\emph{SIAM Rev.}} \bibinfo{volume}{59},
  \bibinfo{number}{1} (\bibinfo{date}{Jan.} \bibinfo{year}{2017}),
  \bibinfo{pages}{65--98}.
\newblock
\showISSN{0036-1445, 1095-7200}
\href{https://doi.org/10.1137/141000671}{doi:\nolinkurl{10.1137/141000671}}


\bibitem[Boldo et~al\mbox{.}(2023)]%
        {fpsurvey}
\bibfield{author}{\bibinfo{person}{Sylvie Boldo},
  \bibinfo{person}{Claude-Pierre Jeannerod}, \bibinfo{person}{Guillaume
  Melquiond}, {and} \bibinfo{person}{Jean-Michel Muller}.}
  \bibinfo{year}{2023}\natexlab{}.
\newblock \showarticletitle{Floating-Point Arithmetic}.
\newblock \bibinfo{journal}{\emph{Acta Numerica}}  \bibinfo{volume}{32}
  (\bibinfo{year}{2023}), \bibinfo{pages}{203--290}.
\newblock
\href{https://doi.org/10.1017/S0962492922000101}{doi:\nolinkurl{10.1017/S0962492922000101}}


\bibitem[Boldo and Muller(2005)]%
        {2fma}
\bibfield{author}{\bibinfo{person}{Sylvie Boldo} {and}
  \bibinfo{person}{Jean{-}Michel Muller}.} \bibinfo{year}{2005}\natexlab{}.
\newblock \showarticletitle{Some Functions Computable with a Fused-MAC}. In
  \bibinfo{booktitle}{\emph{Proceedings of the 17th IEEE Symposium on Computer
  Arithmetic (ARITH)}}. \bibinfo{pages}{52--58}.
\newblock
\href{https://doi.org/10.1109/ARITH.2005.39}{doi:\nolinkurl{10.1109/ARITH.2005.39}}


\bibitem[Boldo and Muller(2011)]%
        {2fma-ext}
\bibfield{author}{\bibinfo{person}{Sylvie Boldo} {and}
  \bibinfo{person}{Jean-Michel Muller}.} \bibinfo{year}{2011}\natexlab{}.
\newblock \showarticletitle{Exact and Approximated Error of the FMA}.
\newblock \bibinfo{journal}{\emph{IEEE Trans. Comput.}} \bibinfo{volume}{60},
  \bibinfo{number}{2} (\bibinfo{year}{2011}), \bibinfo{pages}{157--164}.
\newblock
\href{https://doi.org/10.1109/TC.2010.139}{doi:\nolinkurl{10.1109/TC.2010.139}}


\bibitem[Briggs et~al\mbox{.}(2024)]%
        {megalibm}
\bibfield{author}{\bibinfo{person}{Ian Briggs}, \bibinfo{person}{Yash Lad},
  {and} \bibinfo{person}{Pavel Panchekha}.} \bibinfo{year}{2024}\natexlab{}.
\newblock \showarticletitle{Implementation and {Synthesis} of {Math} {Library}
  {Functions}}.
\newblock \bibinfo{journal}{\emph{Proceedings of the ACM on Programming
  Languages}} \bibinfo{volume}{8}, \bibinfo{number}{POPL}, Article
  \bibinfo{articleno}{32} (\bibinfo{date}{Jan.} \bibinfo{year}{2024}).
\newblock
\showISSN{2475-1421}
\href{https://doi.org/10.1145/3632874}{doi:\nolinkurl{10.1145/3632874}}


\bibitem[Brisebarre and Chevillard(2007)]%
        {fpminimax}
\bibfield{author}{\bibinfo{person}{Nicolas Brisebarre} {and}
  \bibinfo{person}{Sylvain Chevillard}.} \bibinfo{year}{2007}\natexlab{}.
\newblock \showarticletitle{Efficient Polynomial ${L^\infty}$-Approximations}.
  In \bibinfo{booktitle}{\emph{Proceedings of the 18th IEEE Symposium on
  Computer Arithmetic (ARITH)}}. \bibinfo{pages}{169--176}.
\newblock
\showISBNx{978-0-7695-2854-0}
\showISSN{1063-6889}
\href{https://doi.org/10.1109/ARITH.2007.17}{doi:\nolinkurl{10.1109/ARITH.2007.17}}


\bibitem[Brisebarre et~al\mbox{.}(2026)]%
        {crsurvey}
\bibfield{author}{\bibinfo{person}{Nicolas Brisebarre},
  \bibinfo{person}{Guillaume Hanrot}, \bibinfo{person}{Jean-Michel Muller},
  {and} \bibinfo{person}{Paul Zimmermann}.} \bibinfo{year}{2026}\natexlab{}.
\newblock \showarticletitle{Correctly {Rounded} {Evaluation} of a {Function}:
  {Why}, {How}, and at {What} {Cost}?}
\newblock \bibinfo{journal}{\emph{Comput. Surveys}} \bibinfo{volume}{58},
  \bibinfo{number}{1}, Article \bibinfo{articleno}{27} (\bibinfo{date}{Jan.}
  \bibinfo{year}{2026}).
\newblock
\showISSN{0360-0300, 1557-7341}
\href{https://doi.org/10.1145/3747840}{doi:\nolinkurl{10.1145/3747840}}


\bibitem[Brunie et~al\mbox{.}(2015)]%
        {metalibm-arith}
\bibfield{author}{\bibinfo{person}{Nicolas Brunie}, \bibinfo{person}{Florent de
  Dinechin}, \bibinfo{person}{Olga Kupriianova}, {and}
  \bibinfo{person}{Christoph Lauter}.} \bibinfo{year}{2015}\natexlab{}.
\newblock \showarticletitle{Code {Generators} for {Mathematical} {Functions}}.
  In \bibinfo{booktitle}{\emph{Proceedings of the 22nd IEEE Symposium on
  Computer Arithmetic (ARITH)}}. \bibinfo{pages}{66--73}.
\newblock
\showISBNx{978-1-4799-8664-4}
\href{https://doi.org/10.1109/ARITH.2015.22}{doi:\nolinkurl{10.1109/ARITH.2015.22}}


\bibitem[Chevillard et~al\mbox{.}(2010)]%
        {sollya}
\bibfield{author}{\bibinfo{person}{Sylvain Chevillard}, \bibinfo{person}{Mioara
  Joldes}, {and} \bibinfo{person}{Christoph Lauter}.}
  \bibinfo{year}{2010}\natexlab{}.
\newblock \showarticletitle{Sollya: An Environment for the Development of
  Numerical Codes}. In \bibinfo{booktitle}{\emph{Proceedings of the 3rd
  International Congress on Mathematical Software (ICMS)}}.
  \bibinfo{pages}{28--31}.
\newblock
\href{https://doi.org/10.1007/978-3-642-15582-6_5}{doi:\nolinkurl{10.1007/978-3-642-15582-6_5}}


\bibitem[{Cody, Jr.} and Waite(1980)]%
        {codywaite}
\bibfield{author}{\bibinfo{person}{William~J. {Cody, Jr.}} {and}
  \bibinfo{person}{William~M. Waite}.} \bibinfo{year}{1980}\natexlab{}.
\newblock \bibinfo{booktitle}{\emph{Software Manual for the Elementary
  Functions}}.
\newblock \bibinfo{publisher}{Prentice-Hall}.
\newblock
\showISBNx{978-0-13-822064-8}


\bibitem[Daramy-Loirat et~al\mbox{.}(2006)]%
        {crlibm}
\bibfield{author}{\bibinfo{person}{Catherine Daramy-Loirat},
  \bibinfo{person}{David Defour}, \bibinfo{person}{Florent de Dinechin},
  \bibinfo{person}{Matthieu Gallet}, \bibinfo{person}{Nicolas Gast},
  \bibinfo{person}{Christoph Lauter}, {and} \bibinfo{person}{Jean-Michel
  Muller}.} \bibinfo{year}{2006}\natexlab{}.
\newblock \bibinfo{booktitle}{\emph{{CR}-{LIBM}: {A} Library of Correctly
  Rounded Elementary Functions in Double-Precision}}.
\newblock \bibinfo{type}{Research {Report}}. \bibinfo{institution}{LIP}.
\newblock
\urldef\tempurl%
\url{https://ens-lyon.hal.science/ensl-01529804}
\showURL{%
\tempurl}


\bibitem[Dekker(1971)]%
        {doubleword}
\bibfield{author}{\bibinfo{person}{Theodorus~J. Dekker}.}
  \bibinfo{year}{1971}\natexlab{}.
\newblock \showarticletitle{A Floating-Point Technique for Extending the
  Available Precision}.
\newblock \bibinfo{journal}{\emph{Numer. Math.}} \bibinfo{volume}{18},
  \bibinfo{number}{3} (\bibinfo{date}{June} \bibinfo{year}{1971}),
  \bibinfo{pages}{224--242}.
\newblock
\showISSN{0029-599X, 0945-3245}
\href{https://doi.org/10.1007/BF01397083}{doi:\nolinkurl{10.1007/BF01397083}}


\bibitem[DeVito et~al\mbox{.}(2013)]%
        {terra}
\bibfield{author}{\bibinfo{person}{Zachary DeVito}, \bibinfo{person}{James
  Hegarty}, \bibinfo{person}{Alex Aiken}, \bibinfo{person}{Pat Hanrahan}, {and}
  \bibinfo{person}{Jan Vitek}.} \bibinfo{year}{2013}\natexlab{}.
\newblock \showarticletitle{Terra: A Multi-Stage Language for High-Performance
  Computing}. In \bibinfo{booktitle}{\emph{Proceedings of the 34th ACM SIGPLAN
  Conference on Programming Language Design and Implementation (PLDI)}}.
  \bibinfo{pages}{105--116}.
\newblock
\href{https://doi.org/10.1145/2491956.2462166}{doi:\nolinkurl{10.1145/2491956.2462166}}


\bibitem[Estrin(1960)]%
        {estrin}
\bibfield{author}{\bibinfo{person}{Gerald Estrin}.}
  \bibinfo{year}{1960}\natexlab{}.
\newblock \showarticletitle{Organization of Computer Systems: The Fixed Plus
  Variable Structure Computer}. In \bibinfo{booktitle}{\emph{Proceedings of the
  1960 Western Joint IRE-AIEE-ACM Computer Conference}}.
  \bibinfo{pages}{33--40}.
\newblock
\href{https://doi.org/10.1145/1460361.1460365}{doi:\nolinkurl{10.1145/1460361.1460365}}


\bibitem[Faissole et~al\mbox{.}(2024)]%
        {verify-itp24}
\bibfield{author}{\bibinfo{person}{Florian Faissole}, \bibinfo{person}{Paul
  Geneau~de Lamarli{\`e}re}, {and} \bibinfo{person}{Guillaume Melquiond}.}
  \bibinfo{year}{2024}\natexlab{}.
\newblock \showarticletitle{End-to-End Formal Verification of a Fast and
  Accurate Floating-Point Approximation}. In
  \bibinfo{booktitle}{\emph{Proceedings of the 15th International Conference on
  Interactive Theorem Proving (ITP)}}. \bibinfo{pages}{14:1--14:18}.
\newblock
\href{https://doi.org/10.4230/LIPIcs.ITP.2024.14}{doi:\nolinkurl{10.4230/LIPIcs.ITP.2024.14}}


\bibitem[Fousse et~al\mbox{.}(2007)]%
        {mpfr}
\bibfield{author}{\bibinfo{person}{Laurent Fousse}, \bibinfo{person}{Guillaume
  Hanrot}, \bibinfo{person}{Vincent Lef{\`e}vre}, \bibinfo{person}{Patrick
  P{\'e}lissier}, {and} \bibinfo{person}{Paul Zimmermann}.}
  \bibinfo{year}{2007}\natexlab{}.
\newblock \showarticletitle{{MPFR}: {A} Multiple-Precision Binary
  Floating-Point Library with Correct Rounding}.
\newblock \bibinfo{journal}{\emph{ACM Trans. Math. Software}}
  \bibinfo{volume}{33}, \bibinfo{number}{2}, Article \bibinfo{articleno}{13}
  (\bibinfo{date}{June} \bibinfo{year}{2007}).
\newblock
\showISSN{0098-3500, 1557-7295}
\href{https://doi.org/10.1145/1236463.1236468}{doi:\nolinkurl{10.1145/1236463.1236468}}


\bibitem[Frostig et~al\mbox{.}(2018)]%
        {jax}
\bibfield{author}{\bibinfo{person}{Roy Frostig}, \bibinfo{person}{Matthew~James
  Johnson}, {and} \bibinfo{person}{Chris Leary}.}
  \bibinfo{year}{2018}\natexlab{}.
\newblock \showarticletitle{{Compiling Machine Learning Programs via High-Level
  Tracing}}. In \bibinfo{booktitle}{\emph{Proceedings of the 1st Conference on
  Machine Learning and Systems (SysML)}}.
\newblock
\urldef\tempurl%
\url{https://mlsys.org/Conferences/doc/2018/146.pdf}
\showURL{%
\tempurl}


\bibitem[Galassi et~al\mbox{.}(2009)]%
        {gsl}
\bibfield{author}{\bibinfo{person}{Mark Galassi}, \bibinfo{person}{Jim Davies},
  \bibinfo{person}{James Theiler}, \bibinfo{person}{Brian Gough},
  \bibinfo{person}{Gerard Jungman}, \bibinfo{person}{Patrick Alken},
  \bibinfo{person}{Michael Booth}, {and} \bibinfo{person}{Fabrice Rossi}.}
  \bibinfo{year}{2009}\natexlab{}.
\newblock \bibinfo{booktitle}{\emph{GNU Scientific Library Reference Manual}
  (\bibinfo{edition}{third} ed.)}.
\newblock \bibinfo{publisher}{Network Theory Ltd.}
\newblock
\urldef\tempurl%
\url{https://www.gnu.org/software/gsl/}
\showURL{%
\tempurl}


\bibitem[Geneau~de Lamarli{\`{e}}re et~al\mbox{.}(2023)]%
        {verify-arith23}
\bibfield{author}{\bibinfo{person}{Paul Geneau~de Lamarli{\`{e}}re},
  \bibinfo{person}{Guillaume Melquiond}, {and} \bibinfo{person}{Florian
  Faissole}.} \bibinfo{year}{2023}\natexlab{}.
\newblock \showarticletitle{Slimmer Formal Proofs for Mathematical Libraries}.
  In \bibinfo{booktitle}{\emph{Proceedings of the 30th IEEE Symposium on
  Computer Arithmetic (ARITH)}}. \bibinfo{pages}{32--35}.
\newblock
\href{https://doi.org/10.1109/ARITH58626.2023.00026}{doi:\nolinkurl{10.1109/ARITH58626.2023.00026}}


\bibitem[Goodfellow et~al\mbox{.}(2016)]%
        {softmax}
\bibfield{author}{\bibinfo{person}{Ian Goodfellow}, \bibinfo{person}{Yoshua
  Bengio}, {and} \bibinfo{person}{Aaron Courville}.}
  \bibinfo{year}{2016}\natexlab{}.
\newblock \bibinfo{booktitle}{\emph{Deep Learning}}.
\newblock \bibinfo{publisher}{MIT Press}.
\newblock
\urldef\tempurl%
\url{https://www.deeplearningbook.org}
\showURL{%
\tempurl}


\bibitem[Hendrycks and Gimpel(2023)]%
        {gelu}
\bibfield{author}{\bibinfo{person}{Dan Hendrycks} {and} \bibinfo{person}{Kevin
  Gimpel}.} \bibinfo{year}{2023}\natexlab{}.
\newblock \bibinfo{title}{Gaussian {Error} {Linear} {Units} ({GELUs})}.
\newblock
\urldef\tempurl%
\url{https://arxiv.org/abs/1606.08415v5}
\showURL{%
\tempurl}
\newblock
\shownote{arXiv:1606.08415v5 [cs.LG]}.


\bibitem[Hubrecht et~al\mbox{.}(2024)]%
        {powcr}
\bibfield{author}{\bibinfo{person}{Tom Hubrecht},
  \bibinfo{person}{Claude-Pierre Jeannerod}, \bibinfo{person}{Paul Zimmermann},
  \bibinfo{person}{Laurence Rideau}, {and} \bibinfo{person}{Laurent
  Th{\'e}ry}.} \bibinfo{year}{2024}\natexlab{}.
\newblock \bibinfo{title}{Towards a Correctly-Rounded and Fast Power Function
  in {binary64} Arithmetic}.
\newblock
\urldef\tempurl%
\url{https://inria.hal.science/hal-04159652v2}
\showURL{%
\tempurl}
\newblock
\shownote{Extended version of the {ARITH} 2023 paper}.


\bibitem[{IBM}(2001)]%
        {mathlib}
\bibfield{author}{\bibinfo{person}{{IBM}}.} \bibinfo{year}{2001}\natexlab{}.
\newblock \bibinfo{booktitle}{\emph{{IBM Accurate Portable Mathlib}}}.
\newblock
\urldef\tempurl%
\url{https://github.com/dreal-deps/mathlib}
\showURL{%
\tempurl}
\newblock
\shownote{Unofficial copy of the original library}.


\bibitem[{IEEE}(2019)]%
        {ieee-754-2019}
\bibfield{author}{\bibinfo{person}{{IEEE}}.} \bibinfo{year}{2019}\natexlab{}.
\newblock \bibinfo{booktitle}{\emph{{IEEE} Std 754-2019: {IEEE} Standard for
  Floating-Point Arithmetic}}.
\newblock
\href{https://doi.org/10.1109/IEEESTD.2019.8766229}{doi:\nolinkurl{10.1109/IEEESTD.2019.8766229}}


\bibitem[{IEEE}(2026)]%
        {ieee-754-2029}
\bibfield{author}{\bibinfo{person}{{IEEE}}.} \bibinfo{year}{2026}\natexlab{}.
\newblock \bibinfo{booktitle}{\emph{{P754/D3.23}: Draft {IEEE} Standard for
  Floating-Point Arithmetic (November 2026)}}.
\newblock
\urldef\tempurl%
\url{https://listserv.ieee.org/cgi-bin/wa?A2=STDS-754;802b7116.26&S=}
\showURL{%
\tempurl}


\bibitem[{ISO/IEC}(2024)]%
        {c-std-2024}
\bibfield{author}{\bibinfo{person}{{ISO/IEC}}.}
  \bibinfo{year}{2024}\natexlab{}.
\newblock \bibinfo{booktitle}{\emph{{ISO/IEC} 9899:2024: Information Technology
  --- Programming Languages --- {C}}}.
\newblock
\urldef\tempurl%
\url{https://www.iso.org/standard/82075.html}
\showURL{%
\tempurl}


\bibitem[Jeannerod et~al\mbox{.}(2026)]%
        {fast2fma-proof-tight}
\bibfield{author}{\bibinfo{person}{Claude-Pierre Jeannerod},
  \bibinfo{person}{Mioara Joldes}, \bibinfo{person}{Nicolas Louvet}, {and}
  \bibinfo{person}{Jean-Michel Muller}.} \bibinfo{year}{2026}\natexlab{}.
\newblock \showarticletitle{Extended-Precision FMA Under Parameterized
  Double-Word Overlap: Tight Error Bounds and Examples}. In
  \bibinfo{booktitle}{\emph{Proceedings of the 33rd IEEE Symposium on Computer
  Arithmetic (ARITH)}}. \bibinfo{pages}{17--24}.
\newblock
\href{https://doi.org/10.1109/ARITH69512.2026.00012}{doi:\nolinkurl{10.1109/ARITH69512.2026.00012}}


\bibitem[Johansson(2017)]%
        {arb}
\bibfield{author}{\bibinfo{person}{Fredrik Johansson}.}
  \bibinfo{year}{2017}\natexlab{}.
\newblock \showarticletitle{Arb: Efficient Arbitrary-Precision Midpoint-Radius
  Interval Arithmetic}.
\newblock \bibinfo{journal}{\emph{IEEE Trans. Comput.}} \bibinfo{volume}{66},
  \bibinfo{number}{8} (\bibinfo{year}{2017}), \bibinfo{pages}{1281--1292}.
\newblock
\href{https://doi.org/10.1109/TC.2017.2690633}{doi:\nolinkurl{10.1109/TC.2017.2690633}}


\bibitem[Jones et~al\mbox{.}(1993)]%
        {partialeval}
\bibfield{author}{\bibinfo{person}{Neil~D. Jones}, \bibinfo{person}{Carsten~K.
  Gomard}, {and} \bibinfo{person}{Peter Sestoft}.}
  \bibinfo{year}{1993}\natexlab{}.
\newblock \bibinfo{booktitle}{\emph{Partial Evaluation and Automatic Program
  Generation}}.
\newblock \bibinfo{publisher}{Prentice Hall}.
\newblock
\showISBNx{978-0-13-020249-9}
\urldef\tempurl%
\url{https://raspi.itu.dk/people/sestoft/pebook/pebook.html}
\showURL{%
\tempurl}


\bibitem[Kahan(1997)]%
        {2mul-algo}
\bibfield{author}{\bibinfo{person}{William Kahan}.}
  \bibinfo{year}{1997}\natexlab{}.
\newblock \bibinfo{booktitle}{\emph{Lecture Notes on the Status of IEEE
  Standard 754 for Binary Floating-Point Arithmetic}}.
\newblock
\urldef\tempurl%
\url{https://people.eecs.berkeley.edu/~wkahan/ieee754status/IEEE754.PDF}
\showURL{%
\tempurl}
\newblock
\shownote{Accessed: 2026-11-21}.


\bibitem[Knuth(1997)]%
        {taocp}
\bibfield{author}{\bibinfo{person}{Donald~E. Knuth}.}
  \bibinfo{year}{1997}\natexlab{}.
\newblock \bibinfo{booktitle}{\emph{The Art of Computer Programming}
  (\bibinfo{edition}{third} ed.)}. Vol.~\bibinfo{volume}{2: Seminumerical
  Algorithms}.
\newblock \bibinfo{publisher}{Addison-Wesley}.
\newblock
\urldef\tempurl%
\url{https://cs.stanford.edu/~knuth/taocp.html}
\showURL{%
\tempurl}


\bibitem[Koizumi et~al\mbox{.}(2023)]%
        {fast2fma-algo}
\bibfield{author}{\bibinfo{person}{Toru Koizumi}, \bibinfo{person}{Tomoaki
  Tsumura}, \bibinfo{person}{Hidetsugu Irie}, {and} \bibinfo{person}{Shuichi
  Sakai}.} \bibinfo{year}{2023}\natexlab{}.
\newblock \bibinfo{booktitle}{\emph{A Fast Implementation of Perfect-Accuracy
  Double-Precision Exponential Function}}.
\newblock \bibinfo{type}{IPSJ SIG Technical Report} 2023-HPC-192, No.~21.
  \bibinfo{institution}{Information Processing Society of Japan}.
\newblock
\urldef\tempurl%
\url{https://ipsj.ixsq.nii.ac.jp/records/231099}
\showURL{%
\tempurl}
\newblock
\shownote{In Japanese}.


\bibitem[Kupriianova and Lauter(2014)]%
        {metalibm-icms}
\bibfield{author}{\bibinfo{person}{Olga Kupriianova} {and}
  \bibinfo{person}{Christoph Lauter}.} \bibinfo{year}{2014}\natexlab{}.
\newblock \showarticletitle{Metalibm: {A} {Mathematical} {Functions} {Code}
  {Generator}}. In \bibinfo{booktitle}{\emph{Proceedings of the 4th
  International Congress on Mathematical Software (ICMS)}}.
  \bibinfo{pages}{713--717}.
\newblock
\showISBNx{978-3-662-44198-5 978-3-662-44199-2}
\href{https://doi.org/10.1007/978-3-662-44199-2_106}{doi:\nolinkurl{10.1007/978-3-662-44199-2_106}}


\bibitem[Lee et~al\mbox{.}(2016)]%
        {verify-pldi16}
\bibfield{author}{\bibinfo{person}{Wonyeol Lee}, \bibinfo{person}{Rahul
  Sharma}, {and} \bibinfo{person}{Alex Aiken}.}
  \bibinfo{year}{2016}\natexlab{}.
\newblock \showarticletitle{Verifying Bit-Manipulations of Floating-Point}. In
  \bibinfo{booktitle}{\emph{Proceedings of the 37th ACM SIGPLAN Conference on
  Programming Language Design and Implementation (PLDI)}}.
  \bibinfo{pages}{70--84}.
\newblock
\showISBNx{978-1-4503-4261-2}
\href{https://doi.org/10.1145/2908080.2908107}{doi:\nolinkurl{10.1145/2908080.2908107}}


\bibitem[Lee et~al\mbox{.}(2018)]%
        {verify-popl18}
\bibfield{author}{\bibinfo{person}{Wonyeol Lee}, \bibinfo{person}{Rahul
  Sharma}, {and} \bibinfo{person}{Alex Aiken}.}
  \bibinfo{year}{2018}\natexlab{}.
\newblock \showarticletitle{On Automatically Proving the Correctness of math.h
  Implementations}.
\newblock \bibinfo{journal}{\emph{Proceedings of the ACM on Programming
  Languages}} \bibinfo{volume}{2}, \bibinfo{number}{POPL}, Article
  \bibinfo{articleno}{47} (\bibinfo{date}{Jan.} \bibinfo{year}{2018}).
\newblock
\href{https://doi.org/10.1145/3158135}{doi:\nolinkurl{10.1145/3158135}}


\bibitem[Lef{\`e}vre et~al\mbox{.}(2026)]%
        {worsttrig}
\bibfield{author}{\bibinfo{person}{Vincent Lef{\`e}vre}, \bibinfo{person}{Tue
  Ly}, {and} \bibinfo{person}{Paul Zimmermann}.}
  \bibinfo{year}{2026}\natexlab{}.
\newblock \showarticletitle{Computing {Hard}-to-{Round} {Cases} of {sin},
  {cos}, {tan} in {Double} {Precision}}. In
  \bibinfo{booktitle}{\emph{Proceedings of the 33rd IEEE Symposium on Computer
  Arithmetic (ARITH)}}. \bibinfo{pages}{113--120}.
\newblock
\showISBNx{979-8-3315-8448-1}
\href{https://doi.org/10.1109/ARITH69512.2026.00024}{doi:\nolinkurl{10.1109/ARITH69512.2026.00024}}


\bibitem[Lef{\`e}vre and Muller(2001)]%
        {worstelem}
\bibfield{author}{\bibinfo{person}{Vincent Lef{\`e}vre} {and}
  \bibinfo{person}{Jean{-}Michel Muller}.} \bibinfo{year}{2001}\natexlab{}.
\newblock \showarticletitle{Worst Cases for Correct Rounding of the Elementary
  Functions in Double Precision}. In \bibinfo{booktitle}{\emph{Proceedings of
  the 15th IEEE Symposium on Computer Arithmetic (ARITH)}}.
  \bibinfo{pages}{111--118}.
\newblock
\showISBNx{978-0-7695-1150-4}
\href{https://doi.org/10.1109/ARITH.2001.930110}{doi:\nolinkurl{10.1109/ARITH.2001.930110}}


\bibitem[Lilis and Savidis(2019)]%
        {metasurvey}
\bibfield{author}{\bibinfo{person}{Yannis Lilis} {and} \bibinfo{person}{Anthony
  Savidis}.} \bibinfo{year}{2019}\natexlab{}.
\newblock \showarticletitle{A {Survey} of {Metaprogramming} {Languages}}.
\newblock \bibinfo{journal}{\emph{Comput. Surveys}} \bibinfo{volume}{52},
  \bibinfo{number}{6}, Article \bibinfo{articleno}{113} (\bibinfo{date}{Oct.}
  \bibinfo{year}{2019}).
\newblock
\showISSN{0360-0300, 1557-7341}
\href{https://doi.org/10.1145/3354584}{doi:\nolinkurl{10.1145/3354584}}


\bibitem[Lim and Nagarakatte(2021)]%
        {rlibm}
\bibfield{author}{\bibinfo{person}{Jay~P. Lim} {and} \bibinfo{person}{Santosh
  Nagarakatte}.} \bibinfo{year}{2021}\natexlab{}.
\newblock \showarticletitle{High Performance Correctly Rounded Math Libraries
  for 32-bit Floating Point Representations}. In
  \bibinfo{booktitle}{\emph{Proceedings of the 42nd ACM SIGPLAN Conference on
  Programming Language Design and Implementation (PLDI)}}.
  \bibinfo{pages}{359--374}.
\newblock
\showISBNx{978-1-4503-8391-2}
\href{https://doi.org/10.1145/3453483.3454049}{doi:\nolinkurl{10.1145/3453483.3454049}}


\bibitem[{LLVM Developer Group}(2026)]%
        {llvmlibc}
\bibfield{author}{\bibinfo{person}{{LLVM Developer Group}}.}
  \bibinfo{year}{2026}\natexlab{}.
\newblock \bibinfo{booktitle}{\emph{The {LLVM} {C} Library: Math Functions}}.
\newblock
\urldef\tempurl%
\url{https://libc.llvm.org/math/index.html}
\showURL{%
\tempurl}
\newblock
\shownote{Accessed: 2026-11-21}.


\bibitem[Loosemore et~al\mbox{.}(2026)]%
        {c-libm}
\bibfield{author}{\bibinfo{person}{Sandra Loosemore},
  \bibinfo{person}{Richard~M. Stallman}, \bibinfo{person}{Roland McGrath},
  \bibinfo{person}{Andrew Oram}, {and} \bibinfo{person}{Ulrich Drepper}.}
  \bibinfo{year}{2026}\natexlab{}.
\newblock \bibinfo{booktitle}{\emph{The {GNU C} Library Reference Manual:
  Mathematics}}.
\newblock Free Software Foundation.
\newblock
\urldef\tempurl%
\url{https://sourceware.org/glibc/manual/2.44/html_node/Mathematics.html}
\showURL{%
\tempurl}
\newblock
\shownote{For version 2.44}.


\bibitem[Markstein(2000)]%
        {ia64}
\bibfield{author}{\bibinfo{person}{Peter Markstein}.}
  \bibinfo{year}{2000}\natexlab{}.
\newblock \bibinfo{booktitle}{\emph{{IA}-64 and Elementary Functions: Speed and
  Precision}}.
\newblock \bibinfo{publisher}{Prentice Hall}.
\newblock
\showISBNx{978-0-13-018348-4}


\bibitem[M{\o}ller(1965)]%
        {2sum}
\bibfield{author}{\bibinfo{person}{Ole M{\o}ller}.}
  \bibinfo{year}{1965}\natexlab{}.
\newblock \showarticletitle{Quasi Double-Precision in Floating Point Addition}.
\newblock \bibinfo{journal}{\emph{BIT}} \bibinfo{volume}{5},
  \bibinfo{number}{1} (\bibinfo{date}{March} \bibinfo{year}{1965}),
  \bibinfo{pages}{37--50}.
\newblock
\showISSN{0006-3835, 1572-9125}
\href{https://doi.org/10.1007/BF01975722}{doi:\nolinkurl{10.1007/BF01975722}}


\bibitem[Muller(2016)]%
        {elementaryfns}
\bibfield{author}{\bibinfo{person}{Jean-Michel Muller}.}
  \bibinfo{year}{2016}\natexlab{}.
\newblock \bibinfo{booktitle}{\emph{Elementary {Functions}: {Algorithms} and
  {Implementation}} (\bibinfo{edition}{third} ed.)}.
\newblock \bibinfo{publisher}{Birkh{\"a}user}.
\newblock
\showISBNx{978-1-4899-7983-4}
\href{https://doi.org/10.1007/978-1-4899-7983-4}{doi:\nolinkurl{10.1007/978-1-4899-7983-4}}


\bibitem[Muller et~al\mbox{.}(2018)]%
        {fphandbook}
\bibfield{author}{\bibinfo{person}{Jean-Michel Muller},
  \bibinfo{person}{Nicolas Brunie}, \bibinfo{person}{Florent de Dinechin},
  \bibinfo{person}{Claude-Pierre Jeannerod}, \bibinfo{person}{Mioara Joldes},
  \bibinfo{person}{Vincent Lef{\`e}vre}, \bibinfo{person}{Guillaume Melquiond},
  \bibinfo{person}{Nathalie Revol}, {and} \bibinfo{person}{Serge Torres}.}
  \bibinfo{year}{2018}\natexlab{}.
\newblock \bibinfo{booktitle}{\emph{Handbook of Floating-Point Arithmetic}
  (\bibinfo{edition}{second} ed.)}.
\newblock \bibinfo{publisher}{Birkh{\"a}user}.
\newblock
\showISBNx{978-3-319-76526-6}
\href{https://doi.org/10.1007/978-3-319-76526-6}{doi:\nolinkurl{10.1007/978-3-319-76526-6}}


\bibitem[Nielson and Nielson(1988)]%
        {autobta}
\bibfield{author}{\bibinfo{person}{Hanne~R. Nielson} {and}
  \bibinfo{person}{Flemming Nielson}.} \bibinfo{year}{1988}\natexlab{}.
\newblock \showarticletitle{Automatic Binding Time Analysis for a Typed
  $\lambda$-Calculus}. In \bibinfo{booktitle}{\emph{Proceedings of the 15th ACM
  SIGPLAN-SIGACT Symposium on Principles of Programming Languages (POPL)}}.
  \bibinfo{pages}{98--106}.
\newblock
\showISBNx{0897912527}
\href{https://doi.org/10.1145/73560.73569}{doi:\nolinkurl{10.1145/73560.73569}}


\bibitem[Nievergelt(2003)]%
        {2mul-proof}
\bibfield{author}{\bibinfo{person}{Yves Nievergelt}.}
  \bibinfo{year}{2003}\natexlab{}.
\newblock \showarticletitle{Scalar Fused Multiply-Add Instructions Produce
  Floating-Point Matrix Arithmetic Provably Accurate to the Penultimate Digit}.
\newblock \bibinfo{journal}{\emph{ACM Trans. Math. Software}}
  \bibinfo{volume}{29}, \bibinfo{number}{1} (\bibinfo{date}{March}
  \bibinfo{year}{2003}), \bibinfo{pages}{27--48}.
\newblock
\showISSN{0098-3500}
\href{https://doi.org/10.1145/641876.641878}{doi:\nolinkurl{10.1145/641876.641878}}


\bibitem[{Oracle}(2026)]%
        {java-libm}
\bibfield{author}{\bibinfo{person}{{Oracle}}.} \bibinfo{year}{2026}\natexlab{}.
\newblock \bibinfo{booktitle}{\emph{{Java SE} {API} Specification: Class
  {Math}}}.
\newblock
\urldef\tempurl%
\url{https://docs.oracle.com/en/java/javase/27/docs/api/java.base/java/lang/Math.html}
\showURL{%
\tempurl}
\newblock
\shownote{For version 27}.


\bibitem[Paszke et~al\mbox{.}(2019)]%
        {pytorch}
\bibfield{author}{\bibinfo{person}{Adam Paszke}, \bibinfo{person}{Sam Gross},
  \bibinfo{person}{Francisco Massa}, \bibinfo{person}{Adam Lerer},
  \bibinfo{person}{James Bradbury}, \bibinfo{person}{Gregory Chanan},
  \bibinfo{person}{Trevor Killeen}, \bibinfo{person}{Zeming Lin},
  \bibinfo{person}{Natalia Gimelshein}, \bibinfo{person}{Luca Antiga},
  \bibinfo{person}{Alban Desmaison}, \bibinfo{person}{Andreas K{\"o}pf},
  \bibinfo{person}{Edward Yang}, \bibinfo{person}{Zachary DeVito},
  \bibinfo{person}{Martin Raison}, \bibinfo{person}{Alykhan Tejani},
  \bibinfo{person}{Sasank Chilamkurthy}, \bibinfo{person}{Benoit Steiner},
  \bibinfo{person}{Lu Fang}, \bibinfo{person}{Junjie Bai}, {and}
  \bibinfo{person}{Soumith Chintala}.} \bibinfo{year}{2019}\natexlab{}.
\newblock \showarticletitle{{PyTorch}: {An} {Imperative} {Style},
  {High}-{Performance} {Deep} {Learning} {Library}}. In
  \bibinfo{booktitle}{\emph{Proceedings of the 33rd Annual Conference on Neural
  Information Processing Systems (NeurIPS)}}.
\newblock
\urldef\tempurl%
\url{https://proceedings.neurips.cc/paper_files/paper/2019/hash/bdbca288fee7f92f2bfa9f7012727740-Abstract.html}
\showURL{%
\tempurl}


\bibitem[Payne and Hanek(1983)]%
        {paynehanek}
\bibfield{author}{\bibinfo{person}{Mary~H. Payne} {and}
  \bibinfo{person}{Robert~N. Hanek}.} \bibinfo{year}{1983}\natexlab{}.
\newblock \showarticletitle{Radian Reduction for Trigonometric Functions}.
\newblock \bibinfo{journal}{\emph{ACM SIGNUM Newslett.}} \bibinfo{volume}{18},
  \bibinfo{number}{1} (\bibinfo{date}{Jan.} \bibinfo{year}{1983}),
  \bibinfo{pages}{19--24}.
\newblock
\showISSN{0163-5778}
\href{https://doi.org/10.1145/1057600.1057602}{doi:\nolinkurl{10.1145/1057600.1057602}}


\bibitem[{Python Software Foundation}(2026)]%
        {python-libm}
\bibfield{author}{\bibinfo{person}{{Python Software Foundation}}.}
  \bibinfo{year}{2026}\natexlab{}.
\newblock \bibinfo{booktitle}{\emph{{Python} Documentation: {math} ---
  Mathematical Functions}}.
\newblock
\urldef\tempurl%
\url{https://docs.python.org/3.15/library/math.html}
\showURL{%
\tempurl}
\newblock
\shownote{For version 3.15}.


\bibitem[Remez(1934)]%
        {remez}
\bibfield{author}{\bibinfo{person}{Evgeny~Y. Remez}.}
  \bibinfo{year}{1934}\natexlab{}.
\newblock \showarticletitle{Sur un proc{\'e}d{\'e} convergent d'approximations
  successives pour d{\'e}terminer les polyn{\^o}mes d'approximation}.
\newblock \bibinfo{journal}{\emph{C.~R. Acad. Sci. Paris}}
  \bibinfo{volume}{198} (\bibinfo{year}{1934}), \bibinfo{pages}{2063--2065}.
\newblock
\urldef\tempurl%
\url{https://gallica.bnf.fr/ark:/12148/bpt6k31506/f2063.item}
\showURL{%
\tempurl}
\newblock
\shownote{In French}.


\bibitem[Rompf and Odersky(2010)]%
        {lms}
\bibfield{author}{\bibinfo{person}{Tiark Rompf} {and} \bibinfo{person}{Martin
  Odersky}.} \bibinfo{year}{2010}\natexlab{}.
\newblock \showarticletitle{Lightweight Modular Staging: A Pragmatic Approach
  to Runtime Code Generation and Compiled DSLs}. In
  \bibinfo{booktitle}{\emph{Proceedings of the 9th International Conference on
  Generative Programming and Component Engineering (GPCE)}}.
  \bibinfo{pages}{127--136}.
\newblock
\href{https://doi.org/10.1145/1868294.1868314}{doi:\nolinkurl{10.1145/1868294.1868314}}


\bibitem[Shewchuk(1997)]%
        {fast2sum}
\bibfield{author}{\bibinfo{person}{Jonathan~R. Shewchuk}.}
  \bibinfo{year}{1997}\natexlab{}.
\newblock \showarticletitle{Adaptive {Precision} {Floating}-{Point}
  {Arithmetic} and {Fast} {Robust} {Geometric} {Predicates}}.
\newblock \bibinfo{journal}{\emph{Discrete \& Computational Geometry}}
  \bibinfo{volume}{18}, \bibinfo{number}{3} (\bibinfo{date}{Oct.}
  \bibinfo{year}{1997}), \bibinfo{pages}{305--363}.
\newblock
\showISSN{0179-5376}
\href{https://doi.org/10.1007/PL00009321}{doi:\nolinkurl{10.1007/PL00009321}}


\bibitem[Sibidanov et~al\mbox{.}(2022)]%
        {coremath}
\bibfield{author}{\bibinfo{person}{Alexei Sibidanov}, \bibinfo{person}{Paul
  Zimmermann}, {and} \bibinfo{person}{St{\'e}phane Glondu}.}
  \bibinfo{year}{2022}\natexlab{}.
\newblock \showarticletitle{The {CORE}-{MATH} {Project}}. In
  \bibinfo{booktitle}{\emph{Proceedings of the 29th IEEE Symposium on Computer
  Arithmetic (ARITH)}}. \bibinfo{pages}{26--34}.
\newblock
\showISBNx{978-1-6654-7827-4}
\href{https://doi.org/10.1109/ARITH54963.2022.00014}{doi:\nolinkurl{10.1109/ARITH54963.2022.00014}}


\bibitem[{Sun Microsystems}(2004)]%
        {libmcr}
\bibfield{author}{\bibinfo{person}{{Sun Microsystems}}.}
  \bibinfo{year}{2004}\natexlab{}.
\newblock \bibinfo{booktitle}{\emph{{libmcr: Correctly Rounded libm}}}.
\newblock
\urldef\tempurl%
\url{https://github.com/simonbyrne/libmcr}
\showURL{%
\tempurl}
\newblock
\shownote{Unofficial copy of the original library}.


\bibitem[Taha and Sheard(1997)]%
        {metaml}
\bibfield{author}{\bibinfo{person}{Walid Taha} {and} \bibinfo{person}{Tim
  Sheard}.} \bibinfo{year}{1997}\natexlab{}.
\newblock \showarticletitle{Multi-Stage Programming with Explicit Annotations}.
  In \bibinfo{booktitle}{\emph{Proceedings of the 1997 ACM SIGPLAN Symposium on
  Partial Evaluation and Semantics-Based Program Manipulation (PEPM)}}.
  \bibinfo{pages}{203--217}.
\newblock
\showISBNx{978-0-89791-917-3}
\href{https://doi.org/10.1145/258993.259019}{doi:\nolinkurl{10.1145/258993.259019}}


\bibitem[Tang(1989)]%
        {tableexp}
\bibfield{author}{\bibinfo{person}{Ping Tak~Peter Tang}.}
  \bibinfo{year}{1989}\natexlab{}.
\newblock \showarticletitle{Table-Driven Implementation of the Exponential
  Function in {IEEE} Floating-Point Arithmetic}.
\newblock \bibinfo{journal}{\emph{ACM Trans. Math. Software}}
  \bibinfo{volume}{15}, \bibinfo{number}{2} (\bibinfo{date}{June}
  \bibinfo{year}{1989}), \bibinfo{pages}{144--157}.
\newblock
\showISSN{0098-3500, 1557-7295}
\href{https://doi.org/10.1145/63522.214389}{doi:\nolinkurl{10.1145/63522.214389}}


\bibitem[Tang(1991)]%
        {tablebased}
\bibfield{author}{\bibinfo{person}{Ping Tak~Peter Tang}.}
  \bibinfo{year}{1991}\natexlab{}.
\newblock \showarticletitle{Table-Lookup Algorithms for Elementary Functions
  and Their Error Analysis}. In \bibinfo{booktitle}{\emph{Proceedings of the
  10th IEEE Symposium on Computer Arithmetic (ARITH)}}.
  \bibinfo{pages}{232--236}.
\newblock
\showISBNx{978-0-8186-9151-5}
\href{https://doi.org/10.1109/ARITH.1991.145565}{doi:\nolinkurl{10.1109/ARITH.1991.145565}}


\bibitem[{The Julia Project}(2026)]%
        {julia-libm}
\bibfield{author}{\bibinfo{person}{{The Julia Project}}.}
  \bibinfo{year}{2026}\natexlab{}.
\newblock \bibinfo{booktitle}{\emph{The {Julia} Language: Mathematical
  Operations and Elementary Functions}}.
\newblock
\urldef\tempurl%
\url{https://docs.julialang.org/en/v1.13/manual/mathematical-operations/}
\showURL{%
\tempurl}
\newblock
\shownote{For version 1.13}.


\bibitem[{The Sage Developers}(2026)]%
        {sage}
\bibfield{author}{\bibinfo{person}{{The Sage Developers}}.}
  \bibinfo{year}{2026}\natexlab{}.
\newblock \bibinfo{booktitle}{\emph{{SageMath}, the {Sage Mathematics Software
  System}}}.
\newblock
\urldef\tempurl%
\url{https://www.sagemath.org/}
\showURL{%
\tempurl}


\bibitem[Zagier(2007)]%
        {dilog}
\bibfield{author}{\bibinfo{person}{Don Zagier}.}
  \bibinfo{year}{2007}\natexlab{}.
\newblock \showarticletitle{The {Dilogarithm} {Function}}.
\newblock In \bibinfo{booktitle}{\emph{Frontiers in {Number} {Theory},
  {Physics}, and {Geometry} {II}}}, \bibfield{editor}{\bibinfo{person}{Pierre
  Cartier}, \bibinfo{person}{Pierre Moussa}, \bibinfo{person}{Bernard Julia},
  {and} \bibinfo{person}{Pierre Vanhove}} (Eds.).
  \bibinfo{publisher}{Springer}, \bibinfo{pages}{3--65}.
\newblock
\showISBNx{978-3-540-30307-7}
\href{https://doi.org/10.1007/978-3-540-30308-4_1}{doi:\nolinkurl{10.1007/978-3-540-30308-4_1}}


\bibitem[Zimmermann(2026)]%
        {atanhcr}
\bibfield{author}{\bibinfo{person}{Paul Zimmermann}.}
  \bibinfo{year}{2026}\natexlab{}.
\newblock \showarticletitle{The {GNU} {libc} {atanh} Is {Correctly} {Rounded}}.
  In \bibinfo{booktitle}{\emph{Proceedings of the 33rd IEEE Symposium on
  Computer Arithmetic (ARITH)}}. \bibinfo{pages}{97--104}.
\newblock
\showISBNx{979-8-3315-8448-1}
\href{https://doi.org/10.1109/ARITH69512.2026.00022}{doi:\nolinkurl{10.1109/ARITH69512.2026.00022}}


\bibitem[Ziv(1991)]%
        {zivs}
\bibfield{author}{\bibinfo{person}{Abraham Ziv}.}
  \bibinfo{year}{1991}\natexlab{}.
\newblock \showarticletitle{Fast Evaluation of Elementary Mathematical
  Functions with Correctly Rounded Last Bit}.
\newblock \bibinfo{journal}{\emph{ACM Trans. Math. Software}}
  \bibinfo{volume}{17}, \bibinfo{number}{3} (\bibinfo{date}{Sept.}
  \bibinfo{year}{1991}), \bibinfo{pages}{410--423}.
\newblock
\showISSN{0098-3500, 1557-7295}
\href{https://doi.org/10.1145/114697.116813}{doi:\nolinkurl{10.1145/114697.116813}}


\end{thebibliography}
